\documentclass[prd,aps,floats,floatfix,superscriptaddress,preprintnumbers,showpacs,eqsecnum,nofootinbib,twocolumn]{revtex4}
\usepackage{amsfonts,amsmath,epsfig,amssymb,latexsym,color,graphicx,array,subfigure,bm,float }

\definecolor{red}{rgb}{1,0,0}
\definecolor{blue }{rgb}{0,0,1}
\definecolor{green}{rgb}{0,1,0}


\newcommand{\bea}{\begin{eqnarray}}
\newcommand{\ena}{\end{eqnarray}}
\newcommand{\beann}{\begin{eqnarray*}}
\newcommand{\enann}{\end{eqnarray*}}

\newcommand{\dsl}{\pa \kern-0.5em /}

\newcommand{\pa}{\partial}

\newcommand{\nn}{\nonumber\\}

\newcommand{\vect}[1]{\!\!\mbox{ \boldmath $#1$}}
\newcommand{\gsim}{\, \mbox{\raisebox{-1.ex}
{$\stackrel{\textstyle>}{\textstyle\sim}$}}\,}

\begin{document}

\date{\today}

\title{Chaotic von Zeipel-Lidov-Kozai Oscillations of 
Binary System around Rotating Supermassive Black Hole}

\author{Kei-ichi Maeda}
\affiliation{Department of Physics, Waseda University, Shinjuku, Tokyo 169-8555, Japan}
\affiliation{Center for Gravitational Physics and Quantum Information, Yukawa Institute
for Theoretical Physics, Kyoto University, 606-8502, Kyoto, Japan}

\author{Priti Gupta}
\affiliation{Department of Physics, Indian Institute of Science, Bangalore 560012, India}

\author{Hirotada Okawa}
\affiliation{Waseda Institute for Advanced Study {\rm (WIAS)}, 1-21-1 Nishi Waseda, Shinjuku-ku, Tokyo 169-0051, Japan }

\begin{abstract}

In this paper, we investigate the dynamics of a binary system that orbits a rotating supermassive black hole. Our approach employs Fermi-Walker transport to construct a local inertial reference frame, and to set up a Newtonian binary system. We consider a scenario in which a circular geodesic observer is positioned around a Kerr black hole, and thereby derive the equations of motion governing the binary system. To eliminate the interaction terms between the center of mass (CM) of the binary and its relative coordinates, we introduce a small acceleration for the observer. This adjustment leads to the CM closely following the observer's orbit, deviating from a circular geodesic. Here, we first focus on elucidating the stability conditions in a hierarchical triple system. Subsequently, we discuss the phenomenon of von Zeipel-Lidov-Kozai (vZLK) oscillations, which manifest when the binary system is compact and the initial inclination exceeds a critical angle. In hard binary systems, these oscillations exhibit regular behavior, while in soft binary systems, they exhibit a chaotic character, characterized by irregular periods and amplitudes, albeit remaining stable. Additionally, we observe an orbital flip under circumstances of large initial inclination. As for the motion of the CM, we observe deviations from a purely circular orbit that transform into stable yet chaotic oscillations characterized by minute amplitude variations.
\end{abstract}

\maketitle



\section{Introduction}
\label{Introduction}
In the wake of the groundbreaking discovery of gravitational waves (GWs) by the LIGO-Virgo-KAGRA Collaboration~\cite{Abbott_2020,Abbott_2021}, the fields of astronomy and physics have embarked on an unprecedented journey. These detections have ushered in a new era, reshaping our comprehension of celestial phenomena in profound ways. One remarkable outcome has been the identification of astrophysical entities characterized by their astonishingly massive stellar-mass black holes (BHs)~\cite{GW150914}. The synergy between gravitational wave observations and electromagnetic counterparts has provided further compelling evidence affirming Einstein's general theory of relativity, as we have observed that the speed of GWs closely aligns with the speed of light~\cite{GW170817}. The increase in detections anticipated in the coming decade offers an opportunity to probe fundamental questions of the utmost significance --- testing of theories of gravity under the strong field regimes. Additionally, unraveling the redshift distribution of black holes (BHs) and their surrounding environments holds great promise~\cite{test1,test2,test3,test4,test5}. To utilise the potential of detections, it is imperative to accurately model the anticipated GW waveforms.

While current observations predominantly emanate from isolated binary systems, it is crucial to remain open to the possibility that nature may unveil more intricate sources. In this context, our focus turns towards the examination of three-body systems. Within the densely populated environs surrounding supermassive black holes (SMBHs) in galactic nuclei, it is conceivable that binary systems may arise, giving rise to natural hierarchical triple systems \cite{Heggie1975,Hut1993,Samsing2014,Riddle2015,Fabio2016,stephan2019}. Recent compelling evidence from LIGO events has hinted at the possibility that hierarchical systems could serve as a prominent formation channel for the merging binary BHs~\cite{sym13091678,Gayathri_2020,Gerosa_2021}. In light of these developments, this paper explores the dynamics within such hierarchical triple systems, shedding light on their potential as sources of gravitational wave signals.

In a hierarchical triple, the distance between two bodies (forming an ‘inner’ binary) is much less than the distance to the third body. von Zeipel was the first one to explore the dynamics of restricted hierarchical triples in 1910, revealing a remarkable phenomenon~\cite{vonZeipel10},
and in 1962, Lidov and Kozai independently rediscovered the same~\cite{Lidov62,Kozai62}, (known as von Zeipel-Lidov-Kozai (vZLK) resonance) — when the two orbits are inclined relative to each other, there is a periodic exchange between orbital eccentricity and relative inclination in secular timescale~\cite{Shevchenko17}. This phenomenon can rise the eccentricity to unity, in turn, giving rise to the high emission of gravitational waves (GWs) lying in the observable band of future space-based detectors~\cite{Amaro-Seoane2010, Antonini2012,hoang18,Antonini2016,Meiron2017,Robson2018,Lisa2018,Lisa2019,Hoang2019,Loeb2019,Gupta_2020,kuntz2022transverse}.

There has been extensive work on dynamics of such systems based on Newtonian or post-Newtonian approximation~\cite{naoz13b,Naoz12,Naoz16,Naoz2020,tey13,Li15,Will14a,Will14b}. Indirect observation of GW from a triple system is also studied by analyzing the cumulative shift of periastron time of a binary pulsar undergoing vZLK oscillations~\cite{Haruka2019,Suzuki:2020zbg}. 
The presence of a heavier tertiary has been considered in previous studies~\cite{Liu:2019tqr,Liu:2021uam} using double-averaged equations of motion to investigate relativistic effects such as de-Sitter and Lens-Thirring precessions. An additional study also investigated 3-body PN (3BpN) secular effect in a hierarchical system with heavy third body using a multiple scale method~\cite{Lim2020}. As well, they pointed out that 3BpN effects affected the evolution of these triples, resulting in a wider range of eccentricity and inclination. Some research studies explores the impact of supermassive black hole spin on nearby binary black hole systems. SMBH spin (Lense-Thirring precession and gravitomagnetic force in particular) affects BBH eccentricity and orbital inclination, potentially altering BBH merger times~\cite{Fang_2019a,Fang_2019b}. 

In our approach, the binary system is treated as perturbations of SMBH spacetime. When dealing with a single object orbiting the SMBH, it can be regarded as a test particle, subject to the gravitational influence of the central black hole. However, the dynamics become notably more intricate in the case of a binary system because the self-gravitational mutual interaction is much stronger than the gravitational tidal force by SMBH. In order to analyze such a hierarchical system, we first prepare a local inertial frame and set up a binary in this frame. When a binary is tightly bounded but the mutual gravitational interaction is not so strong, the binary motion can be discussed by Newtonian gravitational dynamics.

Using Fermi normal coordinate system or Fermi-Walker transport, 
we can construct a local inertial frame~\cite{1963JMP.....4..735M,Nesterov_1999,Delva:2011abw}. 
Using such a technique, there are several discussions on a tidal force acting on stars near SMBH~\cite{Banerjee_2019,PhysRevD.71.044017,Cheng_2013,Kuntz_2021}, 
and a few works on a binary system have been discussed~\cite{Gorbatsievich_Bobrik,Chen_Zhang,camilloni2023tidal}. 
In the previous paper, assuming that SMBH is described by a 
spherically symmetric Schwarzschild spacetime, 
we analyze dynamics of such a system in detail~\cite{Maeda:2023tao}. We showed that the vZLK oscillations appear even near 
the innermost stable circular orbit (ISCO) radius 
when a binary is compact enough and the inclination angle is larger than the critical value. Although  the oscillations are regular for a highly compact binary, when a binary is softer,  we find the chaotic vZLK oscillations, i.e., the oscillations 
become irregular both in the oscillation period and in the amplitude. Especially, if the initial inclination is large, we find an orbital flip. However SMBH may be rotating in nature.
Hence in this paper, we extend our analysis into a rotating Kerr SMBH.

The paper is organized as follows: We summarize our method 
discussed in the previous paper, which can be applied to 
any background spacetime,  in Sec. II. 
In Sec. III, assuming an observer moving along a circular geodesic in Kerr black hole, 
we derive the equations of motion for a binary system. 
We also discuss the interaction terms between the center of mass (CM) of a binary and its relative coordinates.  Introducing small acceleration of an observer, we remove 
the interaction terms, 
finding the equations of motion for the CM, which gives 
small deviations from a circular geodesic.
In Sec. IV, for our numerical analysis, we rewrite  
the equations of motion in dimensionless form, 
introduce the orbital parameters for analysis, and 
show how to prepare the initial data.
In Sec. VA, we analyze many models numerically 
to find  the stability conditions for a hierarchical triple system.
We then show the properties of binary motions
such as the vZLK oscillations, 
chaotic features, 
and orbital flips. 
Summary and discussion follow in Sec. VI.
In the Appendix A, we 
solve motions of the CM of a binary and show its stability. 
We also present the Lagrange planetary equations of the model 
and write down the equations for the orbital parameters of a binary 
taking averages over inner and outer binary cycles in Appendix B.
We show that this simplified method recovers numerical results obtained by direct integration of the equations of motion in the case of a hard binary.
It also provides the vZLK oscillation time scale and 
the maximum and minimum values of eccentricity. 

\textit{Notation used}: Greek letters range from 0 to 3, while Roman letters run from 1 to 3; Hatted indices denote tetrad components in a proper reference frame rotating along an observer; Bar over symbols correspond to quantities in a static tetrad frame. We use $G = c = 1$ unless specified otherwise.

\section{Binary system in a curved spacetime}
\label{binary_in_curved_ST}
We first summarize how to calculate a binary motion near a supermassive black hole (SMBH), which was described in details  in paper I~\cite{Maeda:2023tao}.
We discuss a binary system in a fixed curved background.
A binary consists of two point particles with the masses $m_1$ and $m_2$.
In order to solve a binary motion around SMBH, we set up a local inertial frame~\cite{Misner1973,1963JMP.....4..735M}, 
and then put a self-gravitating binary system, which follows approximately 
Newtonian  dynamics in a local inertial frame~\cite{Gorbatsievich_Bobrik}.

\subsection{Proper reference frame}
\label{proper reference frame}
The background spacetime  metric, which describes SMBH, 
 is given by
$
d\bar s^2=\bar g_{\mu\nu}dx^\mu dx^\nu
$. 
We consider  an observer, 
whose orbit  is given by 
a world line $ \gamma$ described by 
$
x^\mu=z^\mu(\tau)
$, 
where $\tau$ is a proper time of the observer. 
The 4-velocity of the observer is given by
$
u^\mu(\tau)\equiv {dz^\mu/ d\tau}
$.

We then prepare an orthonormal tetrad system $\{e_{\hat \alpha}^\mu\}$ along   $\gamma$, which is defined  by
the conditions such that
$
e_{\hat \alpha}^{~\mu} e_{\hat \beta \mu}=\eta_{\hat \alpha \hat \beta}\,,~~e_{\hat 0}^{~\mu}=u^\mu
$, 
where $\eta_{\hat \alpha \hat \beta}$ is 
Minkowski spacetime metric.

For a given 4-velocity $u^\mu$, this tetrad system is determined up to three-dimensional rotations.
The tetrad $e_{\hat \alpha}^\mu$  is transported  along  $ \gamma$ 
as

\beann
\frac{D e_{\hat i}^{~\mu}}{d\tau}=-\left(a^\mu  u^\nu-u^\mu a^\nu  +u_\alpha \omega_\beta \epsilon^{\alpha\beta\mu\nu}\right)\,e_{\hat i \,\nu}
\,,
\enann
where 
$
a^\mu \equiv  {Du^\mu/ d\tau}
$
and 
$
\omega_\mu
$
are  the acceleration of the observer
and  the angular velocity of a rotating 
spatial basis vector $e_{(a)}^\mu$, respectively. 

Next, we construct a local coordinate system (the observer's proper reference system) 
near the world line $\gamma$, which is described as
$
(x^{\hat\mu})=(c\tau, x^{\hat a})
$\,,
where the spatial components $x^{\hat a}$ is measured from the point at $\tau$ on the world line $\gamma$ along the spatial hyper-surface $\Sigma(\tau)$ perpendicular to $\gamma$.
\begin{widetext}

The metric form of this proper reference frame 
up to the second order of $x^{\hat a}$ 
is given by
\bea
g_{\hat \mu\hat \nu}=\eta_{\hat \mu \hat \nu}+\varepsilon_{\hat \mu \hat \nu}
+O(|x^{\hat k}|^3),
\label{metric_reference_frame}
\ena 
where
\bea
\varepsilon_{\hat 0 \hat 0}&=&-\frac{1}{c^2}\left[2a_{\hat k}
x^{\hat k}
+\left(c^2
\bar{\cal R}_{\hat 0 \hat k \hat 0 \hat \ell }-
\omega_{\hat j\hat k} \omega^{\hat j}_{~\hat \ell} \right)
x^{\hat k}  x^{\hat \ell} +\frac{\left(a_{\hat k} x^{\hat k}\right)^2}{c^2}   \right],
\label{epsilon00}\\
\varepsilon_{\hat 0 \hat j}&=&-\frac{1}{c^2}\left[c\, \omega_{\hat j\hat k} x^{\hat k} 
 +\frac{2}{3}c^2 \bar{\cal R}_{\hat 0\hat k \hat j  \hat \ell }x^{\hat k}x^{\hat \ell}\right],
\label{epsilon0j}\\
\varepsilon_{\hat i \hat j}&=&-\frac{1}{c^2}\left[\frac{1}{3}c^2\bar{\cal R}_{\hat i \hat k \hat j\hat \ell}x^{\hat k}x^{\hat \ell}
\right]
\,.
\label{epsilonij}
\ena
with
$\bar{\cal R}_{\hat \mu\hat\nu\hat\rho\hat \sigma}$ being the tetrad component of the Riemann curvature of a background spacetime
and
$
\omega_{\hat j\hat k}\equiv \epsilon_{\hat j \hat k \hat \ell}\omega^{\hat \ell}
$~\cite{Misner1973,1963JMP.....4..735M,Gorbatsievich_Bobrik}.
\end{widetext}

The acceleration and angular frequency in the proper reference frame are defined by
\beann
a^{\hat j}&\equiv & e^{\hat j}_{~\mu} \frac{Du^\mu}{d\tau},
\\
\omega^{\hat j}&\equiv &{1\over 2}\stackrel{(3)}{\epsilon}^{\hat j\hat k\hat \ell}e_{\hat \ell \mu}{De_{\hat k}^{~\mu}\over d\tau}
\,,
\enann
where $\stackrel{(3)}{\epsilon}^{\hat j\hat k\hat \ell}\equiv e^{\hat j}_{~\nu}e^{\hat k}_{~\rho}e^{\hat \ell}_{~\sigma}u^\mu \epsilon_{\mu\nu\rho\sigma}$. 
If the observer's orbit  is the geodesic, we recover the Fermi normal coordinates.

\subsection{Self-gravitating binary system}
\label{binary}

Now we discuss a self-gravitating binary system in a fixed curved background spacetime~\cite{Gorbatsievich_Bobrik}. 
We are interested in the case where Newtonian dynamics is valid 
 in the observer's proper reference frame. 
 The necessary condition is 
that the typical scale  $\ell_{\rm binary}$  of a binary system should satisfy 
\beann
\ell_{\rm binary}\ll 
{\rm min} \left[{1\over |a^{\hat j}|}, {1\over |\omega^{\hat j}|}, \ell_{\bar{\cal R}}\right],
\enann
where $\ell_{\bar{\cal R}}$ is the minimum curvature radius defined by
\beann
\ell_{\bar{\cal R}} \equiv {\rm min} \left[|\bar{\cal R}_{\hat \mu\hat \nu\hat \rho\hat \sigma}|^{-{1\over 2}}, |\bar{\cal R}_{\hat \mu\hat \nu\hat \rho\hat \sigma ; \hat \alpha}|^{-{1\over 3}}, |\bar{\cal R}_{\hat \mu\hat \nu\hat \rho\hat \sigma ; \hat \alpha;\hat \beta}|^{-{1\over 4}}
\right]
\,.
\enann

A gravitational interaction in a self-gravitating  
binary system can be described by the metric deviation from 
 a local Minkowski spacetime.
 For example, to discuss 
 Newtonian dynamics of the particle 1, 
it is enough to consider the $00$ component of the metric perturbation, 
$\varphi^{(1)}_{\hat 0\hat 0}=-2\Phi^{(1)}/c^2$, where $\Phi^{(1)}$ is 
the Newtonian potential
of the particle 1, which is
\beann
\Phi^{(1)}(x^{\hat i})=-{Gm_2\over |x_1^{\hat i}-x_2^{\hat i}|}
\,.
\enann

We then obtain the equation of motion for  the particle 1 in
 the observer's proper reference frame
can be derived by the variation with respect to $x_1^{\hat i}$ of the action
\beann
{\cal S}^{(1)}=\int d\tau{\cal L}^{(1)}\,,
\enann
where
\beann
{\cal L}^{(1)}\equiv -m_1c\sqrt{-g^{(1)}_{\hat \mu\hat \nu}{dx_1^{\hat \mu}\over d\tau}{dx_1^{\hat \nu}\over d\tau}}
\,.
\enann 
where the metric $g^{(1)}_{\hat \mu\hat \nu}$ is given by
 \bea
g^{(1)}_{\hat \mu \hat \nu}=\eta_{\hat \mu \hat \nu}+\varepsilon_{\hat \mu \hat \nu}+\varphi^{(1)}_{\hat \mu \hat \nu}.
\label{metric_g+h}
\ena
We also find the similar action for the particle 2. Next, we perform  the post-Newtonian expansion of the total Lagrangian 
${\cal L}={\cal L}^{(1)}+{\cal L}^{(2)}$. 
 The Lagrangian for a binary up to 0.5 PN order is given by
\bea
{\cal L}_{\rm binary}={\cal L}_{\rm N}+{\cal L}_{1/2},
\label{Lagrangian_binary}
\ena
where
\bea
{\cal L}_{\rm N}&\equiv& {1\over 2} \sum_{I=1}^2
m_I \dot{\vect{x}}_I^2
+  {G m_1m_2\over |\vect{x}_1-\vect{x}_2|}
\nn
&&
~~~~~~~~~~
+{\cal L}_{a}+{\cal L}_{\omega}+{\cal L}_{\bar{\cal R}}
\label{Lagrangian_N},
\ena
with
\beann
{\cal L}_{a}
&=&
-\sum_{I=1}^2 m_I a_{\hat k}x_I ^{\hat k},
\\
{\cal L}_{\omega}
&=&
-\sum_{I=1}^2 m_I \left[\epsilon_{\hat j\hat k\hat \ell}\omega^{\hat \ell}
x_I^{\hat k}\dot x_I^{\hat j}-
{1\over 2} \left(\vect{\omega}^2 \vect{x}_I^2 -(\vect{\omega} \cdot \vect{x}_I)^2
\right)\right],
\\
{\cal L}_{\bar{\cal R}}
&=&
-{1\over 2}  \sum_{I=1}^2 m_I \bar{\cal R}_{\hat 0\hat k\hat 0\hat \ell}x_I^{\hat k} x_I^{\hat \ell}
\,,
\enann
and 
\bea
{\cal L}_{1/2}&\equiv &-{2\over 3}\sum_{I=1}^2
m_I c^2\bar{\cal R}_{\hat 0\hat k \hat j  \hat \ell }
x_I^{\hat k}x_I^{\hat \ell}\, {\dot x_I^{\hat j}\over c}
\,.
\label{Lagrangian_0.5PN}
\ena
 A dot ($\cdot$) denotes the proper time derivative $d/d\tau$.


Introducing the center of mass (CM) 
coordinates and the relative coordinates by
\beann
\vect{R}&=&{m_1\vect{x}_1+m_2\vect{x}_2\over m_1+m_2},
\\
\vect{r}&=& \vect{x}_2-\vect{x}_1
\,,
\enann
we find the Newtonian Lagrangian (Eq.~\eqref{Lagrangian_N}) in terms of  $\vect{R}$  and $\vect{r}$ as
\bea
{\cal L}_{\rm N}={\cal L}_{\rm CM}(\vect{R}, \dot{\vect{R}})+
{\cal L}_{\rm rel}(\vect{r}, \dot{\vect{r}})
\label{Lagrangian_N_Rr}
\,,
\ena
\begin{widetext}
where
\beann
{\cal L}_{\rm CM}(\vect{R},\dot{\vect{R}})
&=&{1\over 2} (m_1+m_2)\dot{\vect{R}}^2
+{\cal L}_{{\rm CM}\mathchar`-a}(\vect{R},\dot{\vect{R}})
+{\cal L}_{{\rm CM}\mathchar`-\omega}(\vect{R},\dot{\vect{R}})
+{\cal L}_{{\rm CM}\mathchar`-\bar{\cal R}}(\vect{R},\dot{\vect{R}}),
\enann
with 
\beann
{\cal L}_{{\rm CM}\mathchar`-a}
&=&-(m_1+m_2)a_{\hat k}x_I ^{\hat k},
\\
{\cal L}_{{\rm CM}\mathchar`-\omega}
&=&-(m_1+m_2)\left[
\epsilon_{\hat j\hat k\hat \ell}\omega^{\hat \ell}
R^{\hat k}\dot R^{\hat j}-{1\over 2}\left(\vect{\omega}^2 \vect{R}^2
-\left(\vect{\omega}\cdot \vect{R}\right)^2\right)\right],
\\{\cal L}_{{\rm CM}\mathchar`-\bar{\cal R}}
&=&
-{1\over 2}(m_1+m_2)
\bar{\cal R}_{\hat 0\hat k \hat 0 \hat \ell}R^{\hat k}R^{\hat \ell},
\enann
and
 \beann
{\cal L}_{\rm rel}(\vect{r},\dot{\vect{r}})&=&
{1\over 2}\mu \dot{\vect{r}}^2+  {G m_1m_2\over r}
+{\cal L}_{{\rm rel}\mathchar`-\omega}(\vect{r},\dot{\vect{r}})
+{\cal L}_{{\rm rel}\mathchar`- \bar{\cal R}}(\vect{r},\dot{\vect{r}}),
\enann
with 
\beann
{\cal L}_{{\rm rel}\mathchar`-\omega}
&=&-\mu \left[
\epsilon_{\hat j\hat k\hat \ell}\omega^{\hat \ell}
r^{\hat k}\dot r^{\hat j}-{1\over 2}\left(\vect{\omega}^2 \vect{r}^2
-\left(\vect{\omega}\cdot \vect{r}\right)^2\right)\right]\,,
\\
{\cal L}_{{\rm rel}\mathchar`-\bar{\cal R}}
&=&
-{1\over 2}\mu 
\bar{\cal R}_{\hat 0\hat k \hat 0 \hat \ell}r^{\hat k}r^{\hat \ell}
\,.
\enann
Here, $\mu = m_1 m_2/(m_1+m_2)$ is the reduced mass. When we consider only ${\cal L}_{\rm N}$, we can separate the variables $\vect{R}$ and $\vect{r}$. In particular, when the observer trajectory is a geodesic ($\vect{a}=0$ and $\vect{\omega}=0$),
the orbit of $\vect{R}=0$ becomes a solution of the equation for $\vect{R}$. This implies that the CM follows the observer's geodesic. Consequently, our analytical focus narrows down to the equation dictating the behavior of the relative coordinate $\vect{r}$. However, when we include the 0.5 PN term, it is not the case. The 0.5PN Lagrangian ${\cal L}_{1/2}$  expression is written by use of  $\vect{R}$  and    $\vect{r}$ as follows
\bea
{\cal L}_{1/2}={\cal L}_{1/2\mathchar`-{\rm CM}}(\vect{R}, \dot{\vect{R}})+
{\cal L}_{1/2\mathchar`-{\rm rel}}(\vect{r}, \dot{\vect{r}})
+{\cal L}_{1/2\mathchar`- {\rm int}}(\vect{R}, \dot{\vect{R},}\vect{r}, \dot{\vect{r}}),
\label{Lagrangian_1/2_Rr}
\ena
where
\bea
{\cal L}_{1/2\mathchar`-{\rm CM}}(\vect{R}, \dot{\vect{R}})&=&-{2\over 3}(m_1+m_2)\bar{\cal R}_{\hat 0\hat k \hat j \hat \ell}
R^{\hat k}R^{\hat \ell}\dot R^{\hat j},
\nn
{\cal L}_{1/2\mathchar`-{\rm rel}}(\vect{r}, \dot{\vect{r}})&=&- {2\over 3} \mu{(m_1-m_2)\over (m_1+m_2)}\bar{\cal R}_{\hat 0\hat k \hat j \hat \ell}r^{\hat k}r^{\hat \ell}\dot r^{\hat j},
\nn
{\cal L}_{1/2\mathchar`- {\rm int}}(\vect{R}, \dot{\vect{R},}\vect{r}, \dot{\vect{r}})&=&
- {2\over 3} \mu \bar{\cal R}_{\hat 0\hat k \hat j \hat \ell}
\left[r^{\hat k}r^{\hat \ell}\dot R^{\hat j}
+\left(R^{\hat k}r^{\hat \ell}+r^{\hat k}R^{\hat \ell}
\right)\dot r^{\hat j}\right].
\label{interaction_term}
\ena

The interaction term (Eq.~\eqref{interaction_term}) invalidates the $\vect{R}=0$ orbit as a solution, even when acceleration is absent. The coupling between the CM motion ($\vect{R}(\tau)$) and relative motion ($\vect{r}(\tau)$) makes both binary and CM trajectories intricate, even when the observer follows a geodesic. However, if we introduce an appropriate acceleration $\vect{a}$
in 0.5PN order to cancel the interaction terms, $\vect{R}=0$ will become a solution, i.e., the CM can follow the observer's motion as follows:
Integrating by parts the interaction term, we find
 \beann
{\cal L}_{1/2\mathchar`- {\rm int}}(\vect{R}, \dot{\vect{R},}\vect{r}, \dot{\vect{r}})&=&
-{2\over 3}\mu \bar{\cal R}_{\hat 0\hat k \hat j \hat \ell}
\left[\dot R^{\hat j} r^{\hat k}r^{\hat \ell}
+\dot r^{\hat j}\left(R^{\hat k}r^{\hat \ell}+r^{\hat k}R^{\hat \ell}
\right)\right]
\\
&\approx&
2\mu\left[{1\over 3}{d\bar{\cal R}_{\hat 0\hat k \hat j \hat \ell}\over d\tau}
r^{\hat k}r^{\hat \ell} 
+ \bar{\cal R}_{\hat 0\hat k \hat j \hat \ell}
r^{\hat k}\dot r^{\hat \ell} \right]R^{\hat j}~~{\rm (integration~by~part)}
\,,
\enann
where the time derivative of the curvature is evaluated along the observer's orbit.

If we define the acceleration of the observer by
 \beann
 a_{\hat j}={2\mu \over m_1+m_2}\left[{1\over 3}{d\bar{\cal R}_{\hat 0\hat k \hat j \hat \ell}\over d\tau}
r^{\hat k}r^{\hat \ell} 
+ \bar{\cal R}_{\hat 0\hat k \hat j \hat \ell}
r^{\hat k}\dot r^{\hat \ell} \right]
\label{0.5PN_acceleration}
\,,
 \enann
two terms ${\cal L}_{1/2\mathchar`- {\rm int}}$ and ${\cal L}_{{\rm CM}\mathchar`-a}$ cancel each other.
As a result, the Lagrangians for $\vect{R}$ and $\vect{r}$ are decoupled,
and  $\vect{R}=0$ becomes an exact solution of the equation for $\vect{R}$, which is
derived from the Lagrangian
(${\cal L}_{\rm CM}+{\cal L}_{1/2\mathchar`-{\rm CM}}$).
The CM follows the observer's orbit and therefore, we obtain the decoupled equation for the relative coordinate $\vect{r}$.

To determine the proper observer's orbit, which deviates from a geodesic but remains in proximity to it, we must solve the equation of motion that accounts the small acceleration such that
 \bea
 {Du_{\rm CM}^\mu\over d\tau}=a^\mu=e^{\mu\hat j } a_{\hat j}=
{2\mu \over m_1+m_2}e^{\mu\hat j }\left[{1\over 3}{d\bar{\cal R}_{\hat 0\hat k \hat j \hat \ell}\over d\tau}
r^{\hat k}r^{\hat \ell} 
+ \bar{\cal R}_{\hat 0\hat k \hat j \hat \ell}
r^{\hat k}\dot r^{\hat \ell} \right]
\,.
\label{eq_CM}
 \ena
This equation can also be expressed as:
 \bea
 {Dp_{\rm CM}^\mu\over d\tau}=(m_1+m_2)a^\mu
=
 e^{\mu\hat j } 
\left[
{1\over 2}\bar{\cal R}_{\hat 0\hat j \hat k \hat \ell}
S^{\hat k\hat \ell}
+\bar{\cal R}_{\hat 0\hat k \hat j \hat \ell}
\dot M^{\hat k\hat \ell}+
{2\over 3}{d\bar{\cal R}_{\hat 0\hat k \hat j \hat \ell}\over d\tau}
M^{\hat k\hat \ell} \right]
\,,
\label{eq_CM2}
 \ena
 where $p_{\rm CM}\equiv (m_1+m_2)u_{\rm CM}$ is the CM 4-momentum,
 $S^{\hat k\hat \ell}\equiv r^{\hat k}p^{\hat \ell}-r^{\hat \ell}p^{\hat k}$
 is the angular momentum tensor of a binary, and $M^{\hat k\hat \ell} \equiv \mu r^{\hat k}r^{\hat \ell}$ is the second mass moment, which can be replaced by the mass quadruople moment $Q^{\hat k\hat \ell} \equiv r^{\hat k}r^{\hat \ell}-{1\over 3}r^2\delta^{\hat k \hat \ell}$ in the Ricci-flat vacuum background.
 The first term in the right hand side is the similar to 
 the spin-curvature coupling term 
 appeared in the Mathisson–Papapetrou–Dixon equations of a spinning test particle in a curved spacetime \cite{Mathisson1937,Papapetrou1951,Dixon1974}.
 
Consequently, our initial step involves solving the equation for the relative coordinate, denoted as $\vect{r}$, which is governed solely by the Lagrangian ${\cal L}_{\rm rel}(\vect{r})
 +{\cal L}_{1/2\mathchar`-{\rm rel}}(\vect{r})$.
 Notably, when the masses are equal, i.e., $m_1=m_2$, we have only Newtonian 
 Lagrangian ${\cal L}_{\rm rel}$ because ${\cal L}_{1/2\mathchar`-{\rm rel}}$ vanishes. Once we have obtained the solution for $\vect{r}(\tau)$, we proceed to determine the motion of the center of mass (CM) or the observer within the background spacetime by addressing Eq.~\eqref{eq_CM}. Employing the solution for the relative motion $\vect{r}(\tau)$ in conjunction with the CM motion solution, represented as $x_{\rm CM}^\mu(\tau)$, we can deduce the binary system's trajectory within the specified curved background spacetime, denoted as $x_1^\mu(\tau)$ and $x_2^\mu(\tau)$.

\section{Equation of motion of a binary system in a Kerr spacetime}
\label{EOM_Kerr}
 Now we consider a rotating SMBH as the background spacetime, which
 is given by Kerr solution in Boyer-Lindquist coordinates as
\bea
d\bar s^2
&=&
-{\Delta \over \Sigma}\left(d\mathfrak{t}-a\sin^2\theta d\phi\right)^2+{\sin^2\theta\over \Sigma}\left[
(\mathfrak{r}^2+a^2)d\phi-ad\mathfrak{t}\right]^2+{\Sigma\over \Delta}d\mathfrak{r}^2+\Sigma d\theta^2
\label{Kerr_BL}
\,,
\ena
where
\beann
\Sigma=\mathfrak{r}^2+a^2\sin^2\theta
~~{\rm and}~~
\Delta= \mathfrak{r}^2-2M\mathfrak{r}+a^2
\,.
\enann
 $M$ and $a$ are a gravitational mass and proper angular momentum of a supermassive black hole, respectively.

\end{widetext}

\subsection{A test particle on the equatorial plane}
We consider a circular geodesic on the equatorial plane 
of a test particle with a unit mass.
The proper energy and proper angular momentum are determined by the radius $ \mathfrak{r}_0$ as
\bea
\mathfrak{E}&=&{\mathfrak{r}_0^2-2M\mathfrak{r}_0+a\sigma \sqrt{M\mathfrak{r}_0}\over \mathfrak{r}_0 F_\sigma(\mathfrak{r}_0)},
\label{energy}
\\
\mathfrak{L}&=& {\sigma\sqrt{M\mathfrak{r}_0}
(\mathfrak{r}_0^2+a^2- 2a\sigma\sqrt{M\mathfrak{r}_0})\over \mathfrak{r}_0 F_\sigma(\mathfrak{r}_0)}
\,,
\label{angular_momentum}
\ena
where 
\beann
F_\sigma(\mathfrak{r}_0)\equiv \left(\mathfrak{r}_0^2-3M\mathfrak{r}_0+ 2a\sigma\sqrt{M\mathfrak{r}_0}\right)^{1/2},
\enann
and
$\sigma=+1$ or $-1$, which 
 correspond to prograde and retrograde orbits, respectively.
The existence condition of a circular orbit is 
\beann
\mathfrak{r}_0^2-3M\mathfrak{r}_0+2a\sigma \sqrt{M\mathfrak{r}_0}\geq 0
\,.
\enann
The innermost stable circular orbit (ISCO) is 
obtained by the conditions such that
\beann
{d\mathfrak{E}\over d\mathfrak{r}_0}=0
\,\,,~~
{d\mathfrak{L}\over d\mathfrak{r}_0}=0\,,
\enann
which gives~\cite{Bardeen1972}
\beann
{\mathfrak{r}_{\rm ISCO}\over M}
=3+Z_2-\sigma\left[(3-Z_1)(3+Z_1+2Z_2)\right]^{1/2},
\enann
where
\beann
Z_1&=&1+\left(1-{\chi^2}\right)^{1/3}\left[\left(1-\chi\right)^{1/3}+\left(1+\chi\right)^{1/3}\right],
\\
Z_2&=&\left(3{\chi^2}+Z_1^2\right)^{1/2},
\enann
with 
$
\chi\equiv {a/ M}.
$


Since the energy and angular momentum are defined by
\beann
\mathfrak{E}=-u_{\mathfrak{t}}
\,,~
\mathfrak{L}=u_{\phi}
\,,
\enann
we find
\beann
u^{\mathfrak{t}}&=&{\mathfrak{r}_0^2+a\sigma\sqrt{M\mathfrak{r}_0}\over \mathfrak{r}_0 F_\sigma(\mathfrak{r}_0)},
\\
u^{\phi}&=&{\sigma\sqrt{M\mathfrak{r}_0}\over 
\mathfrak{r}_0 F_\sigma(\mathfrak{r}_0)}.
\enann
From the latter equation, 
we find  the angular velocity of a circular observer measured by the proper time as
\beann
 {d\phi\over d\tau}=u^{\phi}=\sigma  \mathfrak{w}_0,
 \enann
 where
 \beann
 \mathfrak{w}_0\equiv
{\sqrt{M\mathfrak{r}_0}\over 
\mathfrak{r}_0 F_\sigma(\mathfrak{r}_0)}
\,.
\enann

\subsection{A local inertial reference frame}
For the present purpose, there is one convenient  tetrad system of  Kerr spacetime,
which is called Carter's tetrad system such that
\beann
e_{~\bar{\mathfrak{t}}}^\mu&=&{1\over \sqrt{\Sigma\Delta}}\left(r^2+a^2, 0, 0, a\right),
\\
e_{~\bar{\mathfrak{r}}}^\mu&=&\sqrt{\Delta\over \Sigma}\left(0,1,0,0\right),
\\
e_{~\bar \theta}^\mu&=& {1\over\sqrt{\Sigma}}\left(0, 0, 1, 0\right),
\\
e_{~\bar \phi}^\mu&=&{1\over \sqrt{\Sigma}}\left(a\sin\theta, 0, 0, {1\over \sin\theta}\right)
\,.
\enann

Now we construct a local inertial reference frame along the observer.
The transformation matrix from Carter's tetrad to 
a rotating proper reference frame $(\tau, x, y, z)$
is given by
\beann
\Lambda_{\hat 0}^{~\bar \alpha}&=&\left({\mathfrak{E}(\mathfrak{r}_0^2+a^2)-a\mathfrak{L})\over \mathfrak{r}_0\sqrt{\Delta}}\,,0\,,0\,,{\mathfrak{L}-a\mathfrak{E}\over \mathfrak{r}_0}\right),\\
\Lambda_{\hat x}^{~\bar \alpha}&=&{1\over \sqrt{S}}\left(0\,,
\mathfrak{r}_0\Lambda_{\hat 0}^{~\bar 0}\,,0\,,0\right),
\\
\Lambda_{\hat y}^{~\bar \alpha}&=&\sigma\sqrt{K\over S}\left(\Lambda_{\hat 0}^{~\bar 0}\,,0\,,0\,,{S\over K}\Lambda_{\hat 0}^{~\bar \phi}\right),
\\
\Lambda_{\hat z}^{~\bar \alpha}&=&
{\sigma\over \sqrt{K}}\left(0\,,0\,,-\mathfrak{r}_0
\Lambda_{\hat 0}^{~\bar \phi}\,,0\right)
\,,
\enann
where
\beann
K&\equiv&\left(\mathfrak{L}-a\mathfrak{E}\right)^2,
\\
S&\equiv&\mathfrak{r}_0^2+K
\,.
\enann
$K$ is the so-called Carter's constant.

\begin{figure}[htbp]
\begin{center}
\includegraphics[width=4cm]{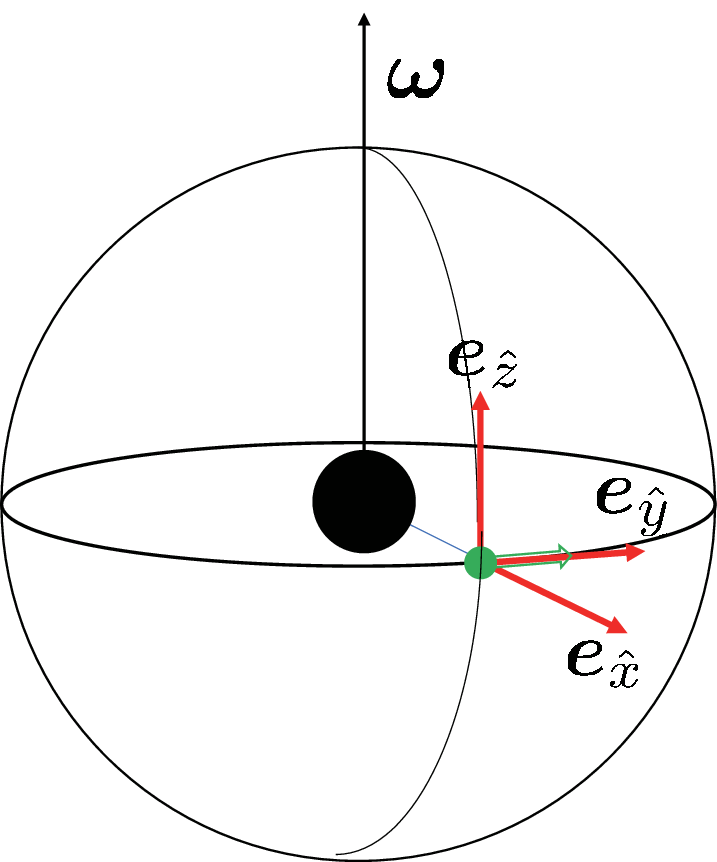}
\caption{A tetrad system $\{
\vect{e}_{\hat x}\,, \vect{e}_{\hat y}\,, \vect{e}_{\hat z} \}$ rotating with an angular velocity $\mathfrak{w}_0$ along a circular orbit. }
\label{tetrad}
\end{center}
\end{figure}

Here, we choose the Descartes coordinate ($x,y,z$)
as shown in Fig. \ref{tetrad}.

Inserting Eqs.(\ref{energy}) and (\ref{angular_momentum}),
we obtain the transformation matrix $\Lambda_{\hat \alpha}^{~\bar \lambda}$ as
\bea
\Lambda_{\hat 0}^{~\bar \lambda}&=&
{1\over F_\sigma(\mathfrak{r}_0)}\left(\sqrt{\Delta(\mathfrak{r}_0)}
\,, 0\,, 0\,,\sigma\sqrt{M\mathfrak{r}_0}-a\right),
~~~~~
\label{Lambda_hat0}\\
\Lambda_{\hat x}^{~\bar \lambda}&=& 
\left(0\,,1\,, 0\,,0\right),
\label{Lambda_hatx}\\
\Lambda_{\hat y}^{~\bar \lambda}&=& {1\over F_\sigma(\mathfrak{r}_0)}
\left(\sigma\sqrt{M\mathfrak{r}_0}-a
\,, 0\,, 0\,,\sqrt{\Delta(\mathfrak{r}_0)}
\right),~~~~~~~~
\label{Lambda_haty}\\
\Lambda_{\hat z}^{~\bar \lambda}&=& \left(0\,, 0\,, -1\,,0\right)
\,,
\label{Lambda_hatz}\ena
which provides the tetrad of the rotating proper reference frame $e_{\hat \alpha}^{~\mu}\equiv 
\Lambda_{\hat \alpha}^{~\bar \lambda}e_{\bar \lambda}^{~\mu}
$ as
\begin{widetext}
\beann
e_{\hat 0}^{~\mu}&\equiv&
{1\over \mathfrak{r}_0F_\sigma(\mathfrak{r}_0)}\left(\mathfrak{r}_0^2+a\sigma\sqrt{M\mathfrak{r}_0}
\,, 0\,, 0\,,\sigma\sqrt{M\mathfrak{r}_0}\right),
\\
e_{\hat x}^{~\mu}&=& 
\left(0\,,{\sqrt{\Delta(\mathfrak{r}_0)}\over \mathfrak{r}_0}\,, 0\,,0\right),
\\
e_{\hat y}^{~\mu}&=& {1\over \mathfrak{r}_0 F_\sigma(\mathfrak{r}_0)\sqrt{\Delta(\mathfrak{r}_0)}}
\left(\sigma\sqrt{M\mathfrak{r}_0}\left[\mathfrak{r}_0^2+a^2-2a\sigma\sqrt{M\mathfrak{r}_0}\right]\,, 0\,, 0\,,
\mathfrak{r}_0^2-2M\mathfrak{r}_0+a\sigma\sqrt{M\mathfrak{r}_0}\right),
\\
e_{\hat z}^{~\mu}&=& \left(0\,, 0\,, -{1\over \mathfrak{r}_0}\,,0\right)
\,.
\enann
\end{widetext}

To discuss the dynamics of a binary,  non-rotating proper reference frame is more convenient. 
We then rotate the $xy$-plane as
($(x,y) \rightarrow (\mathsf{x},\mathsf{y})$), with transformation given by 

\beann
\tilde \Lambda_{\hat {\mathsf{x}}}^{~\bar \alpha}
&=&\Lambda_{\hat x}^{~\bar \alpha}\, \cos \Psi (\tau)-\Lambda_{\hat y}^{~\bar \alpha}\,\sin\Psi(\tau),
\\
\tilde \Lambda_{\hat{\mathsf{y}}}^{~\bar \alpha}
&=&\Lambda_{\hat x}^{~\bar \alpha}\, \sin \Psi (\tau)+\Lambda_{\hat y}^{~\bar \alpha}\,\cos\Psi(\tau)
\,,
\enann
where the rotation angle $\Psi$ satisfies 
the evolution equation such that
\beann
\dot{\Psi}&=&
\sigma
{\sqrt{K}\over \mathfrak{r}_0^2}\left({\mathfrak{E}(\mathfrak{r}_0^2+a^2)-a\mathfrak{L}\over S}
+{a(\mathfrak{L}-a\mathfrak{E})\over K}
\right)
\\[.5em]
&=&\sigma\mathfrak{w}_{\rm R}
\,,
\enann
with
\vskip -.5cm
\beann
\mathfrak{w}_{\rm R} ={M^{1/2}\over \mathfrak{r}_0^{3/2}},
\enann
being the angular frequency of the rotating frame. 
It gives 
\beann
\Psi=\sigma \mathfrak{w}_R\,\tau
\,.
\enann

In order to revert to non-rotating frame, we have to transform back with the angular velocity.
The difference between two angular velocities 
\beann
\mathfrak{w}_{\rm P}\equiv \mathfrak{w}_0-\mathfrak{w}_{\rm R}, 
\enann 
gives rotation of inertial frame.
It also gives the precession of the angular momentum
as we will show it later. It contains two relativistic precessions, i.e., 
the so-called de-Sitter precession and Lense-Thirring precession.
This is evident when we take the limit of $M/\mathfrak{r}_0\ll1$.
Since $a\leq M$, this limit gives $a/\mathfrak{r}_0\ll1$. 
In this limit, we find
\beann
\mathfrak{w}_0&=&{\sqrt{M\mathfrak{r}_0}\over 
\mathfrak{r}_0 \left(\mathfrak{r}_0^2-3M\mathfrak{r}_0+ 2a\sigma\sqrt{M\mathfrak{r}_0}\right)^{1/2}}
\\
&\approx&
{M^{1/2}\over \mathfrak{r}_0^{3/2}}\left[1+{3M\over 2\mathfrak{r}_0}-{a\sigma\over \mathfrak{r}_0} \sqrt{M\over \mathfrak{r}_0}\right]
\\[.5em]
&=&
\mathfrak{w}_{\rm R}
+\mathfrak{w}_{\rm dS}+\mathfrak{w}_{\rm LT}
\,,
\enann
where
\beann
\mathfrak{w}_{\rm dS}&=&{3M\over 2\mathfrak{r}_0}\mathfrak{w}_{\rm R}={3M^{3/2}\over 2\mathfrak{r}_0^{5/2}},
\\
\mathfrak{w}_{\rm LT}&=&-{a\sigma\over \mathfrak{r}_0} \sqrt{M\over \mathfrak{r}_0}\mathfrak{w}_{\rm R}=-{M a\sigma \over \mathfrak{r}_0^3}.
\enann

The frequencies $\mathfrak{w}_{\rm dS}$ and $\mathfrak{w}_{\rm LT}$ correspond to those of de-Sitter  and Lense-Thirring precessions, respectively.
 These two frequencies are quite similar to those discussed in \cite{Liu:2019tqr,Liu:2021uam}.

\begin{figure}[htbp]
\begin{center}
\includegraphics[width=6cm]{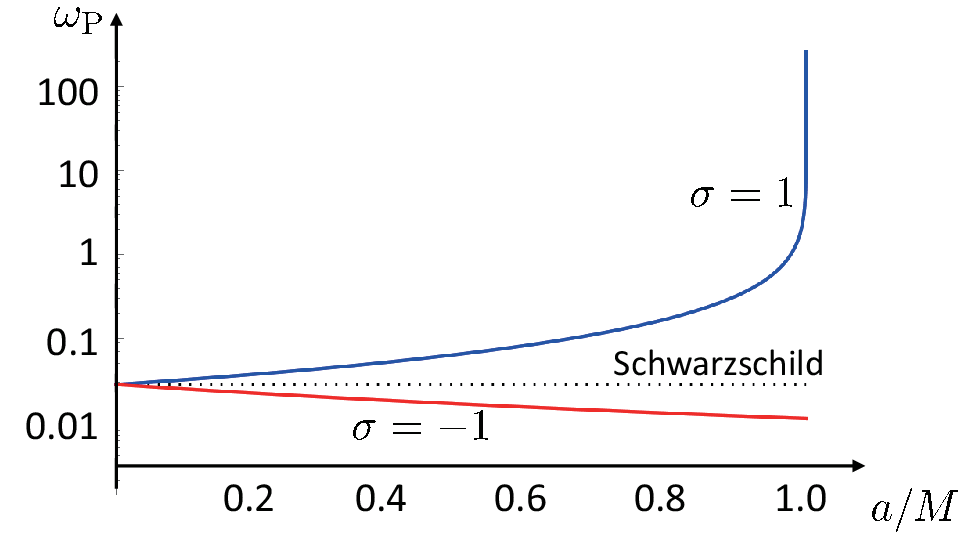}
\\
(a)
\\
\includegraphics[width=6cm]{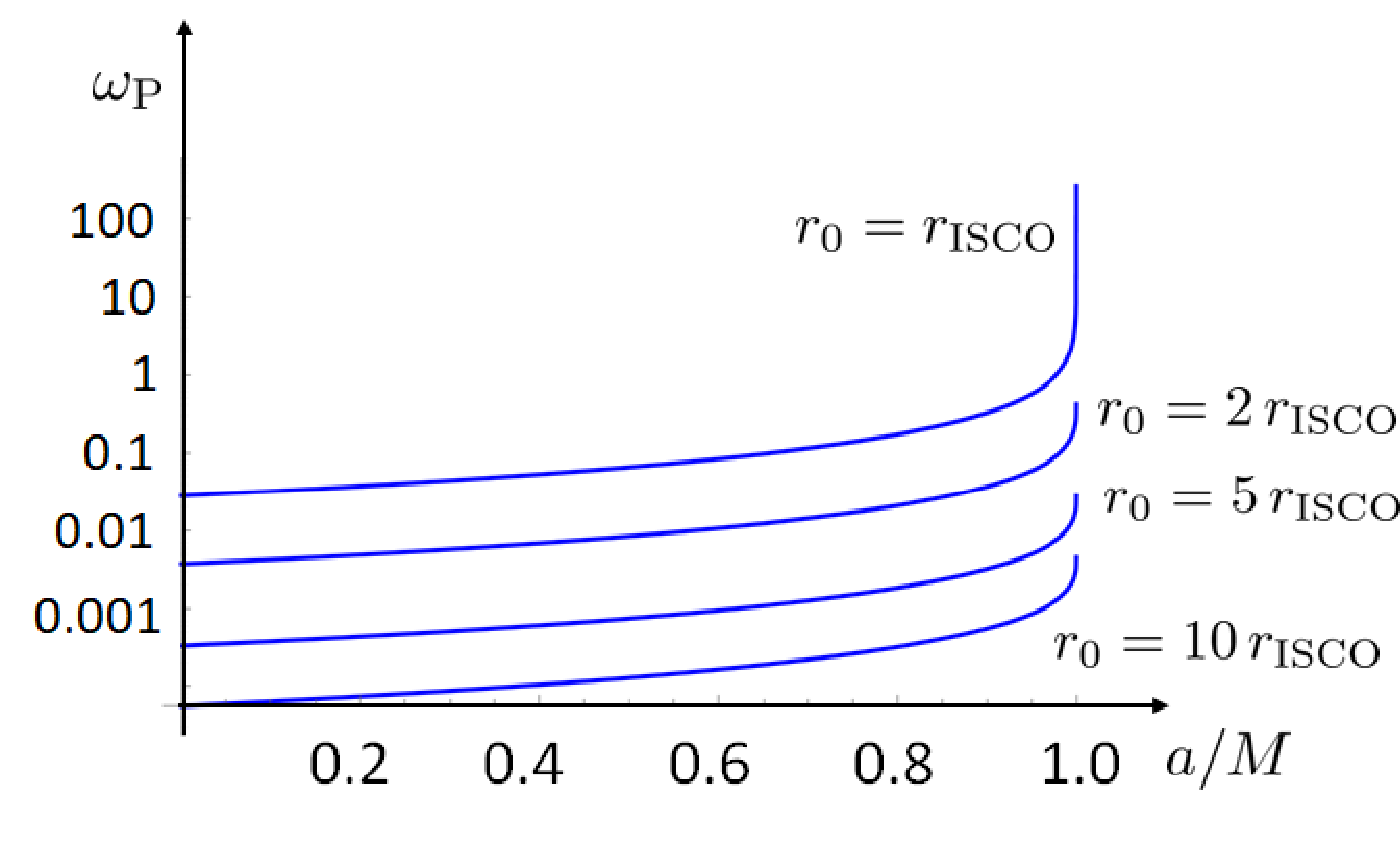}
\\
(b)
\\
\caption{The precession frequency $\mathfrak{w}_{\rm P}$
in terms of the Kerr parameter $a/M$.
(a) The position of the CM $\mathfrak{r}_0$ is chosen at the ISCO radius.
(b) $\mathfrak{r}_0$ for prograde orbits  are chosen as 
$\mathfrak{r}_0= \mathfrak{r}_{\rm ISCO}\,, 2\, \mathfrak{r}_{\rm ISCO}\,, 5\,\mathfrak{r}_{\rm ISCO}$ and $10\,\mathfrak{r}_{\rm ISCO}$. }
\label{fig:precession}
\end{center}
\end{figure}

In Fig.~\ref{fig:precession}, we show the behaviour of 
$\omega_{\rm P}$ in terms of the Kerr parameter $a$.
In  Fig.~\ref{fig:precession}(a), we set $\mathfrak{r}_0=\mathfrak{r}_{\rm ISCO}$, and
in  Fig.~\ref{fig:precession}(b), we change the radius as $\mathfrak{r}_0/\mathfrak{r}_{\rm ISCO}=1, 2, 5,$ and $10$.

We find that $\omega_{\rm P}$ gets larger as $a$ increases and 
this increase become particularly rapid near the extreme limit of $a\rightarrow 1$, which means that the Lense-Thirring precession becomes dominant.
The critical value $a_{\rm cr}$, beyond which the Lense-Thirring precession 
is larger than the de-Sitter precession, is evaluated by
\beann
\mathfrak{w}_{\rm P}(a_{\rm cr})=2\mathfrak{w}_{\rm dS},
\enann
where $\mathfrak{w}_{\rm dS}=\mathfrak{w}_{\rm P}(a=0)$.
In the case of $\mathfrak{r}_0=\mathfrak{r}_{\rm ISCO}$, we find
$a_{\rm cr}\approx 0.43185973M$.

\begin{figure}[htbp]
\begin{center}
\includegraphics[width=6.5cm]{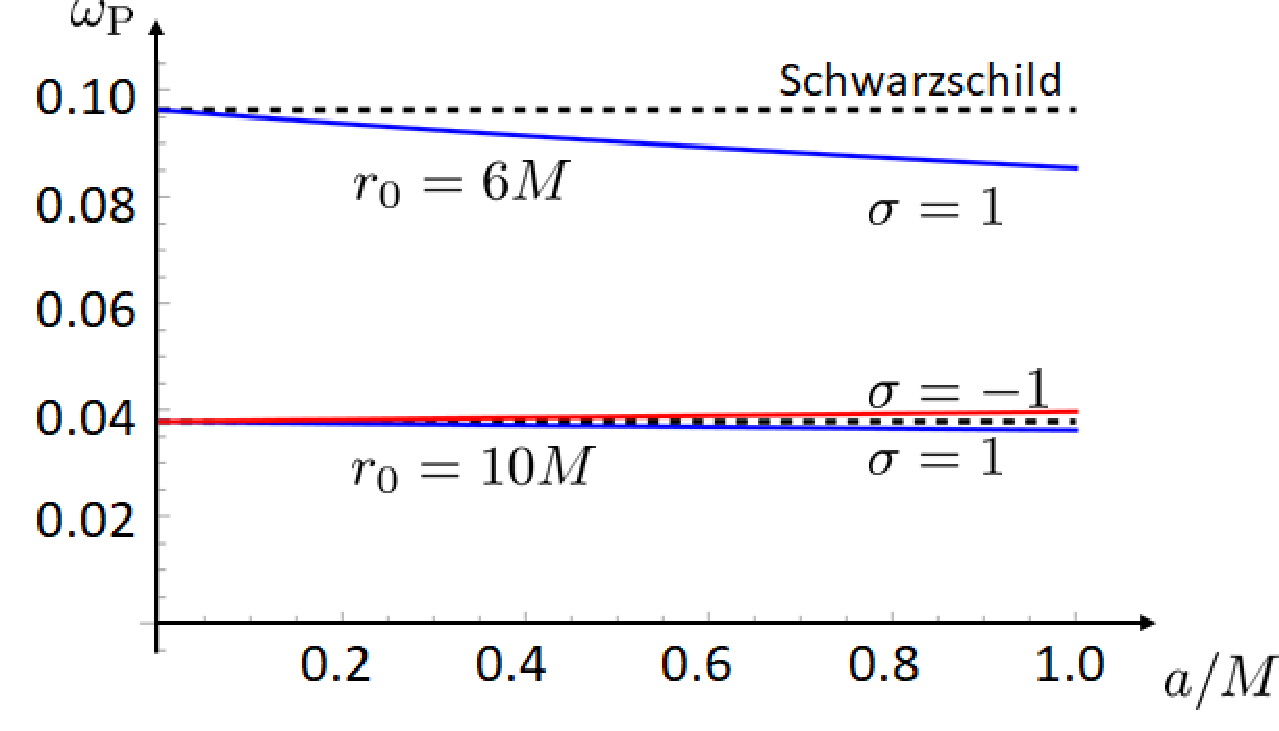}
\caption{The precession frequency  $\mathfrak{w}_{\rm P}$
in terms of $a$. We fix the position of the CM at
 $\mathfrak{r}_0=10M$ and $6M$.
 The blue and red curves denote the prograde and retrograde cases, respectively. The value for $a=0$ is also given  by the black dashed lines as reference.}
\label{fig:precession1}
\end{center}
\end{figure}

However when we fix the radius $\mathfrak{r}_0$, $\omega_{\rm P}$
does not depend on $a$ so much as shown in Fig.~\ref{fig:precession1}.
We show the cases of $\mathfrak{r}_0=6M$ and $\mathfrak{r}_0=10M$.
For $\mathfrak{r}_0=6M$, we give only for the prograde orbits
because the ISCO radius for the retrograde orbits is larger than $6M$.

In the prograde orbits, it decreases as $a$ increases, while it increases for the retrograde orbits.


\subsection{Riemann curvature components in rotating proper reference frame}
The non-trivial components of the Riemann curvature 
in the Carter's tetrad system  are given by

\beann
R_{\bar{\mathfrak{t}}\bar \phi\bar{\mathfrak{t}}\bar \phi}&=&-R_{\bar{\mathfrak{r}}\bar \theta\bar{\mathfrak{r}}\bar \theta}\,=\,-{1\over 2}R_{\bar{\mathfrak{t}}\bar{\mathfrak{r}}\bar{\mathfrak{t}}\bar{\mathfrak{r}}}\,=\,{1\over 2}R_{\bar \theta\bar \phi\bar \theta\bar \phi}
\\
&=&R_{\bar{\mathfrak{t}}\bar \theta\bar{\mathfrak{t}}\bar \theta}\,=\,-R_{\bar{\mathfrak{r}}\bar \phi\bar{\mathfrak{r}}\bar \phi}\,=\, {\cal Q}_1,
\\
R_{\bar{\mathfrak{t}}\bar \phi\bar{\mathfrak{r}}\bar \theta}&=&
{1\over 2}R_{\bar{\mathfrak{t}}\bar{\mathfrak{r}}\bar \phi\bar \theta}
\,=\,R_{\bar{\mathfrak{t}}\bar \theta\bar \phi\bar{\mathfrak{r}}}\,=\,
 -{\cal Q}_2,
\enann
where
\beann
{\cal Q}_1&=&
{M\mathfrak{r}(\mathfrak{r}^2-3a^2\cos^2\theta)\over \Sigma^3}
\\
{\cal Q}_2&=&
{Ma\cos\theta(3\mathfrak{r}^2-a^2\cos^2\theta)\over \Sigma^3}
\,.
\enann
On the equatorial plane ($\theta=\pi/2$), 
we find simpler expression such that 
the non-trivial components are very similar to the Schwarzschild case as
\bea
&&
R_{\bar{\mathfrak{t}}\bar \theta\bar{\mathfrak{t}}\bar \theta}
\,=\,R_{\bar{\mathfrak{t}}\bar \phi\bar{\mathfrak{t}}\bar \phi}
\,=\,-R_{\bar{\mathfrak{r}}\bar \theta\bar{\mathfrak{r}}\bar \theta}
\,=\,-R_{\bar{\mathfrak{r}}\bar \phi\bar{\mathfrak{r}}\bar \phi}
\,=\, {M\over \mathfrak{r}^3}
\nn
&&
R_{\bar \theta\bar \phi\bar \theta\bar \phi}
\,=\,
-R_{\bar{\mathfrak{t}}\bar{\mathfrak{r}}\bar{\mathfrak{t}}\bar{\mathfrak{r}}}
\,=\,{2M\over \mathfrak{r}^3}
\,.
\label{Riemann_Carter}
\ena

\begin{widetext}

Since we know the Riemann curvature components in the Carter's tetrad frame (Eq. (\ref{Riemann_Carter})) and the transformation matrix $\Lambda_{\hat \alpha}^{~\bar \lambda}$ to the rotating 
proper reference frame (Eqs. (\ref{Lambda_hat0})-(\ref{Lambda_hatz})), we can easily find the non-trivial  Riemann curvature components in rotating proper reference frame as
\beann
&&
\bar{\cal R}_{\hat 0\hat x\hat 0\hat x}
=
-\bar{\cal R}_{\hat y\hat z\hat y\hat z}
=-{M\over  F^2_\sigma(\mathfrak{r}_0) \mathfrak{r}_0^3}\left(2\mathfrak{r}_0^2+3a^2-3M\mathfrak{r}_0-2a\sigma\sqrt{M\mathfrak{r}_0}\right)
\,,~
\\
&&
\bar{\cal R}_{\hat 0\hat y\hat 0\hat y}=-\bar{\cal R}_{\hat z\hat x\hat z\hat x}={M\over \mathfrak{r}_0^3}
\,,~
\\
&&
\bar{\cal R}_{\hat 0\hat z\hat 0\hat z}=-\bar{\cal R}_{\hat x\hat y\hat x\hat y}
={M\over  F^2_\sigma(\mathfrak{r}_0) \mathfrak{r}_0^3}\left(\mathfrak{r}_0^2+3a^2-4a\sigma\sqrt{M\mathfrak{r}_0}
\right),
\\
&&
\bar{\cal R}_{\hat 0\hat x\hat y\hat x}=-\bar{\cal R}_{\hat 0\hat z\hat y\hat z}
=-{3 M\over  F^2_\sigma(\mathfrak{r}_0) \mathfrak{r}_0^3}\sqrt{\Delta(\mathfrak{r}_0)}
\left(\sigma\sqrt{M\mathfrak{r}_0}-a\right)
\,
\,.
\enann

\subsection{Equations of motion of a binary}
Since the CM of a binary follows the observer's circular geodesic ($\vect{R}=0$), we have to solve only the equations of motion for the relative coordinate $\vect{r}$.
Using $x=r^{\hat 1}\,,y=r^{\hat 2}\,,z=r^{\hat 3}$,
the relative motion of a binary is
given by the Lagrangian 
\bea
{\cal L}_{\rm rel}(\vect{r},\dot{\vect{r}})&=&
{1\over 2}\mu \dot{\vect{r}}^2+  {G m_1m_2\over r}
+{\cal L}_{{\rm rel} \mathchar`-\mathfrak{w}}(\vect{r},\dot{\vect{r}})
+{\cal L}_{{\rm rel}\mathchar`- \bar{\cal R}}(\vect{r},\dot{\vect{r}}),
\label{EoM_relative_coordinate}
\ena
with
\beann
{\cal L}_{{\rm rel} \mathchar`-\mathfrak{w}}
&=&-\mu\left[\sigma\mathfrak{w}_0\left(x\dot{y}
-y\dot{x}\right)
-{\mathfrak{w}_0^2\over 2}\left(x^2+y^2\right)\right],
\\{\cal L}_{{\rm rel} \mathchar`-\bar{\cal R}}
&=&-{\mu\over 2} 
\left(
\bar{\cal R}_{\hat 0\hat x\hat 0\hat x}x^2+\bar{\cal R}_{\hat 0\hat y\hat 0\hat y}y^2+\bar{\cal R}_{\hat 0\hat z\hat 0\hat z}z^2\right)
\\
&=&-{\mu M\over 2\mathfrak{r}_0^3} 
\left[r^2+
{3\over F^2_\sigma(\mathfrak{r}_0)}\left(-\Delta(\mathfrak{r}_0)x^2+(\sigma\sqrt{M\mathfrak{r}_0}-a)z^2\right)
\right]
\,.
\enann

The first and second terms in ${\cal L}_{{\rm rel}\mathchar`-\mathfrak{w}}$ describe the Coriolis force and 
the centrifugal force, respectively.
The first half terms in ${\cal L}_{{\rm rel}\mathchar`- \bar{\cal R}}$ are the same 
as those in Newtonian hierarchical triple  system under quadrupole approximation.
Note that in the present approach (approximation up to the second order of $r^{\hat a}$),
we cannot go beyond quadrupole approximation.

In order to analyze the relative motion of a binary,
it is better to work in a non-rotating initial reference frame.
Since the angular frequency of a rotating tetrad frame is $\mathfrak{w}_{\rm R}$, 
the position $(x, y, z)$ in the rotating frame can be replaced by the position $(\mathsf{x},\mathsf{y},\mathsf{z})$  in a non-rotating Descartes' coordinate system by use of the following transformation; 
\beann
x&=&\mathsf{x}\cos\sigma\mathfrak{w}_{\rm R}   \tau -\mathsf{y} \sin \sigma\mathfrak{w}_{\rm R}  \tau,
\\
y&=&
\mathsf{x} \sin\sigma\mathfrak{w}_{\rm R}   \tau +\mathsf{y} \cos \sigma\mathfrak{w}_{\rm R}    \tau,
 \\
z&=&\mathsf{z}
\,.
\enann

The Lagrangian ${\cal L}_{\rm rel}$ in a non-rotating proper reference frame 
is given by
\bea
{\cal L}_{\rm rel}={1\over 2}\mu \left({ d {\vect{\mathsf{r} }}\over d\tau}\right)^2+  {G m_1m_2\over \mathsf{r}}
+{\cal L}_{\rm rel\mathchar`-P}(\vect{\mathsf{r}},\dot{\vect{\mathsf{r}}})
+{\cal L}_{{\rm rel}\mathchar`-\bar{\cal R}}(\vect{\mathsf{r}},\tau)
\,,
\label{EoM_relative_non-rotating}
\ena
where
\beann
{\cal L}_{\rm rel\mathchar`-P}(\vect{\mathsf{r}},\dot{\vect{\mathsf{r}}})&=&\mu \sigma\mathfrak{w}_{\rm P}\left(\dot{\mathsf{x}}\mathsf{y}
-\dot{\mathsf{y}}\mathsf{x}\right)+{\mu\over 2} \mathfrak{w}_{\rm P}^2\left(\mathsf{x}^2+\mathsf{y}^2\right),
\\
{\cal L}_{{\rm rel}\mathchar`-\bar{\cal R}}(\vect{\mathsf{r}},\tau)
&=&-{\mu M\over 2\mathfrak{r}_0^3} 
\left[\mathsf{r}^2+{3\over F^2_\sigma(\mathfrak{r}_0)}\left(-\Delta(\mathfrak{r}_0)\left(\mathsf{x}\cos\mathfrak{w}_{\rm R}   \tau -\mathsf{y} \sin \mathfrak{w}_{\rm R}  \tau\right)^2
+(\sigma\sqrt{M\mathfrak{r}_0}-a)^2 \mathsf{z}^2\right)
 \right]\,.
\enann
Since the momentum is defined by
\beann
\mathsf{p}_{\mathsf{x}}&\equiv &{\partial {\cal L}\over \partial \dot{\mathsf{x}}}=\mu \dot{\mathsf{x}}+\mu\sigma\mathfrak{w}_{\rm P} \mathsf{y},
\\
\mathsf{p}_{\mathsf{y}}&\equiv &{\partial {\cal L}\over \partial \dot{\mathsf{y}}}=\mu \dot{\mathsf{y}}-\mu\sigma\mathfrak{w}_{\rm P} \mathsf{x},
\\
\mathsf{p}_{\mathsf{z}}&\equiv &{\partial {\cal L}\over \partial \dot{\mathsf{z}}}=\mu \dot{\mathsf{z}}
\,,
\enann
we obtain the Hamiltonian as
\bea
{\cal H}_{\rm rel}={\cal H}_0+{\cal H}_1
\label{Hamiltonian_non-rotating}
\,,
\ena
where
\beann
{\cal H}_0
&=&{1\over 2\mu} {\vect{\mathsf{p}}}^2- {G m_1m_2\over \mathsf{r}},
\\
{\cal H}_1
&=&{\cal H}_{1\mathchar`-{\rm P}}+{\cal H}_{1\mathchar`-\bar{\cal R}},
\enann
with
\beann
{\cal H}_{1\mathchar`-{\rm P}}&=&
\sigma\mathfrak{w}_{\rm P}\left(\mathsf{p}_{\mathsf{y}}\mathsf{x}-\mathsf{p}_{\mathsf{x}}\mathsf{y}\right),
\\
{\cal H}_{1\mathchar`-\bar{\cal R}}&=&{\mu M\over 2\mathfrak{r}_0^3} 
\left[\mathsf{r}^2+{3\over F^2_\sigma(\mathfrak{r}_0)}\left(-\Delta(\mathfrak{r}_0)\left(\mathsf{x}\cos\mathfrak{w}_{\rm R}   \tau -\mathsf{y} \sin \mathfrak{w}_{\rm R}  \tau\right)^2
+(\sigma\sqrt{M\mathfrak{r}_0}-a)^2 \mathsf{z}^2\right)
 \right]\,.
\enann

The equations of motion  are given as
\bea
\dot{\mathsf{x}}&=&{\partial{\cal H}\over \partial\mathsf{p}_{\mathsf{x}}}
={\mathsf{p}_{\mathsf{x}}\over \mu}-\sigma\mathfrak{w}_{\rm P}\mathsf{y},
\label{px}
\\
\dot{\mathsf{y}}&=&{\partial{\cal H}\over \partial\mathsf{p}_{\mathsf{y}}}
={\mathsf{p}_{\mathsf{y}}\over \mu}+\sigma\mathfrak{w}_{\rm P}\mathsf{x},
\label{py}
\\
\dot{\mathsf{z}}&=&{\partial{\cal H}\over \partial\mathsf{p}_{\mathsf{z}}}
={\mathsf{p}_{\mathsf{z}}\over \mu},
\label{pz}
\ena
and 
\bea
\dot{\mathsf{p}}_{\mathsf{x}}&=&-{\partial{\cal H}\over \partial\mathsf{x}}=-{Gm_1m_2\over \mathsf{r}^3}\mathsf{x}-\sigma\mathfrak{w}_{\rm P}\mathsf{p}_{\mathsf{y}}
-{\mu M\over \mathfrak{r}_0^3}
\left[\mathsf{x}-3\lambda(\mathsf{x}\cos\mathfrak{w}_{\rm R}   \tau -\mathsf{y} \sin \mathfrak{w}_{\rm R}  \tau)\cos\mathfrak{w}_{\rm R}   \tau\right]\,,
\label{eq_x}
\\
\dot{\mathsf{p}}_{\mathsf{y}}&=&-{\partial{\cal H}\over \partial\mathsf{y}}=-{Gm_1m_2\over \mathsf{r}^3}\mathsf{y}+\sigma\mathfrak{w}_{\rm P}\mathsf{p}_{\mathsf{x}}
-{\mu M\over \mathfrak{r}_0^3}
\left[\mathsf{y}+3\lambda(\mathsf{x}\cos\mathfrak{w}_{\rm R}   \tau -\mathsf{y} \sin \mathfrak{w}_{\rm R}  \tau)\sin\mathfrak{w}_{\rm R}   \tau\right],
\label{eq_y}
\\
\dot{\mathsf{p}}_{\mathsf{z}}&=&-{\partial{\cal H}\over \partial\mathsf{z}}=-{Gm_1m_2\over \mathsf{r}^3}\mathsf{z}
-{\mu M\over \mathfrak{r}_0^3}\left[-2+3\lambda\right]\mathsf{z}
\label{eq_z}
\,,
\ena
where
\bea
\lambda\equiv {\Delta(\mathfrak{r}_0)\over F_\sigma^2(\mathfrak{r}_0)}={
\mathfrak{r}_0^2-2M\mathfrak{r}_0+a^2
\over 
\mathfrak{r}_0^2-3M\mathfrak{r}_0
+2a\sigma  \sqrt{M\mathfrak{r}_0}.
}
\label{def_lambda}
\ena

\end{widetext}
\section{preliminary considerations}
\label{preliminary_considerations}
For numerical analysis in next section, we shall first rewrite
 the basic equations using dimensionless variables, 
 introduce the orbital parameters, show 
 how to set up initial data, and discuss validity of the present model and stability condition. 

\subsection{Normalization}
In this paper we will analyze a binary model with $m_1=m_2$.
We have to solve Eqs. (\ref{px}) -(\ref{eq_z}).
In order to solve these basic equations, we shall introduce dimensionless variables as follows: 
Since we consider a hierarchical triple system, the initial motion of a 
binary can be approximated by an elliptic orbit.
Hence the length scale of a binary is normalized by an initial semi-major axis $\mathfrak{a}_0$, while 
time scale is normalized by an initial binary mean motion $n_0$, which is defined by
\beann
n_0\equiv \bigg({G(m_1+m_2)\over \mathfrak{a}_0^3}\bigg)^{1/2}
\,.
\enann
Note that the initial binary period is given by $P_{\rm in}=2\pi/n_0$.
\\

Introducing 
\beann
&&
\tilde \tau=n_0\tau,
\\
&&
\tilde{\mathsf{x}}={\mathsf{x}\over \mathfrak{a}_0}\,,~
\tilde{\mathsf{y}}={\mathsf{y}\over \mathfrak{a}_0}\,,~
\tilde{\mathsf{z}}={\mathsf{z}\over \mathfrak{a}_0}\,,~
\tilde{\mathsf{r}}={\mathsf{r}\over \mathfrak{a}_0},
\\
&&
\tilde{\mathsf{p}}_\mathsf{x}={\mathsf{p}_\mathsf{x}\over \mu \mathfrak{a}_0 n_0}\,,~
\tilde{\mathsf{p}}_\mathsf{y}={\mathsf{p}_\mathsf{y}\over \mu \mathfrak{a}_0 n_0}\,,~
\tilde{\mathsf{p}}_\mathsf{z}={\mathsf{p}_\mathsf{z}\over \mu \mathfrak{a}_0 n_0 }
\,,
\enann
we find 
\bea
{d\tilde{\mathsf{x}}\over d\tilde \tau}&=&\tilde{\mathsf{p}}_\mathsf{x}-\sigma\tilde{\mathfrak{w}}_{\rm P}\tilde{\mathsf{y}},
\label{npx}
\\
{d\tilde{\mathsf{y}}\over d\tilde \tau}&=&\tilde{\mathsf{p}}_\mathsf{y}+\sigma\tilde{\mathfrak{w}}_{\rm P}\tilde{\mathsf{x}},
\label{npy}
\\
{d\tilde{\mathsf{z}}\over d\tilde \tau}&=&\tilde{\mathsf{p}}_\mathsf{z},
\label{npz}
\ena
\begin{widetext}
and 
\bea
{d\tilde{\mathsf{p}}_{\mathsf{x}}\over d\tilde \tau}&=&
-{\tilde{\mathsf{x}}\over \tilde{\mathsf{r}}^3}-\sigma\tilde{\mathfrak{w}}_{\rm P}\tilde{\mathsf{p}}_{\mathsf{y}}
-{1\over \mathfrak{f}}
\left[\tilde{\mathsf{x}}-3\lambda
(\tilde{\mathsf{x}}\cos\tilde{\mathfrak{w}}_{\rm R}\tilde{\tau} -\tilde{\mathsf{y}} \sin \tilde{\mathfrak{w}}_{\rm R}\tilde{\tau})\cos \tilde{\mathfrak{w}}_{\rm R}\tilde{\tau}\right]
\label{neq_x},
\\
{d\tilde{\mathsf{p}}_{\mathsf{y}}\over d\tilde \tau}&=&
-{\tilde{\mathsf{y}}\over \tilde{\mathsf{r}}^3}
+\sigma\tilde{\mathfrak{w}}_{\rm P}\tilde{\mathsf{p}}_{\mathsf{x}}
-{1\over \mathfrak{f}}
\left[\tilde{\mathsf{y}}+3\lambda (\tilde{\mathsf{x}}\cos\tilde{\mathfrak{w}}_{\rm R} \tilde{\tau} -\tilde{\mathsf{y}} \sin \tilde{\mathfrak{w}}_{\rm R} \tilde{\tau})\sin\tilde{\mathfrak{w}}_{\rm R} \tilde{\tau}\right],
\label{neq_y}
\\
{d\tilde{\mathsf{p}}_{\mathsf{z}}\over d\tilde \tau}&=&
-{\tilde{\mathsf{z}}\over \tilde{\mathsf{r}}^3}
-{1\over \mathfrak{f}}
\left(-2+3\lambda\right)\tilde{\mathsf{z}},
\label{neq_z}
\ena
\end{widetext}
where
\beann
\tilde{\mathfrak{w}}_{\rm P} &\equiv &{\mathfrak{w}_{\rm P}\over n_0},
\\
\tilde{\mathfrak{w}}_{\rm R} &\equiv &{\mathfrak{w}_{\rm R}\over n_0},
\enann
and
\beann
\lambda
={
\mathfrak{r}_0^2-2M\mathfrak{r}_0+a^2
\over 
\mathfrak{r}_0^2-3M\mathfrak{r}_0
+2a\sigma  \sqrt{M\mathfrak{r}_0}
}
\,.
\enann

$\mathfrak{f}$ denotes the firmness parameter of a binary,
which  is defined by
\beann
\mathfrak{f}&\equiv& {{\rm gravitational~force}\over {\rm tidal~force~by~SMBH}}={{Gm_1m_2/\mathfrak{a}_0^2}\over
 G{\mu M \,\mathfrak{a}_0 / \mathfrak{r}_0^3}}
\\&=&\left({ m_1+m_2\over M}\right)\left({\mathfrak{r}_0\over \mathfrak{a}_0}\right)^3
\,.
\enann

The initial semi-major axis $\mathfrak{a}_0$ is given by
\bea
\mathfrak{a}_0=\mathfrak{f}^{-{1\over 3}}
\,\left({ m_1+m_2\over M}\right)^{1/3}\, \mathfrak{r}_0\,.
\label{semi-major_axis}
\ena
Using $\mathfrak{f}$, we find
\bea
\tilde{\mathfrak{w}}_{\rm P} &=&\nu\, \mathfrak{f}^{-{1\over 2}}
\,,
\\
\tilde{\mathfrak{w}}_{\rm R} &=&\mathfrak{f}^{-{1\over 2}}
\,,
\ena
where
\bea
\nu&\equiv& {\mathfrak{r}_0\over F_\sigma(\mathfrak{r}_0)}-1
\nn
&=&{
\mathfrak{r}_0
\over 
\sqrt{\mathfrak{r}_0^2-3M\mathfrak{r}_0
+2a\sigma  \sqrt{M\mathfrak{r}_0}}
}
-1
\label{def_nu}
\ena

The basic equations (\ref{npx})-(\ref{neq_z}) contain three independent 
parameters;  $\lambda$, $\nu$ and 
the firmness $\mathfrak{f}$.
$\lambda$ changes from $1$ to $4/3$, while $\nu$ runs 
from $0$ to $\sqrt{2}-1$ for $a=0$ and to $\infty$ for $a=M$.

As for the firmness $\mathfrak{f}$, as we will discuss it later,
 $\mathfrak{f}\gg 1$ is required for stability.
 In the limit of $\mathfrak{f}\rightarrow \infty$, we find 
 an integrable system. The orbit shows 
 the precession with the period $\tilde{\mathfrak{w}}_{\rm P}$.

\subsection{Orbital Parameters}
 In order to discuss the properties of 
a binary orbit, it is more convenient to use the orbital parameters. We may assume
 that the binary motion is close to an elliptic orbit, which 
 is described by
\beann
\mathsf{r}={\mathfrak{a}(1-e^2)\over 1+e\cos f}
\,,
\enann
where $\mathfrak{a}$ is a semi-major axis, $e$ is the eccentricity, and $f$ is true anomaly.
Since the orbital plane is not, in general, $\mathsf{z}=0$, we have to introduce 
three angular variables; the argument of periapsis $\omega$, the ascending node $\Omega$ and the inclination angle $I$.
We have the relations between the position $\bf{\mathsf{r}}=(\mathsf{x},\mathsf{y},\mathsf{z})$ of the component of a binary and the orbital parameters $(\omega\,,\Omega\,,  \mathfrak{a}\,, e\,, I\,, f)$ as
\begin{widetext}
\bea
\begin{pmatrix}
\mathsf{x} \\
\mathsf{y} \\
\mathsf{z} \\
\end{pmatrix}
&=&
\mathsf{r}\begin{pmatrix}
\cos \Omega\cos(\omega+f)-\sin\Omega\sin(\omega+f)\cos I \\
\sin \Omega\cos(\omega+f)+\cos\Omega\sin(\omega+f)\cos I\\
\sin(\omega+f)\sin I \\
\end{pmatrix}
\label{orbital_parameters}
\ena
\end{widetext}

In order to extract the orbital parameters 
from the orbit given by the Cartesian 
coordinates, one can use the osculating orbit when the orbit is close to an ellipse. 
The magnitude of the 
normalized Laplace-Runge-Lenz vector,
which is defined by
\bea
\vect{e}\equiv \tilde{\vect{\mathsf{p}}}\times (\tilde{\vect{\mathsf{r}}}
\times \tilde{\vect{\mathsf{p}}})-{\tilde{\vect{\mathsf{r}}}\over \tilde{\mathsf{r}}}\,,
\label{LRL_vector}
\ena
is commonly used for a measure of orbital eccentricity.

The inclination angle $I$ is defined as mutual inclination between angular momenta of the inner and outer binary.
\bea
I=\cos^{-1}\left({\tilde L_{\mathsf{z}} \over |\tilde{\vect{L}}|}\right)
\,,
\label{def_incrination}
\ena
where $\tilde{\vect{L}}\equiv \tilde{\vect{\mathsf{r}}}\times \tilde{\vect{\mathsf{p}}}$ is the angular momentum of a binary.

The other two essential angles $\Omega$ and $\omega$ governing the orientation of the orbital plane. The line that marks the intersection of the orbital plane with the reference plane (the equatorial plane in the present case) is called the node line, and the point on the node line where the orbit passes above the reference plane from below is called the ascending node. The angle between the reference axis (say $\mathsf{x}$-axis) and node line vector $\vect{N}$ is the longitude of ascending node $\Omega$.  First, node line is defined as,
\bea
\vect{N} = \hat{\vect{\mathsf{z}}}\times \tilde{\vect{L}} \,.
\label{def_node_line}
\ena
where $\hat{\vect{\mathsf{z}}}$ is normal to the reference plane (the unit vector in the $\mathsf{z}$ direction). Thus, $\Omega$ is computed as,
\bea
\Omega = \cos^{-1} (N_{\mathsf{x}}/N) \,.
\label{def_Omega}
\ena
The argument of periapsis $\omega$ is the angle between node line and periapsis measured in the direction of motion. Therefore,
\bea
\omega = \cos^{-1} \bigg(\frac{\vect{N} \cdot \vect{e}}{N\,e}\bigg) \,.
\label{def_omega}
\ena

However, one must be careful
with the definitions of orbital elements when using the osculating method. 
For instance, 
It may show an ``apparent" rise in eccentricity or unphysical rapid oscillations especially when the eccentricity is very small~\cite{Will_2019}. In such a  case, it is better to 
define the eccentricity by the averaged one over one cycle
as
\beann
\langle e \rangle \equiv {\mathsf{r}_{\rm max}-\mathsf{r}_{\rm min}\over 
\mathsf{r}_{\rm max}+\mathsf{r}_{\rm min}},
\enann
where $\mathsf{r}_{\rm max}$ and $\mathsf{r}_{\rm min}$  correspond to orbital separation at adjacent turning points of an eccentric orbit.

When the orbit can be approximated well by the osculating one, 
$\vect{e}$ is given by the normalized Laplace-Runge-Lenz vector (\ref{LRL_vector}).
Otherwise, we define the averaged eccentricity vector by 
\beann
\langle \vect{e}\rangle  \equiv -{\left(\vect{\mathsf{r}}_{\rm min}+\vect{\mathsf{r}}_{\rm max}\right)\over \left(\mathsf{r}_{\rm min}+\mathsf{r}_{\rm max}\right)}
\enann
pointing towards the periapsis,
where  $\vect{\mathsf{r}}_{\rm max}$ and $\vect{\mathsf{r}}_{\rm min}$
are numerical data of position vector.
We have used both definitions and found that most results agree well.


\subsection{ Initial Data}

In order to provide the initial data of a binary, i.e., $\tilde{\mathsf{x}}_0\,,\tilde{\mathsf{y}}_0\,,\tilde{\mathsf{z}}_0$ and $\tilde{\mathsf{p}}_{\mathsf{x}0}\,,\tilde{\mathsf{p}}_{\mathsf{y}0}\,,\tilde{\mathsf{p}}_{\mathsf{z}0}$, 
we shall give the initial orbital parameters  $(\omega_0\,,\Omega_0\,,  \mathfrak{a}_0\,, e_0)$.
From (\ref{orbital_parameters}), assuming $f=0$ at $\tau=0$, 
we find
\beann
\tilde{\mathsf{x}}_0
&=&(1-e_0)\left[\cos \Omega_0\cos\omega_0-\sin\Omega_0\sin\omega_0\cos I_0\right]\,,
\\
\tilde{\mathsf{y}}_0
&=&(1-e_0)\left[\sin \Omega_0\cos\omega_0+\cos\Omega_0\sin\omega_0\cos I_0\right]\,,
\\
\tilde{\mathsf{z}}_0
&=&(1-e_0)\sin\omega_0\sin I_0
\,.
\enann
As for the momentum $\tilde{\mathsf{p}}_{\mathsf{x}0}\,,\tilde{\mathsf{p}}_{\mathsf{y}0}\,,
\tilde{\mathsf{p}}_{\mathsf{z}0}$, 
we use the definitions of the orbital parameters 
of the osculating orbit, i.e., Eqs. (\ref{LRL_vector}), (\ref{def_incrination}), (\ref{def_omega}) and 
(\ref{def_Omega}) with (\ref{def_node_line}).

From Eq. (\ref{def_node_line}), we find
\beann
N_{\mathsf{x}0}=-\tilde L_{\mathsf{y}0}
\,,~
N_{\mathsf{y}0}=\tilde L_{\mathsf{x}0}
\,,~
N_{\mathsf{z}0}=0\,.
\enann
From Eq. (\ref{def_Omega}), we obtain
\beann
N_{\mathsf{x}0}=N_0\cos\Omega_0.
\enann
Since $N_0^2=N_{\mathsf{x}0}^2+N_{\mathsf{y}0}^2$, we find
\beann
N_{\mathsf{y}0}=N_0\sin\Omega_0\,.
\enann
Hence, from Eq. (\ref{def_omega}), we find
\beann
e_0\cos\omega_0 
=  \bigg(\frac{\vect{N}_0}{N_0} \cdot \vect{e}_0\bigg) 
=\cos\Omega_0 e_{\mathsf{x}0}+\sin\Omega_0 e_{\mathsf{y}0}\,,
\enann
where
\beann
e_{\mathsf{x}0}&=&\tilde{\mathsf{p}}_{\mathsf{y}0} \tilde L_{\mathsf{z}0}-\tilde{\mathsf{p}}_{\mathsf{z}0} \tilde L_{\mathsf{y}0}-{\tilde{\mathsf{x}}_0\over \tilde{\mathsf{r}}_0},
\\
e_{\mathsf{y}0}&=&\tilde{\mathsf{p}}_{\mathsf{z}0} \tilde L_{\mathsf{x}0}-\tilde{\mathsf{p}}_{\mathsf{x}0} \tilde L_{\mathsf{z}0}-{\tilde{\mathsf{y}}_0\over \tilde{\mathsf{r}}_0},
\enann
with
\beann
\tilde L_{\mathsf{x}0}&=&\tilde{\mathsf{y}}_0\tilde{\mathsf{p}}_{\mathsf{z}0}
-\tilde{\mathsf{z}}_0\tilde{\mathsf{p}}_{\mathsf{y}0},
\\
\tilde L_{\mathsf{y}0}&=&\tilde{\mathsf{z}}_0\tilde{\mathsf{p}}_{\mathsf{x}0}
-\tilde{\mathsf{x}}_0\tilde{\mathsf{p}}_{\mathsf{z}0}
\,.
\enann

Since
\beann
\cos\Omega_0&=&{N_{\mathsf{x}0}\over N_0}=-{\tilde L_{\mathsf{y}0}\over \sqrt{\tilde L_{\mathsf{x}0}^2+\tilde L_{\mathsf{y}0}^2}},
\\
\sin\Omega_0&=&{N_{\mathsf{y}0}\over N_0}={\tilde L_{\mathsf{x}0}\over \sqrt{\tilde L_{\mathsf{x}0}^2+\tilde L_{\mathsf{y}0}^2}},
\enann
and
\beann
\tilde L_{\mathsf{z}0}=\tilde L_0\cos I_0
\,,
\enann
we find that
\beann
\tilde L_{\mathsf{x}0}=\tilde L_0\sin I_0 \sin\Omega_0\,,~
\tilde L_{\mathsf{y}0}=-\tilde L_0\sin I_0 \cos\Omega_0.
\enann
From the 
normalized Laplace-Runge-Lenz vector $\vect{e}_0$, we obtain
\beann
e_0^2=\left[\tilde{\mathsf{p}}_0^2 \tilde{\mathsf{r}}_0^2-(\tilde{\vect{\mathsf{p}}}_0\cdot \tilde{\vect{\mathsf{r}}}_0)^2\right]\left(\tilde{\mathsf{p}}_0^2-{2\over \tilde{\mathsf{r}}_0}\right)+1.
\enann
Hence, when we prepare 
the initial orbital parameters ($e_0\,, I_0\,, \omega_0\,, \Omega_0$), 
we can provide the initial data
of $(\tilde{\vect{\mathsf{r}}}_0, \tilde{\vect{\mathsf{p}}}_0)$ for the normalized evolution equations (\ref{npx})-(\ref{neq_z}). 
Note that since we choose the initial point at the periapsis
($f=0$ at $\tau=0$), we find 
\beann
( \tilde{\vect{\mathsf{p}}}_0\cdot\tilde{\vect{\mathsf{r}}}_0)=0
\,,
\enann
which can be used to find the initial data.

\subsection{Validity and Stability}
\label{validity}
Before showing our numerical results, 
we discuss validity of the present approach and 
the stability conditions. The minimum curvature radius at the radius $\mathfrak{r}_0$ is evaluated as 
\beann
 \ell_{\bar{\cal R}} &\equiv& {\rm min} \left[|\bar{\cal R}_{\hat \mu\hat \nu\hat \rho\hat \sigma}|^{-{1\over 2}}, |\bar{\cal R}_{\hat \mu\hat \nu\hat \rho\hat \sigma ; \hat \alpha}|^{-{1\over 3}}, |\bar{\cal R}_{\hat \mu\hat \nu\hat \rho\hat \sigma ; \hat \alpha;\hat \beta}|^{-{1\over 4}}
\right]
\\
&\sim&
 {\rm min} \left[\left({M\over \mathfrak{r}_0^3}\right)^{-{1\over 2}}\,,\left({M\over \mathfrak{r}_0^4}\right)^{-{1\over 3}}\,,
\left( {M\over \mathfrak{r}_0^5}\right)^{-{1\over 4}}
\right]
\\
&\sim &\mathfrak{r}_0 \left( {\mathfrak{r}_0 \over M}\right)^{1/4}
\,.
\enann
When we put a binary at $\mathfrak{r}=\mathfrak{r}_0$,
 the  binary size $\ell_{\rm binary}$ should satisfy
\beann
\ell_{\rm binary}\ll  \ell_{\bar{\cal R}} 
\enann

The relativistic effect in a binary is not  important when
\beann
\ell_{\rm binary} &\gg& {G(m_1+m_2)\over c^2}
\,.
\enann

As for stability of a binary, the mutual gravitational interaction between a binary 
should be much larger than the tidal force by a third body.
The condition is given by
\beann
{Gm_1m_2\over r^2}\gg {\mu M  \over \mathfrak{r}_0^3}\,r
\,,
\enann
which corresponds to 
the condition on the firmness parameter as
$\mathfrak{f}\gg 1$.
It gives the constraint on a binary size $\ell_{\rm binary}$ as
\bea
\ell_{\rm binary} 
&\ll &
\left({m_1+m_2\over M}\right)^{1\over 3}\,\mathfrak{r}_0.
\label{stability_tidal}
\ena
Hence, 
for a binary with the size of
\beann
{G(m_1+m_2)\over c^2 }\ll \ell_{\rm binary} 
 \ll 
 \left({m_1+m_2\over M}\right)^{1\over 3}\,\mathfrak{r}_0\,,
 \enann
 we may apply the present Newtonian approach.
 
If we are interested in the orbit near the ISCO radius 
of near extreme Kerr BH ($\mathfrak{r}_0 =M$), 
we find the condition as
\beann
 &&
 2\times 10^{-7}{\rm au}\left({m_1+m_2\over20 M_\odot }\right) \ll \ell_{\rm binary} 
 \\
 &&~~~~
 \ll 6 \times 10^{-3} {\rm au}\left({m_1+m_2\over20 M_\odot }\right)^{1\over 3}\left({M\over 10^8 M_\odot}\right)^{{2\over 3}}\,.
 \enann
Note that  $\ell_{\bar{\cal R}}\sim 1\,{\rm au}\left({M/10^8 M_\odot}\right)$ in this limit.

We also have another criterion for stability.
In order to avoid a chaotic energy exchange 
instability, we may have to impose 
the condition for the ratio of the circular radius $\mathfrak{r}_0$ to 
the binary size $\ell_{\rm binary} $ 
such that
\beann
{\mathfrak{r}_0\over \ell_{\rm binary} }
\gsim C_{\rm chaotic}\left({M\over m_1+m_2}\right)^p,
\enann
when $M\gg m_1, m_2$. 
Two parameters in this inequality 
are evaluated by $N$-body simulations of two groups
\cite{mardling2001tidal,myllari2018stability}
as
\beann
&
C_{\rm chaotic}\sim 2.8~~~~~{\rm and}~p={2\over 5}
&{\rm (criterion\,1)},
\\
&
C_{\rm chaotic}\sim 5.2 \mathsf{f} ^{1\over 3}~{\rm and}~p={1\over 3}&{\rm (criterion\,2}.
)
\enann
Here $\mathsf{f} $ is a complicated function of inner eccentricity 
$e$ and inclination $I$ such that
\begin{widetext}
\beann
\mathsf{f} &=&1 - {2\over 3}e(1 - e^2) 
- 
  0.3 \cos I \left[1 - {e\over 2} + 2\cos I \left(1 - {5\over 2} e^{3/2} 
  - \cos I\right)\right]\,.
\enann
\end{widetext}
It takes the value in the range of $0$ to $2.25$, 
but mostly between $0.6$ and $1.0$.

Since the above stability condition is only obtained 
for stellar masses triples and the
direct $N$-body integration is a reliable test of stability in such a setting, we will check such chaotic instability condition in our model.

\section{Numerical Analysis}
\label{numerical_analysis}
In a hierarchical triple system, there are several
important features. One is the so-called 
von Zeipel-Lidov-Kozai (vZLK) oscillations.
If the system is inclined more than some critical angle, there appears an oscillation between the eccentricity and inclination angle.
The second interesting feature is an orbital flip, which  may appears when 
the inclination angle evolves into near $90^\circ$.
The last one which we show is a chaotic feature in the long-time evolution.

Before showing the dynamical evolution of the present system, 
we first discuss chaotic instability condition in our model.

\subsection{Chaotic Instability}
\label{chaotic_instability}

In this subsection, we show our results of stability analysis 
for several values of the initial orbital parameters
(the semi-major axis $\mathfrak{a}_0$, 
eccentricity $e_0$, inclination $I_0$), the circular radius $\mathfrak{r}_0$,
 and Kerr rotation parameter $a$.
We choose $a=0.1, 0.3, 0.5, 0.7, 0.9, 0.999 M$, 
$\mathfrak{a}_0=0.005, 0.015, 0.025 M$, 
$e_0=0.01, 0.9$ and $I_0=0 ({\rm coplanar}), 85^\circ$.
 We fix the argument of periapsis and 
 the ascending node as $\omega_0=60^\circ\,, \Omega_0=30^\circ$ for simplicity.
 
 We perform the simulation until $\tilde \tau=10^4$, which is $\tau
 =10^4  n_0\approx 1600 P_0$, where $P_0$
  is initial orbital period of a binary. 
Since a binary is broken when the system is unstable, 
we judge stability at the end point of the simulation.
However, since the present system is non-integrable and show 
some chaotic features as we will show later, the boundary values 
between stability and instability is not sharp.
In fact, fixing the orbital parameters of a binary and Kerr parameter $a$,
even when  we find a stable orbit for some value of $\mathfrak{r}_0$,
we obtain an unstable orbit for a slightly larger value of $\mathfrak{r}_0$.
Changing the values of $\mathfrak{r}_0$, the stable and unstable orbits appear randomly. Fortunately, there exists  the minimum value of 
$\mathfrak{r}_0$ for stable orbits, below which a binary is broken 
before $\tilde \tau\approx 1000$ (mostly within 
a few dynamical time scale).

\begin{figure}[htbp]
\begin{center}
\includegraphics[width=8cm]{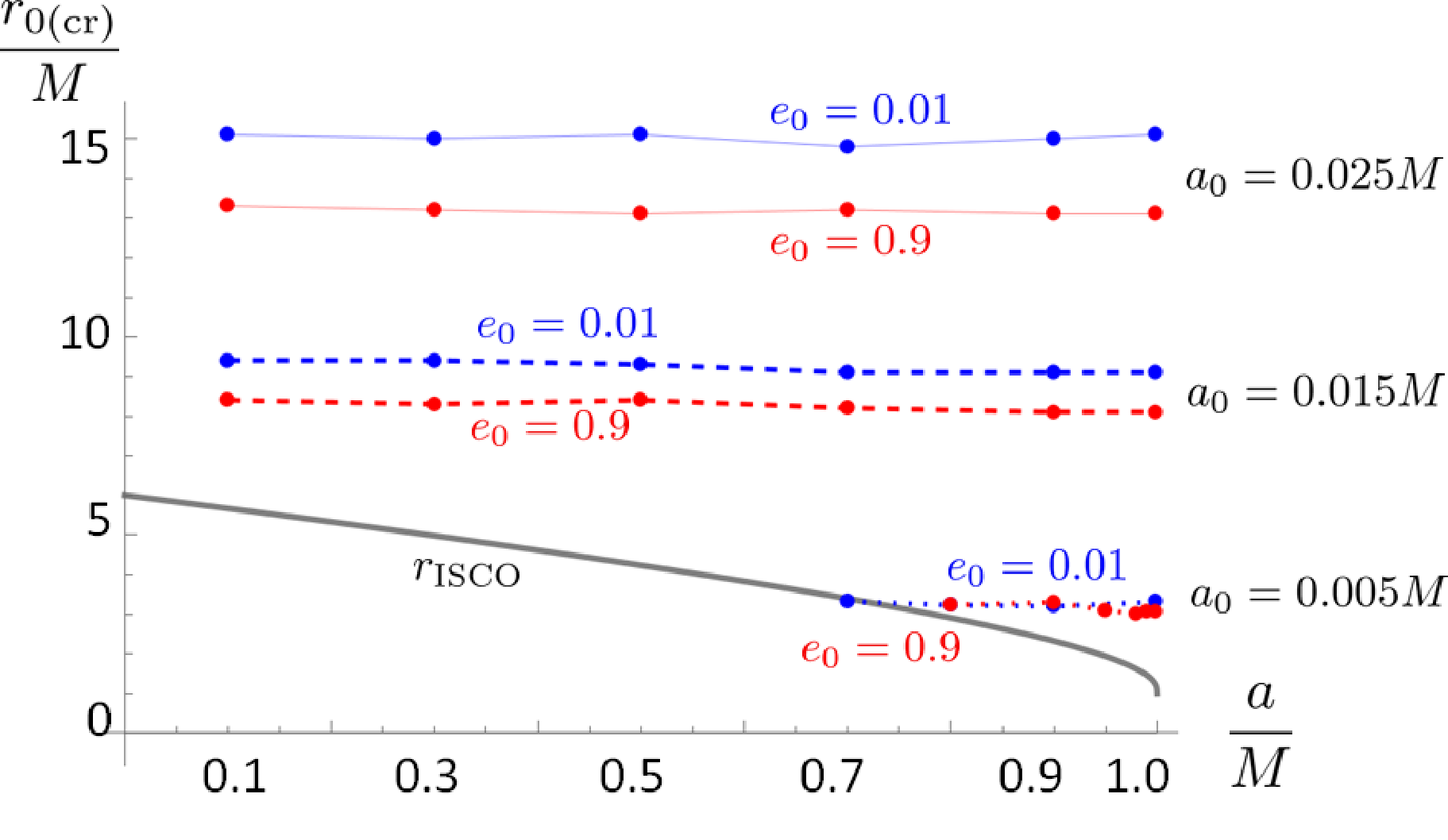}
\caption{The minimum radius ($\mathfrak{r}_{0 {\rm (cr)}}$) 
 in terms of Kerr rotation parameter $a$ for 
given values of the semi-major axis $\mathfrak{a}_0=0.025M ({\rm thin~line})$,  $0.015M ({\rm dashed~line})$, $0.005 M ({\rm dotted~line})$.
The blue and red correspond to the cases of 
the initial eccentricity $e_0=0.01$ and  $0.9$, respectively.
We fix the other initial orbital parameters as $e_0=0.01\,, I_0=85^\circ\,,\omega_0=60^\circ\,, \Omega_0=30^\circ$. Below the critical radius  
$\mathfrak{r}_{0 {\rm (cr)}}$, we always find unstable binary.  Beyond the critical radius, stable orbits and unstable ones appear randomly.
 We also show the ISCO radius (black), 
 below which the circular orbit of the CM becomes unstable.}
\label{fig:r0cr85}
\end{center}
\end{figure}

In Fig. \ref{fig:r0cr85}, we show this minimum value 
($\mathfrak{r}_{0 {\rm (cr)}}$) for given orbital parameters
 as a reference of chaotic instability.
We choose the initial orbital parameters of a binary as
the semi-major axis $\mathfrak{a}_0=0.005$ (dotted line), 0.015
(dashed line) , and $0.025 M$ (thin line). 
The initial eccentricity $e_0$ is chosen $0.01$(blue) and $0.9$(red).
We fix the other initial orbital parameters as $ I_0=85^\circ\,, \omega_0=60^\circ\,, \Omega_0=30^\circ$. 

Below the critical radius $\mathfrak{r}_{0 {\rm (cr)}}$, we always find unstable binary within $\tilde \tau=10^3$. While,
 beyond 
the critical radius, 
 stable and unstable orbits  appear randomly.
 In this case, however, unstable orbit can evolve beyond
 $\tilde \tau =10^3$, a binary is broken before $\tilde \tau =10^4$.
Since the instability appears after many orbital cycles, 
 it is not dynamical instability but may be caused by chaotic instability.
 It is consistent with the fact that stable and unstable orbits  appear randomly beyond the critical radius. 
 The appearance of unstable orbits 
 becomes less frequent as $\mathfrak{r}_{0}$ increases.
 We expect that the system becomes much stable when 
 $\mathfrak{r}_{0}$ is large enough as we show later.
 
 Fig. \ref{fig:r0cr85} shows that the critical radius $\mathfrak{r}_{0 {\rm (cr)}}$ is almost independent  of the Kerr parameter $a$. 
We also find that its dependence on the initial eccentricity is rather small.
The compactness of a binary (the initial semi-major axis $\mathfrak{a}_0$) is the most important factor. 
We find that the more compact binary can have stable orbits 
closer to SMBH.
In fact, a stable binary with $\mathfrak{a}_0=0.005 M$ 
can exist near the ISCO radius.

\begin{figure}[htbp]
\begin{center}
\includegraphics[width=8cm]{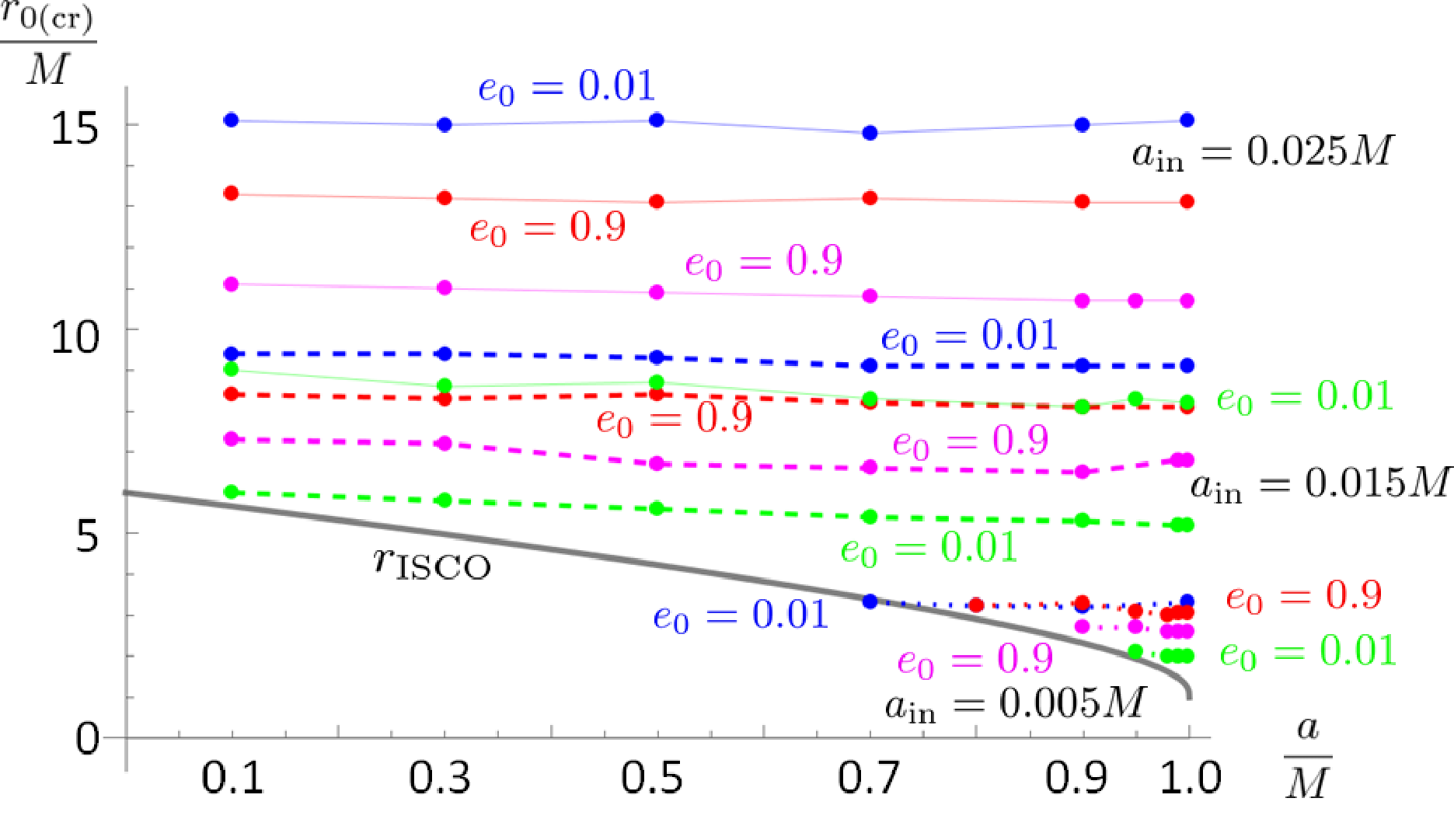}
\caption{The critical radius ($\mathfrak{r}_{0 {\rm (cr)}}$) 
 in terms of Kerr rotation parameter $a$ for 
given values of the semi-major axis $\mathfrak{a}_0=0.025M ({\rm thin~line})$,  $0.015M ({\rm dashed~line})$, $0.005 M ({\rm dotted~line})$.
We add the coplanar cases ($I_0=0^\circ$)
in Fig. \ref{fig:r0cr85}.
The green and magenta colored lines correspond to the cases of 
the initial eccentricity $e_0=0.01$ and  $0.9$, respectively. 
We fix the other initial orbital parameters as  $\omega_0=60^\circ\,, \Omega_0=30^\circ$. }
\label{fig:r0cr}
\end{center}
\end{figure}

The inclination dependence is also important as shown in 
Fig. \ref{fig:r0cr}, in which we include the coplanar cases
($I_0=0^\circ$). 
The green and magenta colored lines correspond to 
the cases with the initial eccentricity $e_0=0.01$ and $0.9$, 
respectively.

Fig.~\ref{fig:r0cr} shows that the critical radius in the coplanar case 
is smaller than that in the highly inclined orbit.
The coplanar binary is more stable than the highly inclined binary.
It may be because the vZLK oscillation appears in the highly inclined orbit.

\begin{figure}[htbp]
\begin{center}
\includegraphics[width=7cm]{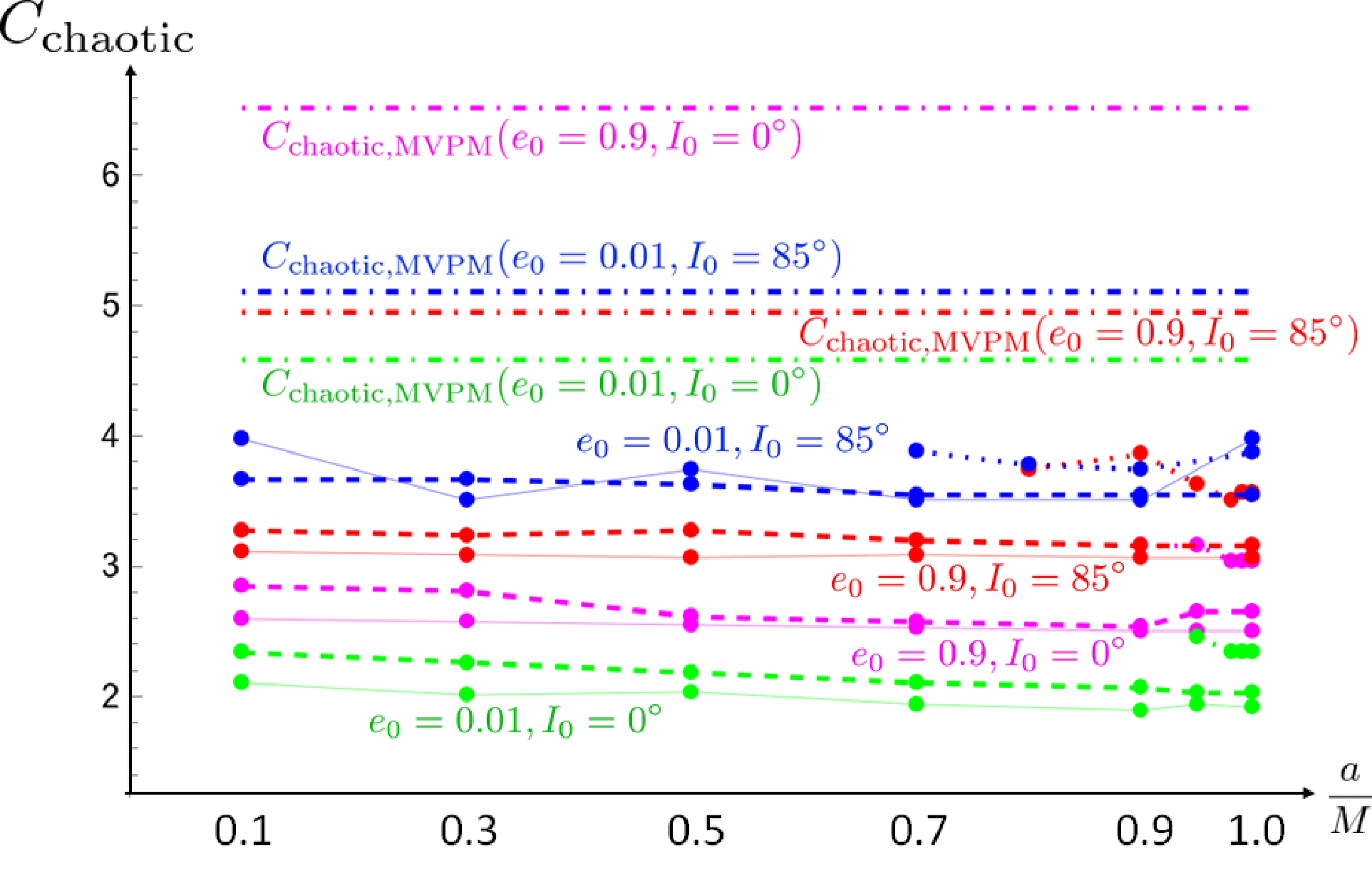}
\caption{$C_{\rm chaotic}$ is shown for various values of the parameters. 
The colors and types of lines denote the same models as those in Fig. \ref{fig:r0cr}. The values of $C_{\rm chaotic}$ for the highly inclined 
orbits is larger than those for the coplanar orbits. 
As references, we also show the values of $C_{\rm chaotic}$ 
in the criterion 2 by the dot-dashed lines.}
\label{fig:C_chaotic}
\end{center}
\end{figure}

As we discussed in Sec. \ref{validity}, the criterion of chaotic instability 
of three body system can be described by two parameters, 
$C_{\rm chaotic}$ and $p$.
In the present model, we can also evaluate those parameters. 
We find $p=1/3$ and $C_{\rm chaotic}=\mathfrak{f}_{\rm (cr)}^{1/3}$, 
because the critical value of the firmness parameter $\mathfrak{f}$ 
depend slightly on the initial eccentricity $e_0$ and inclination $I_0$, 
but are almost independent of the other parameters, 
which means the criterion 2 is more suitable to the present model.
We show our results of  $C_{\rm chaotic}$ in Fig. \ref{fig:C_chaotic}. 

The values of $C_{\rm chaotic}$ for the highly inclined 
orbits is  always larger than those for the coplanar orbits. 
The values of $C_{\rm chaotic}$  in our model
 is always smaller than those in the criterion 2, which is  given by Myll{\"a}ri et al (MVPM)\cite{myllari2018stability}.
 However, since 
the system is chaotic near
$C_{\rm chaotic}$, there is no clear critical value. In fact,
a binary with some initial data with the firmness
$\mathfrak{f} > \mathfrak{f}_{\rm cr} \equiv  C_{\rm chaotic}^3$
 shows instability. In this sense,
our result is consistent 
with the criterion 2.
\subsection{vZLK Oscillations}
\label{KL_oscillation}
In a hierarchical triple system, 
when the inclination is larger than some critical angle, a stable binary orbit shows vZLK oscillations.
\subsubsection{\rm Regular vZLK Oscillations}
\label{vZLK0185_005_10}
\begin{figure}[htbp]
\begin{center}
\includegraphics[width=6cm]{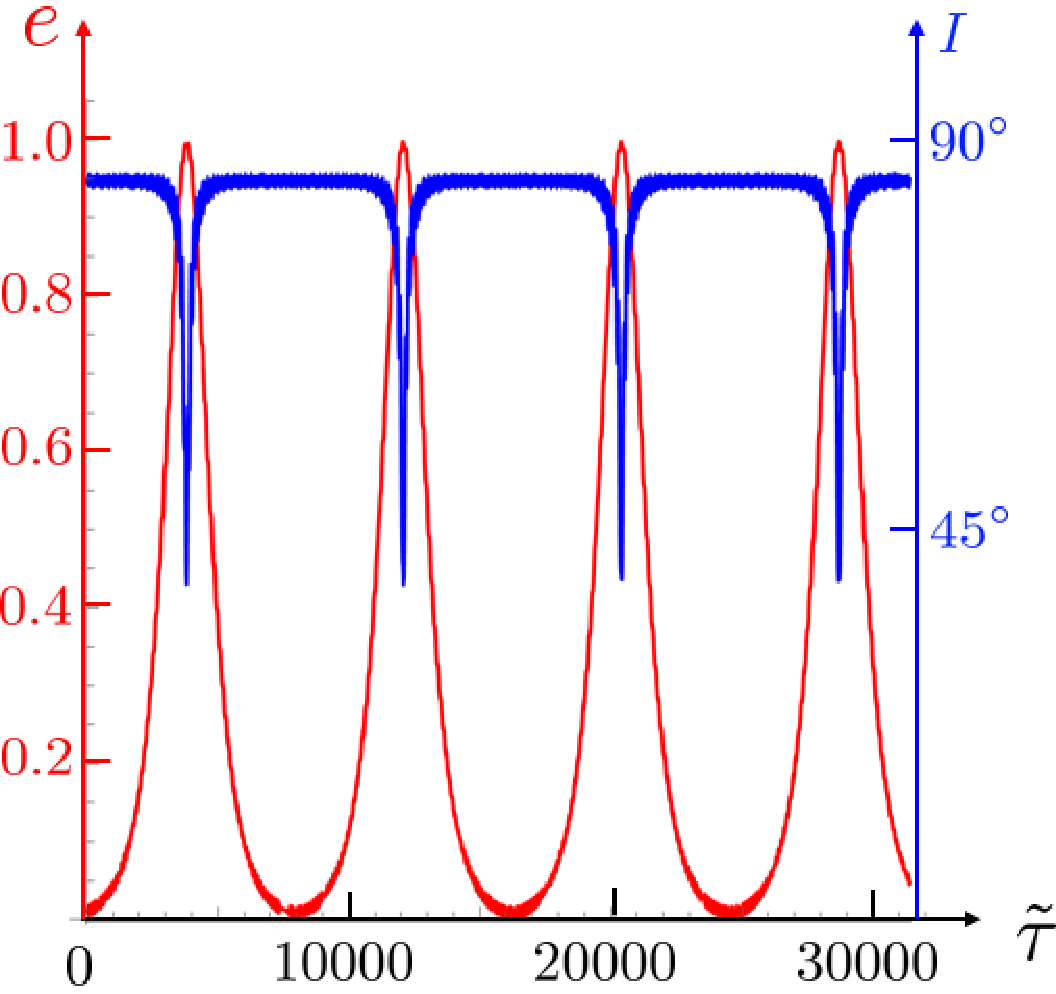}
\\
(a) prograde ($\sigma=1$)
\\[.5em]
\includegraphics[width=6cm]{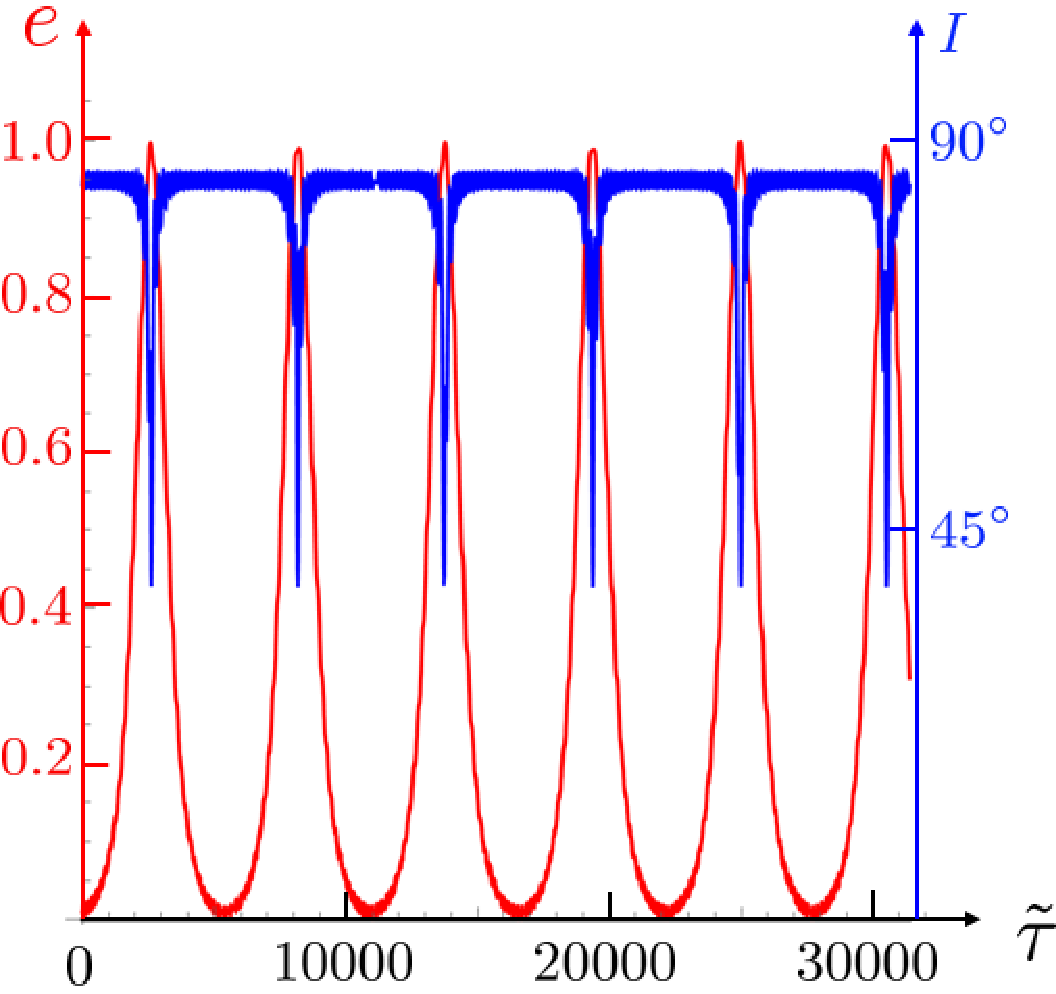}
\\
(b) retrograde ($\sigma=-1$)
\caption{Time evolution of the eccentricity $e$ and inclination angle $I$ 
with $\mathfrak{a}_0=0.005M$ and $\mathfrak{r}_0=10 M$. 
The Kerr parameter is $a=0.9M$, and the other 
initial orbital parameters of a binary are $
e_0=0.01\,, I_0=85^\circ, \,\omega_0=60^\circ,\,$ and $\Omega_0=30^\circ$.
 }
\label{fig:vZLK0185_005_10}
\end{center}
\end{figure}

We first show one example of the regular vZLK oscillations 
in Fig. \ref{fig:vZLK0185_005_10}. 
We choose 
the Kerr rotation parameter as $a=0.9M$, 
the initial orbital parameters of a binary as $\mathfrak{a}_0=0.005M\,,
e_0=0.01\,, I_0=85^\circ\,, \omega_0=60^\circ\,, \Omega_0=30^\circ$
and the circular radius as $\mathfrak{r}_0=10 M$.
We call it Model I.

We can clearly see the vZLK oscillations 
(a periodic exchange between the eccentricity $e$ and inclination $I$).
The eccentricity can reach almost unity.
Since the critical radius for 
$\mathfrak{a}_0=0.005M$ is 
$\mathfrak{r}_{0{\rm (cr)}}=3.2 M$, this model with $\mathfrak{r}_0=10 M$ should be highly stable.
In fact, as shown in Fig. \ref{fig:vZLK0185_005_10}, 
the oscillations  
are highly regular.

\subsubsection{{\rm Comparison with the double-averaging approach}}
\label{comparison_with_DA_approach}
In Newtonian hierarchical triple system, 
one sometimes use the double-averaging (DA) 
approach, in which we take averages of the 
Lagrange 
planetary equations over two orbital periods and analyze 
the averaged equations for the orbital parameters
in order to find their long-time behaviours.
In Appendix \ref{planetary_equations}, 
we discuss the planetary equations
in the present models and 
analyze the DA equations.

\begin{figure}[htbp]
\begin{center}
\includegraphics[width=6cm]{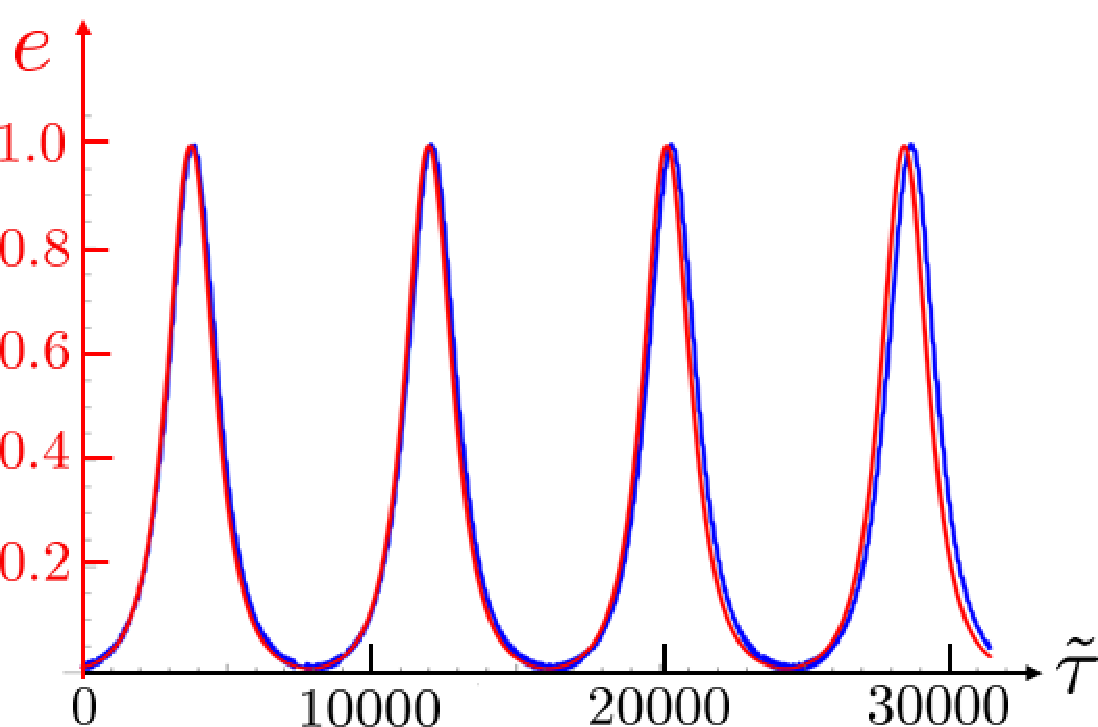}
\\[.5em]
(a) eccentricity evolution
\\[1em]
\includegraphics[width=6cm]{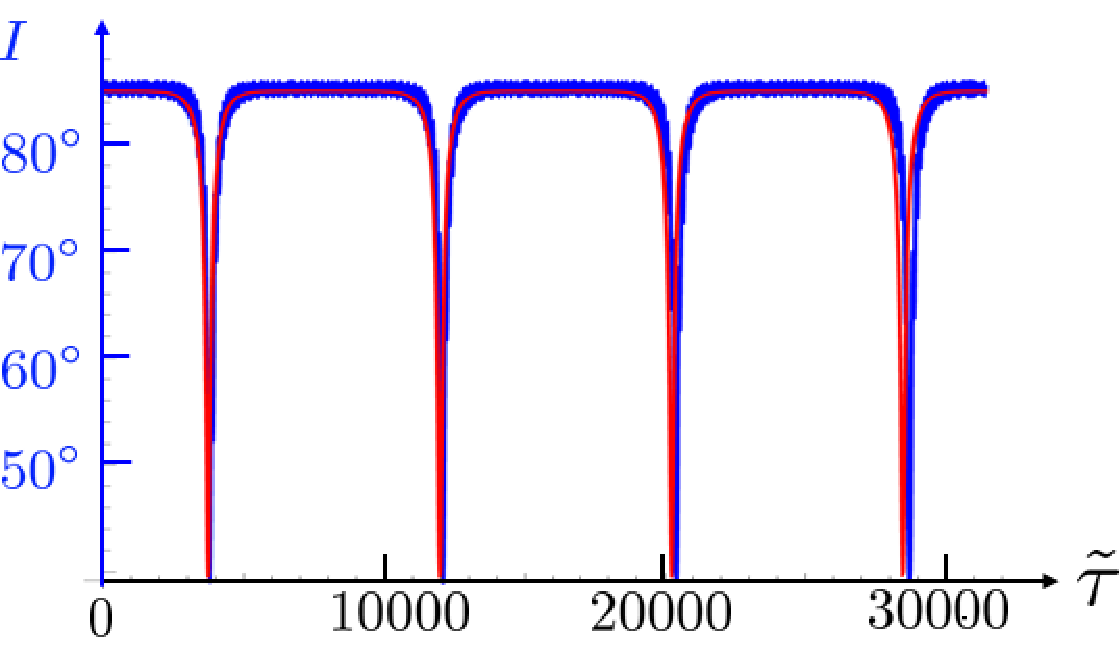}
\\[.5em]
(b) inclination evolution
\\
\caption{Comparison of the direct integration (blue) and the double-averaging approximation (red) of (a)  the eccentricity $e$ 
and  (b) inclination angle $I$.
The model is the same model as that in Fig. \ref{fig:vZLK0185_005_10}.
The results are completely overlapped.
}
\label{fig:DAvsDI_0185_005}
\end{center}
\end{figure}

In the present Model I, 
the DA approach gives very 
good approximation as
shown in Fig. \ref{fig:DAvsDI_0185_005},
which is the result for the prograde orbit.
We also find the almost same figures for the retrograde orbit.

\subsubsection{{\rm Oscillation period}}
\label{Oscillation_period}
One of the important properties of vZLK oscillations 
for observation is the oscillation period 
$T_{\rm vZLK}$.
In Newtonian vZLK oscillations. It is 
approximately given by
\beann
T_{\rm vZLK}&\sim& {P_{\rm out}^2\over P_{\rm in}}=
{(m_1+m_2)\over M}\left({\mathfrak{r}_0\over \mathfrak{a}_0}\right)^3P_{\rm in}=\mathfrak{f}
\,P_{\rm in}
\,.
\enann
As we discussed in the previous subsection, 
when we find regular vZLK oscillations,
the DA approach gives a good approximation. 
Using the DA approximation, we can easily evaluate 
the oscillation period analytically as shown in Appendix \ref{planetary_equations}.
It shows that 
the relativistic effects 
reduce the vZLK time scale by a factor $0.1-0.25$.

One interesting observation in Fig.~\ref{fig:vZLK0185_005_10} is that the oscillation period for the retrograde orbit 
is smaller than that for the prograde orbit.
It is confirmed by Fig.~\ref{fig:vZLK_timescale}
 obtained by the DA approach.

\subsubsection{{\rm Critical inclination angle}}
\label{critical_inclination_angle}

The critical inclination angle, beyond which the vZLK oscillation occurs, 
can be obtained by evaluating the maximum value of the eccentricity, $e_{\rm max}$, 
in the models with very small initial eccentricity.

\begin{figure}[htbp]
\begin{center}
\includegraphics[width=7cm]{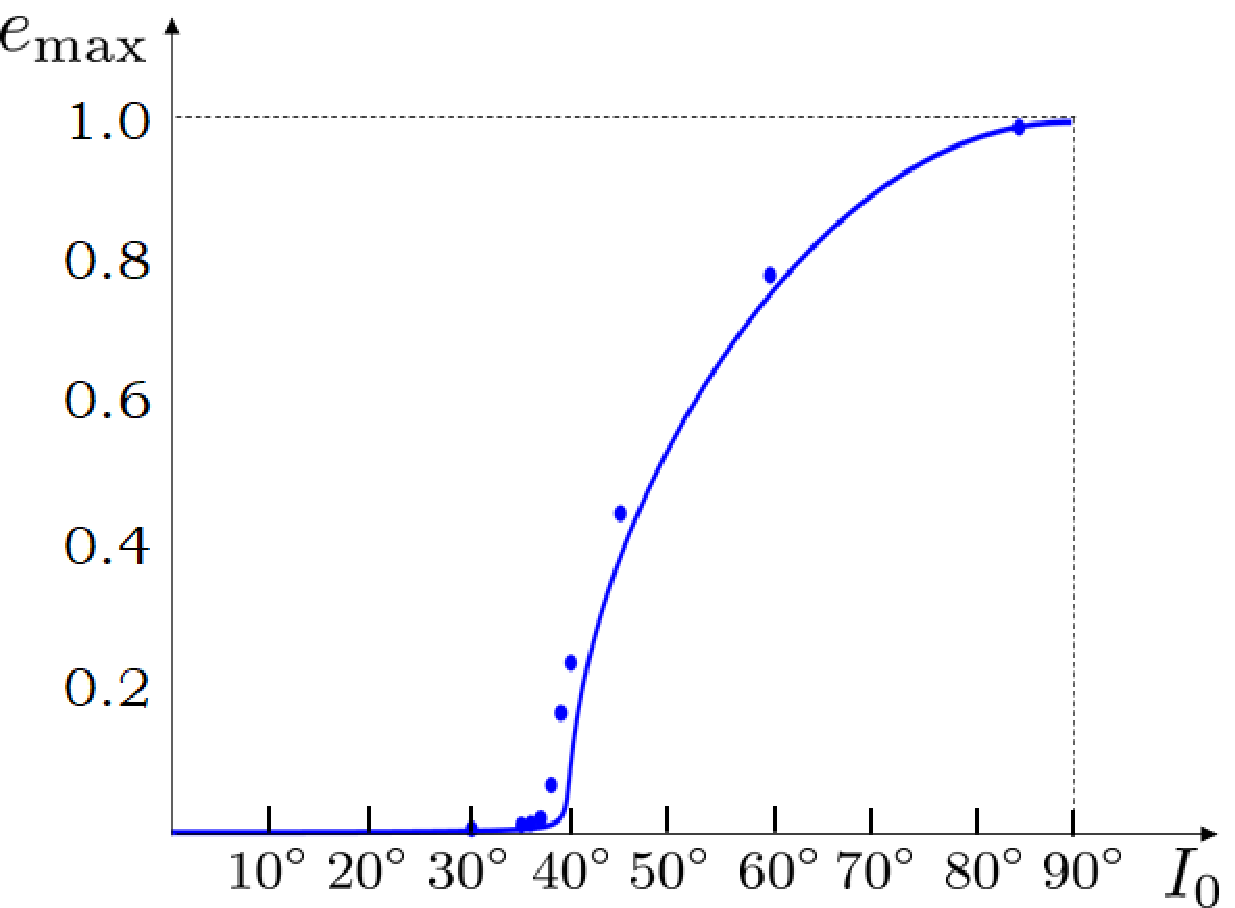}
\caption{The maximum value of the eccentricity in vZLK oscillations in terms of the initial inclination angle $I_0$. We assume $e_0=0.01$, $\mathfrak{a}_0=0.005M$, $\mathfrak{r}_0=10 M$, and $a=0.9 M$. The solid curve denotes $e_{\rm max}$ obtained by the DA approximation.}
\label{fig:I0vsemax}
\end{center}
\end{figure}

In Fig. \ref{fig:I0vsemax}, we show $e_{\rm max}$ in terms of 
the initial inclination angle $I_0$.
We choose $e_0=0.01$, $\mathfrak{a}_0=0.005M$ 
and $\mathfrak{r}_0=10 M$
to find regular vZLK oscillations.
The solid curve denotes $e_{\rm max}$ obtained by the DA approximation.
Although we show  the case of $a=0.9M$
 in Fig. \ref{fig:I0vsemax}, the results do not depend 
 on $a$  at all.
This figure shows that the critical 
inclination angle is a little less than 40$^\circ$.

The direct integration gives a little bit smaller value of the critical inclination angle than the double-averaging (DA) approximation,
but the DA scheme is not too contrasting.
In the extreme limit 
such that $\mathfrak{r}_0\rightarrow r_{\rm ISCO}$ 
as well as $a\rightarrow M$,
the larger critical inclination angle such as $60^\circ$ 
could be found as discussed in Appendix \ref{critical_inclination_angle_in_DA}. 
However, it turns out that 
when we restrict to stable binary orbits, 
the critical value is much smaller
even in the extreme limit. 
This is confirmed by the direct integration
(red dots in Fig.~\ref{fig:I0cr_DA999}), which 
provides almost the same as Fig.~\ref{fig:I0vsemax}.

\subsection{Chaotic features}
\label{chaotic_behaviour}
Since a three-body system is non-integrable, 
we may find chaotic features in binary orbits, 
especially near the critical boundary 
between stable and unstable orbits.
As we discussed in Sec. \ref{chaotic_instability}, 
there exists a critical radius $\mathfrak{r}_{0{\rm (cr)}}$
for given orbital parameters of a binary, below which 
a binary is broken before $\tilde \tau=10^3$.
We then expect some chaotic features appear 
in a stable binary orbit with a slightly larger radius than 
 $\mathfrak{r}_{0{\rm (cr)}}$. Here we give three models (Model II, Model III, and Model IV) 
listed in Table \ref{TableI}.\\
\begin{table}[h]
\begin{center}
\begin{tabular}{|c||c|c|c|c|c|c|c|}
\hline 
&&&&&&&\\[-.5em]
Model &$a/M$&$\mathfrak{a}_0/M $&$\mathfrak{r}_0/M$&$e_0$&$I_0$ &$\omega_0$&$\Omega_0$ 
\\
\hline
I&$0.9$&$0.005$&$10$&$0.01$&$85^\circ$&$60^\circ$&$30^\circ$ 
\\
\hline
II&$0.9$&$0.005$&$2.9$&$0.01$&$60^\circ$&$60^\circ$&$30^\circ$ 
\\
\hline
III&$0.9$&$0.005$&$3.2$&$0.01$&$85^\circ$&$60^\circ$&$30^\circ$ 
\\
\hline
IV&$0.9$&$0.015$&$10$&$0.01$&$85^\circ$&$60^\circ$&$30^\circ$ 
\\
\hline
 \end{tabular}
\caption{Parameters of the models. The Kerr rotation parameter $a$, the argument of periapsis $\omega_0$, and 
 the ascending node 
 $\Omega_0$ are fixed. }
\label{TableI}
\end{center}
\end{table}

 In Model II, 
 we change the circular radius $\mathfrak{r}_0$
from $10M$ (Model I) to the critical radius $\mathfrak{r}_{0{\rm (cr)}}=2.9M$ as well as
the initial inclination angle $I_0$ from $85^\circ$ to $60^\circ$,
while 
in Model III we change only  the circular radius $\mathfrak{r}_0$ from $10M$ 
to $\mathfrak{r}_{0{\rm (cr)}}=3.2M$.
In the model IV, 
we change only the semi-major axis $\mathfrak{a}_0$ from 
$0.005M$ in Model I to $0.015M$. 
In Model II and Model III, 
the parameters are just 
on the the critical boundary for chaotic instability, while 
in Model IV, they are near the instability boundary.

\begin{figure}[htbp]
\begin{center}
\includegraphics[width=6cm]{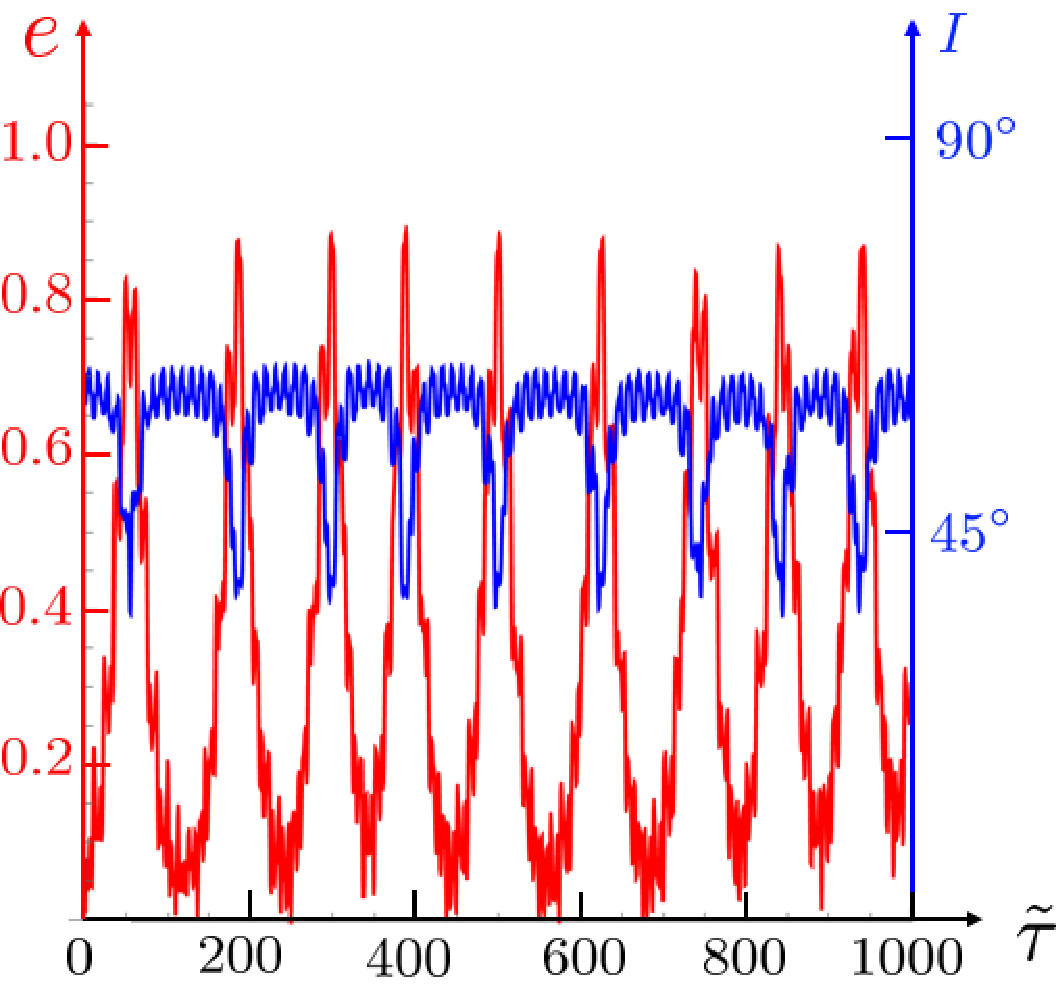}
\\[.5em]
(a) Model II 
\\[2em]
\includegraphics[width=6cm]{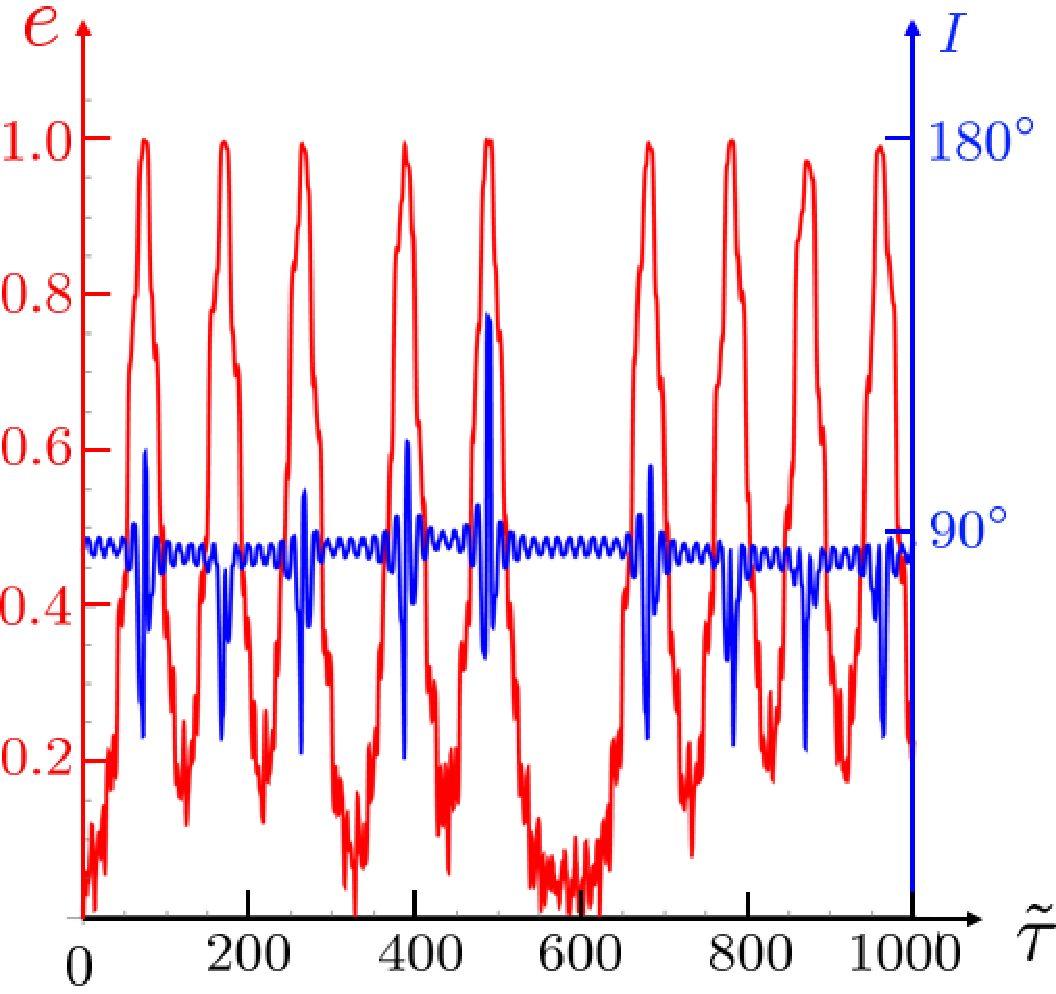}
\\[.5em]
(b) Model III 
\caption{Time evolution of eccentricity and inclination for a) Model II and b) Model III, respectively. The model parameters are given in Table~\ref{TableI}.
}
\label{fig:vZLK01_cr}
\end{center}
\end{figure}

\begin{figure}[htbp]
\begin{center}
\includegraphics[width=6cm]{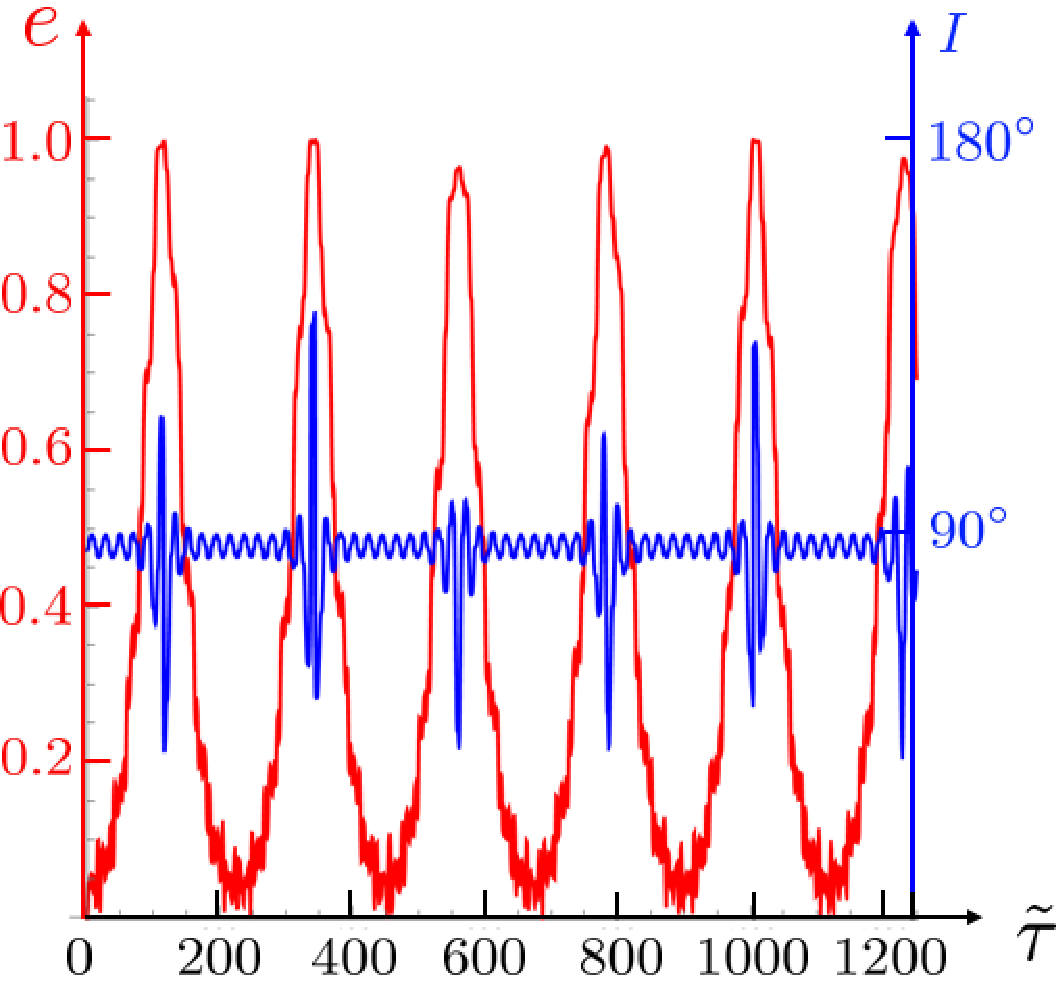}
\caption{Time evolution of eccentricity and inclination for Model IV. The model parameters are given in Table~\ref{TableI}.
}
\label{fig:vZLK0185_015_10}
\end{center}
\end{figure}
We find two typical chaotic features in those models:
irregular vZLK oscillations and orbital flip.

\subsubsection{\rm Irregular oscillations}
\label{irregular_oscillations}
In Figs. \ref{fig:vZLK01_cr} and \ref{fig:vZLK0185_015_10}, we show 
time evolution of the eccentricity
(red curve) and inclination (blue curve).
We find the vZLK oscillations are stable 
but quite irregular.
We can easily find that the oscillation 
period is not constant 
in Fig. \ref{fig:vZLK01_cr} (b).
We also see that the maximum and minimum values of the eccentricities are changing in time.

Since the oscillation period looks regular, 
Model IV is less chaotic than 
Model II and Model III.
It is because  
$\mathfrak{r}_0=10M$ of Model IV 
is slightly larger than the critical radius 
$\mathfrak{r}_{0{\rm (cr)}}=9.1 M$
for $\mathfrak{a}_0=0.015M$, while 
the parameters of Model II and Model III
are just on the critical boundary for chaotic instability.

\subsubsection{\rm Orbital Flip}
\label{orbital_flip}
When the 
initial inclination angle 
is close to $90^\circ$, we may expect a phenomenon of orbital flip, in which the inclination angle 
will evolve beyond $90^\circ$. We do not find such phenomenon 
when the initial inclination angle is not so large, e.g., $I_0=60^\circ$
as shown in Fig. \ref{fig:vZLK01_cr} (a).
Even the initial inclination angle is close to $90^\circ$, 
the orbital flip does not occur 
in regular 
vZLK oscillations as shown in Fig. \ref{fig:vZLK0185_005_10}.

When the initial inclination angle is 
close to $I_0=90^\circ$ and the vZLK oscillations are chaotic
(Model III and Model IV),
we find the orbital flips as shown in
Figs.~\ref{fig:vZLK01_cr}(b) and \ref{fig:vZLK0185_015_10}.
Although it is quite random when the orbital flip occurs, it happens when the eccentricity gets large.
It is interesting because in a regular vZLK oscillations, the eccentricity is larger when 
the inclination angle becomes smaller, 
while in the orbital flip 
models, 
the eccentricity becomes very large
not only when the inclination angle 
becomes smaller but also 
when the inclination angle goes beyond $90^\circ$.

\subsection{Rotation Dependence}
\label{rotation_dependence}
As we mentioned before, the properties of dynamics of a binary are little dependent of a SMBH rotation parameter $a$ 
when the circular radius $\mathfrak{r}_0$ and the initial semi-major axis $\mathfrak{a}_0$ are fixed.
It is because the Riemann curvature on the equatorial plane only depends on $\mathfrak{r}_0$ just as Schwarzschild BH and 
the procession frequency $\mathfrak{w}_{\rm P}$ at the fixed $\mathfrak{r}_0$ very weakly depends on $a$ as shown in Fig.\ref{fig:precession1}. 

\begin{figure}[htbp]
\begin{center}
\includegraphics[width=7cm]{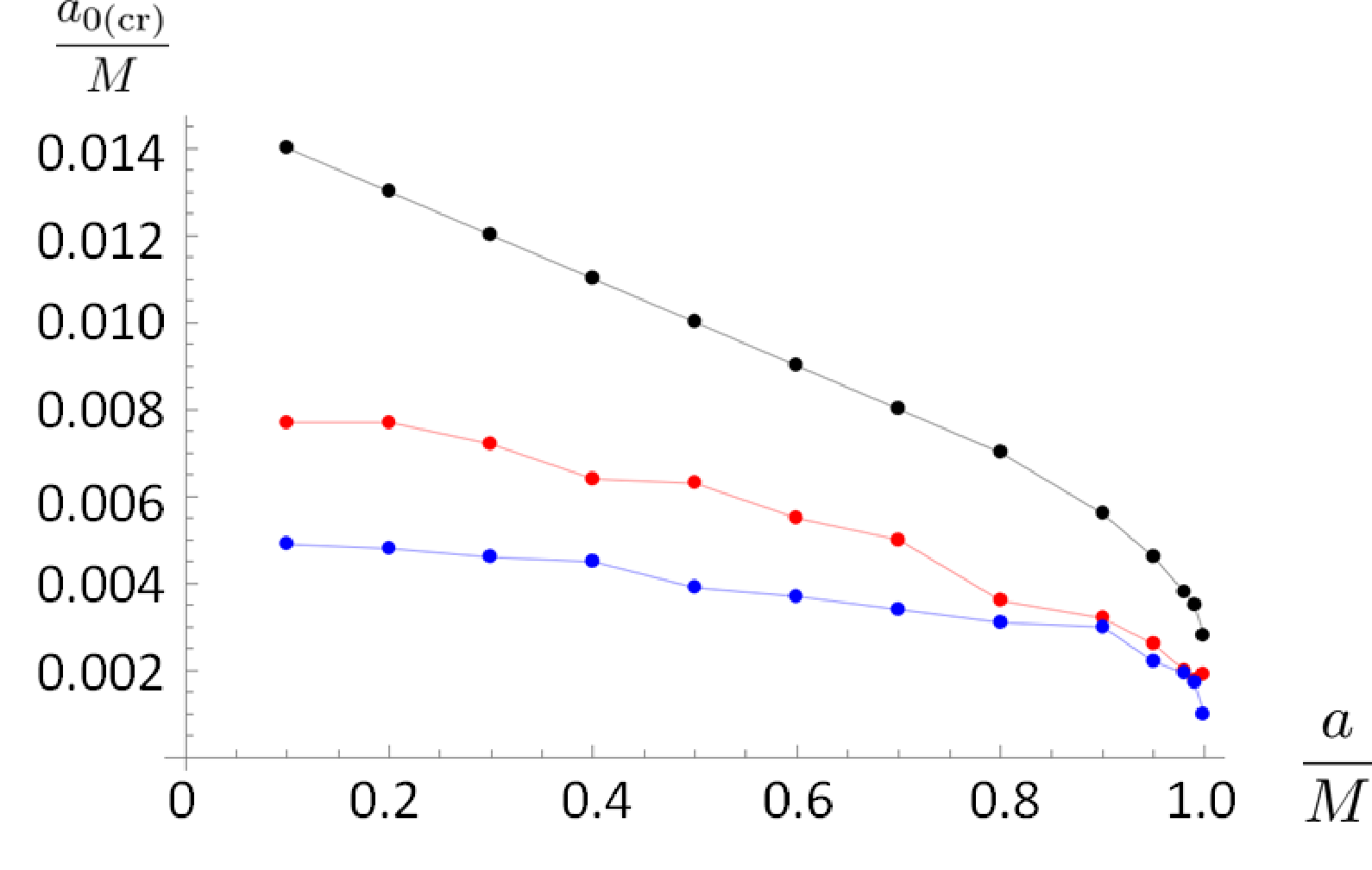}
\includegraphics[width=7cm]{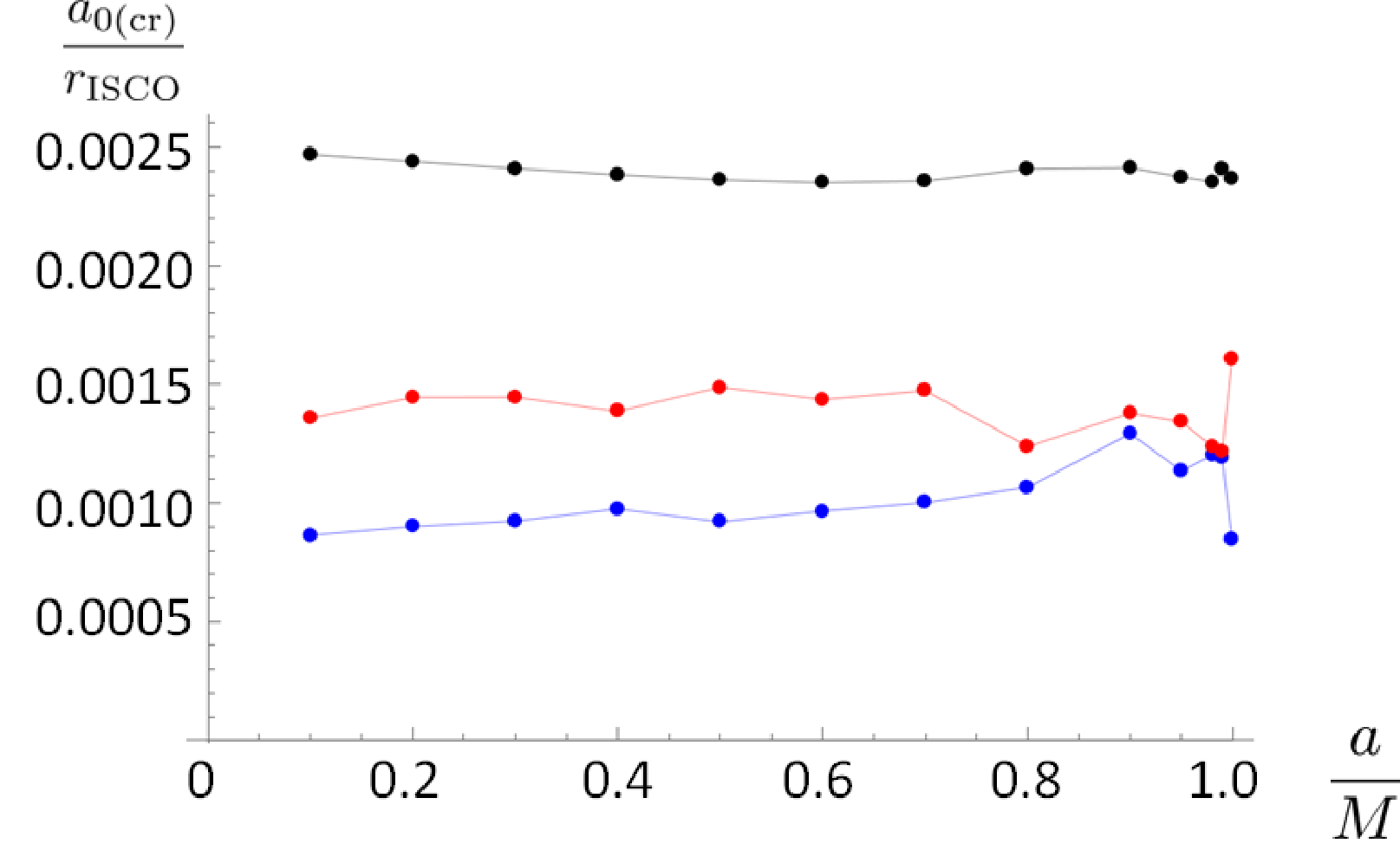}
\caption{(Top) Critical semi-major axis $\mathfrak{a}_{0{\rm (cr)}}$ at $\mathfrak{r}_{\rm ISCO}$ in terms of the Kerr rotation parameter $a$. 
(Bottom) $\mathfrak{a}_{0{\rm (cr)}}/\mathfrak{r}_{\rm ISCO}$ in terms of $a$.
The initial eccentricity is $e_{0}=0.01$, while the initial inclination angle are chosen as
$I_0=0$ (black)  and $I_0=85^\circ$ (red and blue).
For the coplanar case ($I_0=0$), all orbits are stable below black curves.
For $I_0=85^\circ$, all orbits are stable below blue curves, while there exist stable and unstable orbits between blue curves and red curves.
}
\label{aincr_rISCO}
\end{center}
\end{figure}

However, the ISCO radius, below which the CM of a binary becomes unstable as shown in Appendix \ref{CM_motion}, highly depends on $a$. 
As a result, a highly compact binary can exist at more inner circular radius for a more rapidly rotating SMBH.
In the top figure of 
Fig. \ref{aincr_rISCO}, 
We show the critical semi-major axis $\mathfrak{a}_{0{\rm (cr)}}$ below which a binary is stable. The circular radius is chosen at $\mathfrak{r}_0=
\mathfrak{r}_{\rm ISCO}$.
The initial eccentricity is $e_{0}=0.01$, while the initial inclination angle is chosen as
$I_0=0$ (coplanar)  and $I_0=85^\circ$.
The coplanar case ($I_0=0$) is shown by black curves.
Since it is a regular orbit, the boundary between stable and unstable orbit is clear.
On the other hand, in highly inclined case ($I_0=85^\circ$), since the system is chaotic, 
stability highly depends on the initial conditions. 
As a result, there are two critical radii; one is
that all orbits are stable below some critical radius (blue curves), 
and the other is that 
there exist stable and unstable orbits between blue curves and 
another critical radius (red curves).

In the bottom figure of 
Fig. \ref{aincr_rISCO}, 
we show $\mathfrak{a}_{0{\rm (cr)}}/\mathfrak{r}_{\rm ISCO}$,
which is little dependent of 
the Kerr rotation parameter $a$.
In the extreme limit of $a\rightarrow M$, there appears 
strange behaviours especially for $I_0=85^\circ$. It is because $\mathfrak{w}_{\rm P}$
becomes very large near the ISCO radius (see Fig. \ref{fig:precession}).

\section{Summary and Discussion}
\label{summary}
In this paper, we discuss dynamics of a binary system orbiting around 
a rotating SMBH.
Assuming an observer rotating on a nearly circular orbit
around a Kerr SMBH, 
we write down the equations of motion of a binary
in the observer's local inertial frame.
Using Fermi-Walker transport with small acceleration, 
which removes the interaction terms between the CM of a binary and its relative coordinates,
we set up Newtonian self-gravitating system in the local 
proper reference frame.
As a result, the CM of a binary 
follows the observer's orbit, 
but its motion deviates from an exact geodesic.
Since the relative motion is decoupled from the system, 
we first solve it, and  then 
find the motion of the CM by the perturbation equations with the small acceleration, which is given by the relative motion. 

To present our results, we first discuss the stability conditions.
In the hierarchical triple system, there are two widely used criteria for 
chaotic instability, which 
are evaluated by $N$-body simulations by two independent groups
\cite{mardling2001tidal,myllari2018stability}.  
The criterion given by \cite{myllari2018stability}
fits our model because the dependence of mass parameter is the same 
as $p=1/3$.
The critical value for chaotic instability ($C_{\rm chaotic}$)
in our model is slightly 
smaller than the result by  \cite{myllari2018stability}.
However, the system is chaotic near $C_{\rm chaotic}$, 
there is no clear critical value. In fact, 
a binary with 
some initial data with the firmness 
$\mathfrak{f}>\mathfrak{f}_{\rm cr}\equiv C_{\rm chaotic}^{3}$
shows instability.
In this sense, our result is consistent with that by  \cite{myllari2018stability}.

We then analyze the properties of stable vZLK oscillations.
Although we show the result only for the case of $a=0.9M$,
the behaviours of a binary are almost the same when we fix 
the circular radius $\mathfrak{r}_0$ and the initial 
semi-major axis $\mathfrak{a}_0$.

For highly compact binaries with large firmness parameter 
($\mathfrak{f}\gg \mathfrak{f}_{\rm cr}$), 
the vZLK oscillations is quite regular and stable.
The double-averaging method gives a good approximation in this parameter space.
The critical inclination angle for vZLK oscillations 
is about $40^\circ$, 
which is also almost independent of the rotation parameter $a$.

For the binary with the firmness parameter 
slightly larger than the critical value 
($\mathfrak{f}\gsim \mathfrak{f}_{\rm cr}$), 
since the system is chaotic, we find 
chaotic vZLK oscillations, which
become irregular both in the oscillation period and in the amplitude.
If the initial inclination is large, we find an orbital flip, which also appears randomly.

Next important and interesting subject is the gravitational waves
from a binary system orbiting around SMBH. 
There are two GW sources in a hierarchical triple system: One is the GWs from an inner binary and the other is those from the outer binary.

The mereger time scale of circular binary due to emission of gravitational waves is evaluated as~\cite{MMbook}
 \beann
T_{\rm GW}={5\over 256}{c^5 R_0^4\over G^3 
 m^2\mu}
 \,,
  \enann
  where $m=m_1+m_2$ and $\mu=m_1m_2/(m_1+m_2)$ and $R_0$ is the initial distance.
  Hence, the ratio of the time scale of an outer binary with the radius $\mathfrak{r}_0$ to that of an inner binary with the semi-major axis 
  $\mathfrak{a}_0$ is
  \beann
  {T_{\rm outer}\over T_{\rm inner}}&=&{m_1m_2\over M^2}
  \left({\mathfrak{r}_0\over \mathfrak{a}_0}\right)^4
  \\
  &=&
  {m_1m_2\over M^2}\left({M\over m_1+m_2}\right)^{4\over 3}\mathfrak{f}^{4\over 3}
    \,.
  \enann
For the present model ($m_1=m_2=10M_\odot, M=10^8M_\odot$), 
since
$ {T_{\rm outer}/T_{\rm inner}}\approx 8.55\times 10^{-6}\,\mathfrak{f}^{4\over 3}$, we usually find 
$T_{\rm outer}\ll T_{\rm inner}$ except for a highly compact binary
such as $\mathfrak{f}>6000$.
Hence, in most cases,  gravitational waves from the inner binary are less effective compared with those from the outer orbit.

However if the inner binary has large eccentricity, it may not be the case.
The emission time scale is 
reduced by the factor~\cite{Peters-Mathews} 
\beann
F(e_{\rm in})\approx  {768\over  429 }
\left(1-e_{\rm in}^2\right)^{7/2},
\enann
when $e_{\rm in}\approx 1$.
As a result, the GWs from the inner binary may become larger than those from the outer binary.

\begin{figure}[htbp]
\begin{center}
\includegraphics[width=6cm]{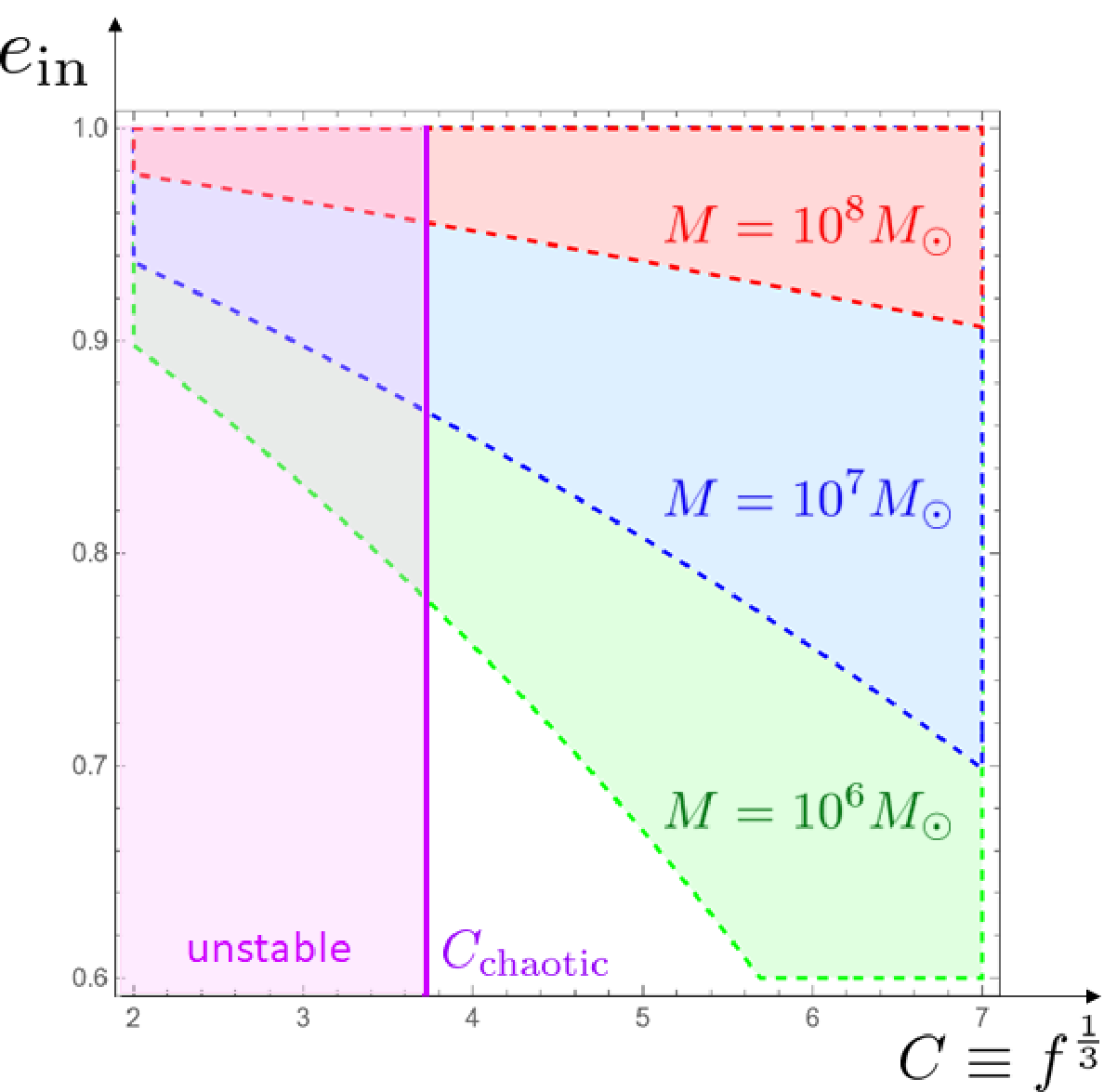}
\caption{The parameter region of the firmness $\mathfrak{f}$ and 
the inner eccentricity for the condition $T_{\rm outer}\geq T_{\rm inner}$.
We assume $m_1=m_2=10M_\odot$, and $M=10^6 M_\odot ({\rm green})\,,10^7 M_\odot ({\rm blue})\,$ or $10^8 M_\odot ({\rm red})$. 
Here we adopt $C_{\rm chaotic}\approx 3.74$ of Model III.}
\label{fig:coalescence_time}
\end{center}
\end{figure}

In Fig. \ref{fig:coalescence_time}, we show the 
parameter region of the firmness $\mathfrak{f}$ and 
the inner eccentricity $e_{\rm in}$, which satisfies 
the condition $T_{\rm outer}\gsim T_{\rm inner}$, which is 
rewritten as
\beann
{\mathfrak{f}^{4\over 3}\over (1-e_{\rm in}^2)^{7\over 2}}\gsim {M^2\over m_1m_2}\left({M\over m_1+m_2}\right)^{4\over 3}
\enann
We assume $m_1=m_2=10M_\odot$, and $M=10^6\,,10^7$ or $10^8 M_\odot$. 
As for the chaotic instability condition, 
$C_{\rm chaotic}\approx 3-4$ in the present model, 
although it depends on binary configuration.
In this figure, we adopt the value of Model III ($C_{\rm chaotic}
\approx 3.74$).

From this figure, in the present model 
($m_1=m_2=10M_\odot\,,M=10^8 M_\odot$), we find that 
if $e_{\rm in} \gsim 0.96$, $T_{\rm outer}\gsim T_{\rm inner}$.
In this case, the GWs from inner binary becomes more important.
The binary can be orbiting near the ISCO radius.

It may happen for an almost coplanar binary with large initial eccentricity.
However, the more interesting case is 
when the vZLK oscillations is found in a binary motion.
We expect a large amount of the GW emission because the eccentricity becomes large.
The large eccentricity also provides much higher frequencies than that from a circular binary~\cite{Lisa2019,Hoang2019,Gupta_2020}.
Another interesting point on the GWs from the vZLK oscillations is 
that the large amount of the GW emission repeats periodically
with the vZLK oscillation time scale.
It is a good advantage in the observations because we have 
a certain preparation time for next observations.

In recent years, three-body systems and the emission of GWs from them have received significant attention~\cite{Amaro-Seoane2010, Antonini2012,hoang18,Antonini2016,Meiron2017,Robson2018,Lisa2018,Lisa2019,Hoang2019,Loeb2019,Gupta_2020}. Our future work will involve evaluating the GWs from the present hierarchical triple setting using the black hole perturbation approach, since near the ISCO radius the quadrupole formula may not be valid~\cite{PhysRevD.103.L081501}.

In this paper, we assume that 
the CM of a binary moves along an almost circular orbit, but 
an eccentric orbit is interesting to be studied since the vZLK oscillation may be modulated on a longer timescale\cite{Lithwick:2011hh,Katz:2011hn,Naoz:2011mb,Li2014ApJ,Liu2015MNRAS}.
However, for such a highly eccentric orbit, 
the present proper reference frame expanded 
up to the second order of the spatial coordinates $x^{\hat a}$
may not be sufficient.
We may need higher-order terms in the metric,
where the derivatives of the Riemann curvature appear\cite{Nesterov_1999,Delva:2011abw}.
Although the basic equations are very complicated, 
such an extension is straightforward.


\begin{acknowledgments}
We thank  Luc Blanchet, Vitor Cardoso, Eric Gourgoulhon and Haruka Suzuki for
useful discussions. This work was supported in part by JSPS KAKENHI
Grant Numbers  JP17H06359, JP19K03857 (K.M.), and by
JP20K03953 (H.O.). P.G. acknowledges support from C.V. Raman fellowship grant. K.M.
would like to acknowledges the Yukawa Institute for Theoretical
Physics at Kyoto University, where the present work was begun during
the Visitors Program of FY2022. 
He would also thank Niels Bohr Institute/Niels Bohr International Academy, The Max Planck Institute for Gravitational Physics (Albert Einstein Institute), The laboratory AstroParticle and Cosmology (APC), and CENTRA/Instituto Superior T\'ecnico, where most of this work was performed.

\end{acknowledgments}

\appendix
\begin{widetext}
\section{Motion with 0.5 PN correction term}
\label{CM_motion}

As we discussed in the text, in order to analyze the motion of the CM of a binary system, we have to consider 0.5 PN terms.
As shown before, we can assume  $\vect{R}=0$  by introduction of the acceleration given by Eq. (\ref{0.5PN_acceleration}).
We first solve the relative coordinates $\vect{r}$, and then 
the motion of the observer (or the CM).

\subsection{\bf Equations of motion for relative coordinates}
The equation of motion for relative coordinates $\vect{r}$ of a binary is now
given by 
\beann
\tilde {\cal L}_{\rm rel}(\vect{r}, \dot{\vect{r}})&=&{\cal L}_{\rm rel}(\vect{r}, \dot{\vect{r}})+{\cal L}_{1/2\mathchar`-{\rm rel}}(\vect{r}, \dot{\vect{r}})
\,,
\enann
where 
${\cal L}_{\rm rel}$  is given by Eq. (\ref{EoM_relative_coordinate}), while 
\beann
{\cal L}_{1/2\mathchar`-{\rm rel}}(\vect{r}, \dot{\vect{r}})&=&-\mu {2(m_1-m_2)\over 3(m_1+m_2)}\left(
\bar{\cal R}_{\hat 0\hat x\hat y\hat x}x(x\dot y-y\dot x)
+\bar{\cal R}_{\hat 0\hat z\hat y\hat z}z(z\dot y-y\dot z)\right)
\\
&=&-\mu {2(m_1-m_2)M\over (m_1+m_2)\mathfrak{r}_0^3}
{\sqrt{\Delta(\mathfrak{r}_0)}(\sigma\sqrt{M\mathfrak{r}_0}-a)\over  F^2_\sigma(\mathfrak{r}_0)}
\left(
-x(x\dot y-y\dot x)
+z(z\dot y-y\dot z)\right)
\,.
\enann

 In non-rotating Fermi-Walker coordinates, 
 we find
 ${\cal L}_{\rm rel}$ is given by Eq. (\ref{EoM_relative_non-rotating}),
 while 
 \beann
{\cal L}_{1/2\mathchar`-{\rm rel}}(\vect{\mathsf{r}}, \dot{\vect{\mathsf{r}}})&=&
\mu {2(m_1-m_2)M\over (m_1+m_2)\mathfrak{r}_0^3}
{\sqrt{\Delta(\mathfrak{r}_0)}(\sigma\sqrt{M\mathfrak{r}_0}-a)\over  F^2_\sigma(\mathfrak{r}_0)}\times 
\\
&&
\Big{\{}\cos \mathfrak{w}_{\rm R}   \tau \left[\mathsf{x}\left(\mathsf{x}\dot{\mathsf{y}}-\mathsf{y}\dot{\mathsf{x}}\right)+\mathsf{z}\left(\mathsf{y}\dot{\mathsf{z}}-\mathsf{z}\dot{\mathsf{y}})+\mathfrak{w}_{\rm R}\mathsf{x}(\mathsf{x}^2+\mathsf{y}^2 -\mathsf{z}^2\right)\right]
\\
&&
-\sin \mathfrak{w}_{\rm R}   \tau\left[\mathsf{y}\left(\mathsf{x}\dot{\mathsf{y}}-\mathsf{y}\dot{\mathsf{x}}\right)+\mathsf{z}\left(\mathsf{z}\dot{\mathsf{x}}-\mathsf{x}\dot{\mathsf{z}})+\mathfrak{w}_{\rm R}\mathsf{y}(\mathsf{x}^2+\mathsf{y}^2 -\mathsf{z}^2\right)\right]\Big{\}}.
\enann

The momentum is obtained from the Lagrangian $\tilde {\cal L}_{\rm rel}(\vect{\mathsf{r}}, \dot{\vect{\mathsf{r}}})$ as 
\beann
p_{\mathsf{x}}&=&\mu \dot{\mathsf{x}} +\mu \mathfrak{w}_{\rm P}\mathsf{y}
+\mu {2(m_1-m_2)M\over (m_1+m_2)\mathfrak{r}_0^3}
{\sqrt{\Delta(\mathfrak{r}_0)}(\sigma\sqrt{M\mathfrak{r}_0}-a)\over  F^2_\sigma(\mathfrak{r}_0)}
\left(-\mathsf{x}\mathsf{y}\cos \mathfrak{w}_{\rm R} \tau+(\mathsf{y}^2-\mathsf{z}^2) \sin \mathfrak{w}_{\rm R}\tau\right),
\\
p_{\mathsf{y}}&=&\mu \dot{\mathsf{y}} -\mu \mathfrak{w}_{\rm P}\mathsf{x}
+\mu {2(m_1-m_2)M\over (m_1+m_2)\mathfrak{r}_0^3}
{\sqrt{\Delta(\mathfrak{r}_0)}(\sigma\sqrt{M\mathfrak{r}_0}-a)\over  F^2_\sigma(\mathfrak{r}_0)}
\left(-\mathsf{x}\mathsf{y}\sin\mathfrak{w}_{\rm R} \tau+(\mathsf{x}^2-\mathsf{z}^2) \cos \mathfrak{w}_{\rm R}\tau\right),
\\
p_{\mathsf{z}}&=&\mu \dot{\mathsf{z}} 
+\mu {2(m_1-m_2)M\over (m_1+m_2)\mathfrak{r}_0^3}
{\sqrt{\Delta(\mathfrak{r}_0)}(\sigma\sqrt{M\mathfrak{r}_0}-a)\over  F^2_\sigma(\mathfrak{r}_0)}\, 
\mathsf{z}\left(\mathsf{y}\cos \mathfrak{w}_{\rm R} \tau+\mathsf{x} \sin \mathfrak{w}_{\rm R}\tau\right).
\enann
The Hamiltonian is given by
\beann
\tilde {\cal H}_{\rm rel}(\vect{\mathsf{r}}, \vect{\mathsf{p}})&=&{\cal H}_{\rm rel}(\vect{\mathsf{r}}, \vect{\mathsf{p}})+{\cal H}_{1/2\mathchar`-{\rm rel}}(\vect{\mathsf{r}}, \vect{\mathsf{p}})
\,,
\enann
where 
${\cal H}_{\rm rel}(\vect{\mathsf{r}}, \vect{\mathsf{p}})$  is given by Eq. (\ref{Hamiltonian_non-rotating}), while 
\beann
{\cal H}_{1/2\mathchar`-{\rm rel}}(\vect{\mathsf{r}}, \vect{\mathsf{p}})&=&
- {2\mu (m_1-m_2)^2 M^2\over (m_1+m_2)^2 \mathfrak{r}_0^6}
{\Delta(\mathfrak{r}_0)(\sigma\sqrt{M\mathfrak{r}_0}-a)^2\over  F^4_\sigma(\mathfrak{r}_0)}\Big[\left(-\mathsf{x}\mathsf{y}\cos \mathfrak{w}_{\rm R} \tau+(\mathsf{y}^2-\mathsf{z}^2) \sin \mathfrak{w}_{\rm R}\tau\right)^2
\\
&&
~~~~~~~~~~~~
+\left(-\mathsf{x}\mathsf{y}\sin\mathfrak{w}_{\rm R} \tau+(\mathsf{x}^2-\mathsf{z}^2) \cos \mathfrak{w}_{\rm R}\tau\right)^2+\mathsf{z}^2\left(\mathsf{y}\cos \mathfrak{w}_{\rm R} \tau+\mathsf{x} \sin \mathfrak{w}_{\rm R}\tau\right)^2\Big]
\,.
\enann
This Hamiltonian is complicated, but it should not be considered because it is beyond quadrupole approximation, although the momenta of the particles  are modified. For an equal mass binary ($m_1=m_2$), the 0.5PN correction term vanishes and the momenta are also the same as the Newtonian ones. As a result, the Newtonian solution is also valid.

\subsection{\bf Motion of the CM of a binary and its stability}

In order to study stability of the CM of  a binary system, we analyze Eq. (\ref{eq_CM}).
Since $\vect{R}$ is measured by the circular observer at $\mathfrak{r}=\mathfrak{r}_0$, we can split the 4-velocity $u^\mu$ 
as
\beann
u^\mu&=&u^\mu_{(0)}+u^\mu_{(1)},
\enann
where
\beann
u^\mu_{(0)}&=&{d\mathfrak{x}^\mu_{(0)}\over d\tau}=(u^0_{(0)}\,,0\,,0\,,u^3_{(0)})={1\over \mathfrak{r}_0F_\sigma(\mathfrak{r}_0)}\left(\mathfrak{r}_0^2+a\sigma\sqrt{M\mathfrak{r}_0}
\,, 0\,, 0\,,\sigma\sqrt{M\mathfrak{r}_0}\right)
\,,
\\
u^\mu_{(1)}&=&{d\mathfrak{x}^\mu_{(1)}\over d\tau},
\enann
with 
\beann
\mathfrak{x}^\mu_{(0)}&=&\left({\mathfrak{r}_0^2+a\sigma\sqrt{M\mathfrak{r}_0}\over \mathfrak{r}_0F_\sigma(\mathfrak{r}_0)} \tau\,,\mathfrak{r}_0\,,{\pi\over 2}\,,{\sigma\sqrt{M\mathfrak{r}_0}\over \mathfrak{r}_0F_\sigma(\mathfrak{r}_0)}
\tau\right),
\\
\mathfrak{x}^\mu_{(1)}&\equiv& e^{\mu}_{~\hat \ell}R^{\hat \ell}.
\enann
The acceleration $a^\mu$ is given by the motion of a binary $x^{\hat \mu}(\tau)$ 
in a rotating frame as
\beann
a^\mu&=&
{6\mu\over m_1+m_2}{\sqrt{\Delta}(\sigma\sqrt{M\mathfrak{r}_0 }-a)\over F^2_\sigma(\mathfrak{r}_0)}{M\over \mathfrak{r}_0^3}
\Big{[}\delta^\mu_1{ \sqrt{\Delta}\over \mathfrak{r}_0} \dot y x\, +\delta^\mu_2{1\over \mathfrak{r}_0}\dot y z +\\
&&
+{1\over \mathfrak{r}_0 F_\sigma(\mathfrak{r}_0) \sqrt{\Delta}}
\left(
\delta^\mu_0\sigma\sqrt{M\mathfrak{r}_0}(\mathfrak{r}_0^2+a^2-2a\sigma\sqrt{M\mathfrak{r}_0})
+\delta^\mu_3(\mathfrak{r}_0^2-2M\mathfrak{r}_0+a\sigma\sqrt{M\mathfrak{r}_0})
\right)
(-\dot x x+\dot z z)\,
\Big{]}.
\enann

Here, we assume that the deviation from a circular orbit is small, i.e.,  $\mathfrak{x}^\mu_{(1)}$ and $u^\mu_{(1)}$ are small perturbations.
Ignoring non-linear deviation terms in the equations of motion 
$\displaystyle{{Du^\mu\over d\tau}=a^\mu}$, because the circular orbit $\mathfrak{x}^\mu_{(0)}(\tau)$ is a geodesic,
we obtain
 a linear differential equation as
\beann
{du^\mu_{(1)}\over d\tau}
+2\Gamma^{\mu}_{~\rho\sigma}(\mathfrak{r}_0)u^\rho_{(0)} u^\sigma_{(1)}+{\partial \Gamma^{\mu}_{~\rho\sigma}
\over \partial \mathfrak{x}^\alpha}(\mathfrak{r}_0)\, \mathfrak{x}_{(1)}^\alpha 
\, u^\rho_{(0)} u^\sigma_{(0)}
=a^\mu,
\enann
where $a^\mu$ acts as an external force.
Describing the deviation as 
\beann
\mathfrak{x}^\mu_{(1)}=(\mathfrak{t}_{(1)},\mathfrak{r}_{(1)},\mathfrak{\theta}_{(1)},\mathfrak{\varphi}_{(1)})
\,,
\enann
we find
\bea
&&
{d^2\mathfrak{t}_{(1)}\over d\tau^2}
+{2M\over \mathfrak{r}_0 \Delta (\mathfrak{r}_0)
F_\sigma(\mathfrak{r}_0)}\left(\mathfrak{r}_0^2+a^2-2a\sigma\sqrt{M\mathfrak{r}_0}\right)
{d\mathfrak{r}_{(1)}\over d\tau}
\nonumber \\
&&~~~~~~~~~~=a^0=-{6\mu\over m_1+m_2}
{M\over \mathfrak{r}_0^4 F^3_\sigma(\mathfrak{r}_0) }
(M\mathfrak{r}_0 -a\sigma\sqrt{M\mathfrak{r}_0})(\mathfrak{r}_0^2+a^2-2a\sigma\sqrt{M\mathfrak{r}_0})
(\dot x x-\dot z z),
\label{eq_t1}
\\
&&
{d^2\mathfrak{r}_{(1)}\over d\tau^2}
-{3M\Delta(\mathfrak{r}_0)\over \mathfrak{r}_0^3 F^2_\sigma(\mathfrak{r}_0)}
\mathfrak{r}_{(1)}
+ {2M\Delta(\mathfrak{r}_0) \over \mathfrak{r}_0^3 F_\sigma(\mathfrak{r}_0)}
{d\mathfrak{t}_{(1)}\over d\tau}
-{2\Delta(\mathfrak{r}_0)\over \mathfrak{r}_0^3F_\sigma(\mathfrak{r}_0)}\left(Ma+\sigma\mathfrak{r}_0\sqrt{M\mathfrak{r}_0}\right)
{d\mathfrak{\varphi}_{(1)}\over d\tau}
\nonumber \\
&&~~~~~~~~~~
=a^1=
{6\mu\over m_1+m_2}{M\Delta(\sigma\sqrt{M\mathfrak{r}_0 }-a)\over \mathfrak{r}_0^4 F^2_\sigma(\mathfrak{r}_0)}
x \dot y\,,
~~~~~~
\label{eq_r1}
\\
&&
{d^2\mathfrak{\theta}_{(1)}\over d\tau^2}
+{M\over \mathfrak{r}_0^3 F^2_\sigma(\mathfrak{r}_0) }\left(\mathfrak{r}_0^2+3a^2-4a\sigma\sqrt{M\mathfrak{r}_0}\right)
\mathfrak{\theta}_{(1)}=a^2=
{6\mu\over m_1+m_2}{M\sqrt{\Delta}(\sigma\sqrt{M\mathfrak{r}_0 }-a)\over \mathfrak{r}_0^4 F^2_\sigma(\mathfrak{r}_0)}
\, z \dot y,
\label{eq_th1}
\\
&&
{d^2\mathfrak{\varphi}_{(1)}\over d\tau^2}
+{2\over \mathfrak{r}_0\Delta(\mathfrak{r}_0) F_\sigma(\mathfrak{r}_0) }\left(Ma+\sigma\sqrt{M\mathfrak{r}_0}(\mathfrak{r}_0-2M)\right)
{d\mathfrak{r}_{(1)}\over d\tau}
\nonumber \\
&&~~~~~~~~~~ =a^3=-
{6\mu\over m_1+m_2}{M\over \mathfrak{r}_0^4 F^3_\sigma(\mathfrak{r}_0)} 
(\sigma\sqrt{M\mathfrak{r}_0 }-a) \left(
\mathfrak{r}_0^2-2M\mathfrak{r}_0+a\sigma\sqrt{M\mathfrak{r}_0}
\right)(\dot x x-\dot z z),
\label{eq_ph1}
\ena

Integrating Eqs. (\ref{eq_t1}) and (\ref{eq_ph1}), we obtain 
\bea
&&
{d\mathfrak{t}_{(1)}\over d\tau}
=-{2M\over \mathfrak{r}_0 \Delta (\mathfrak{r}_0)
F_\sigma(\mathfrak{r}_0)}\left(\mathfrak{r}_0^2+a^2-2a\sigma\sqrt{M\mathfrak{r}_0}\right)
\mathfrak{r}_{(1)}
\nonumber \\
&&~~~~~~~~~~-{3\mu\over m_1+m_2}
{M\over \mathfrak{r}_0^4 F^3_\sigma(\mathfrak{r}_0) }
(M\mathfrak{r}_0 -a\sigma\sqrt{M\mathfrak{r}_0})(\mathfrak{r}_0^2+a^2-2a\sigma\sqrt{M\mathfrak{r}_0})
(x^2- z^2)\,,
\label{eq_t2}
\\
&&
{d\mathfrak{\varphi}_{(1)}\over d\tau}
=-{2\over \mathfrak{r}_0\Delta(\mathfrak{r}_0) F_\sigma(\mathfrak{r}_0) }\left(Ma+\sigma\sqrt{M\mathfrak{r}_0}(\mathfrak{r}_0-2M)\right)
\mathfrak{r}_{(1)}
\nonumber \\
&&~~~~~~~~~~ -
{3\mu\over m_1+m_2}{M\over \mathfrak{r}_0^4 F^3_\sigma(\mathfrak{r}_0)} 
(\sigma\sqrt{M\mathfrak{r}_0 }-a) \left(
\mathfrak{r}_0^2-2M\mathfrak{r}_0+a\sigma\sqrt{M\mathfrak{r}_0}
\right)
(x^2- z^2)\,,
\label{eq_ph2}
\ena
where we set the integration constants as zero.
Plugging Eqs. (\ref{eq_t2}) and (\ref{eq_ph2}) into Eq. (\ref{eq_r1}), we obtain the perturbation equation for the radial coordinates 
$\mathfrak{r}_{(1)}$ as
\bea
{d^2\mathfrak{r}_{(1)}\over d\tau^2}
+ k_{\mathfrak{r}}^2\mathfrak{r}_{(1)}
+A\left(x^2-z^2\right)
+B\, \dot y x=0
\,,
\label{eq:radial_perturbation}
\ena
where
\bea
k_{\mathfrak{r}}^2&\equiv &
{M\over \mathfrak{r}_0^3 F^2_\sigma(\mathfrak{r}_0)}\left(
\mathfrak{r}_0^2-6M\mathfrak{r}_0 -3a^2+8a\sigma\sqrt{M\mathfrak{r}_0}\right)
\label{coeff_kr2}
\\
A&\equiv&{6\mu M\Delta \over (m_1+m_2)\mathfrak{r}_0^5 F^4_\sigma(\mathfrak{r}_0)}
\left(\mathfrak{r}_0^2-3M\mathfrak{r}_0-2a^2\right)
\label{coeff_Ar}
\\
B&\equiv&-{6\mu M\Delta \over (m_1+m_2)\mathfrak{r}_0^4 F^2_\sigma(\mathfrak{r}_0)}\left(\sigma\sqrt{M\mathfrak{r}_0}-a\right)\,.
\label{coeff_Br}
\ena
We also rewrite the equation for $\mathfrak{\theta}_{(1)}$ as
\bea
{d^2\mathfrak{\theta}_{(1)}\over d\tau^2}
+k_{\theta}^2 \, 
\mathfrak{\theta}_{(1)}
+B\,  \dot y z =0\,,
\label{eq:angular_perturbation}
\ena
where
\bea
k_{\theta}^2&\equiv&
{M\over \mathfrak{r}_0^3 F^2_\sigma(\mathfrak{r}_0) }\left(\mathfrak{r}_0^2+3a^2-4a\sigma\sqrt{M\mathfrak{r}_0}\right)
\label{coeff_kth2}
\ena

We find that $k_{\mathfrak{r}}^2>0$  and 
 $k_{\theta}^2>0$ when 
$ \mathfrak{r}_0>\mathfrak{r}_{\rm ISCO}$, 
which guarantees stability 
against homogeneous perturbations.
This fact is consistent with the stability of a circular geodesic. 
In order to understand the oscillations of the deviation $\mathfrak{r}_{(1)}$ and $\mathfrak{\theta}_{(1)}$, we expand the oscillation frequencies 
$k_{\mathfrak{r}}^2$  and 
 $k_{\theta}^2$ in the limit of $M/\mathfrak{r}_0 \ll 1$ as
 \beann
k_{\mathfrak{r}}&\approx&
{M^{1/2}\over \mathfrak{r}_0^{3/2}}\left[1-{3M\over 2\mathfrak{r}_0}+{3a\sigma\over \mathfrak{r}_0} \sqrt{M\over \mathfrak{r}_0}\right]
=\mathfrak{w}_{\rm R}
-\mathfrak{w}_{\rm dS}-\widetilde{\mathfrak{w}}_{\rm LT}
\\
 k_{\theta}&\approx&
{M^{1/2}\over \mathfrak{r}_0^{3/2}}\left[1+{3M\over 2\mathfrak{r}_0}-{3a\sigma\over \mathfrak{r}_0} \sqrt{M\over \mathfrak{r}_0}\right]
=\mathfrak{w}_{\rm R}
+\mathfrak{w}_{\rm dS}+\widetilde{\mathfrak{w}}_{\rm LT}
\enann
where
$\widetilde{\mathfrak{w}}_{\rm LT}=-{3M a\sigma /\mathfrak{r}_0^3}$.
The meaning of the $\mathfrak{\theta}_{(1)}$ oscillations is more clear since it describes the deviation from the equatorial plane. There are three origins of the oscillations: One is the angular frequency of the rotating frame, the second is 
caused by the de Sitter precession, and the third is related to the Lense-Thirring precession between the rotation of the CM and the BH spin, which is slightly different from that of a binary angular momentum and the BH spin\cite{Liu:2019tqr,Liu:2021uam}.

Including the inhomogeneous perturbations caused by a binary motion, we find formally general solutions for Eqs. (\ref{eq:radial_perturbation}) and (\ref{eq:angular_perturbation})
as
\bea
\mathfrak{r}_{(1)}&=&a_{\mathfrak{r}}\cos[k_{\mathfrak{r}}\tau]
+b_{\mathfrak{r}}\sin[k_{\mathfrak{r}}\tau]
-{1\over k_{\mathfrak{r}}}\int_0^\tau d\tau'
\left[A(x(\tau')^2-z(\tau')^2)+B\dot y(\tau') x(\tau')\right]\sin[k_{\mathfrak{r}}(\tau-\tau')],
~~~~
\\
\mathfrak{\theta}_{(1)}&=&a_\theta\cos[k_{\theta}\tau]
+b_\theta\sin[k_{\mathfrak{r}}\tau]
-{B\over k_{\theta}}\int_0^\tau d\tau'
\dot y(\tau') z(\tau')\sin[k_{\mathfrak{r}}(\tau-\tau')]
\,,
\ena
where $a_{\mathfrak{r}}, b_{\mathfrak{r}}, a_\theta$ and $b_\theta$
are arbitrary constants, which are determined by initial conditions.
Since $\dot y x$ and $\dot y z$ may oscillate 
around zero, the integration 
with those terms
may not induce 
instability. 
On the other hand, the integration with $(x^2-z^2)$ 
does not 
have definite sign. 
We have to check its stability numerically 
by use of numerical solution of a binary motion.\\[1em]

\begin{figure}[htbp]
\begin{center}
\includegraphics[width=7cm]{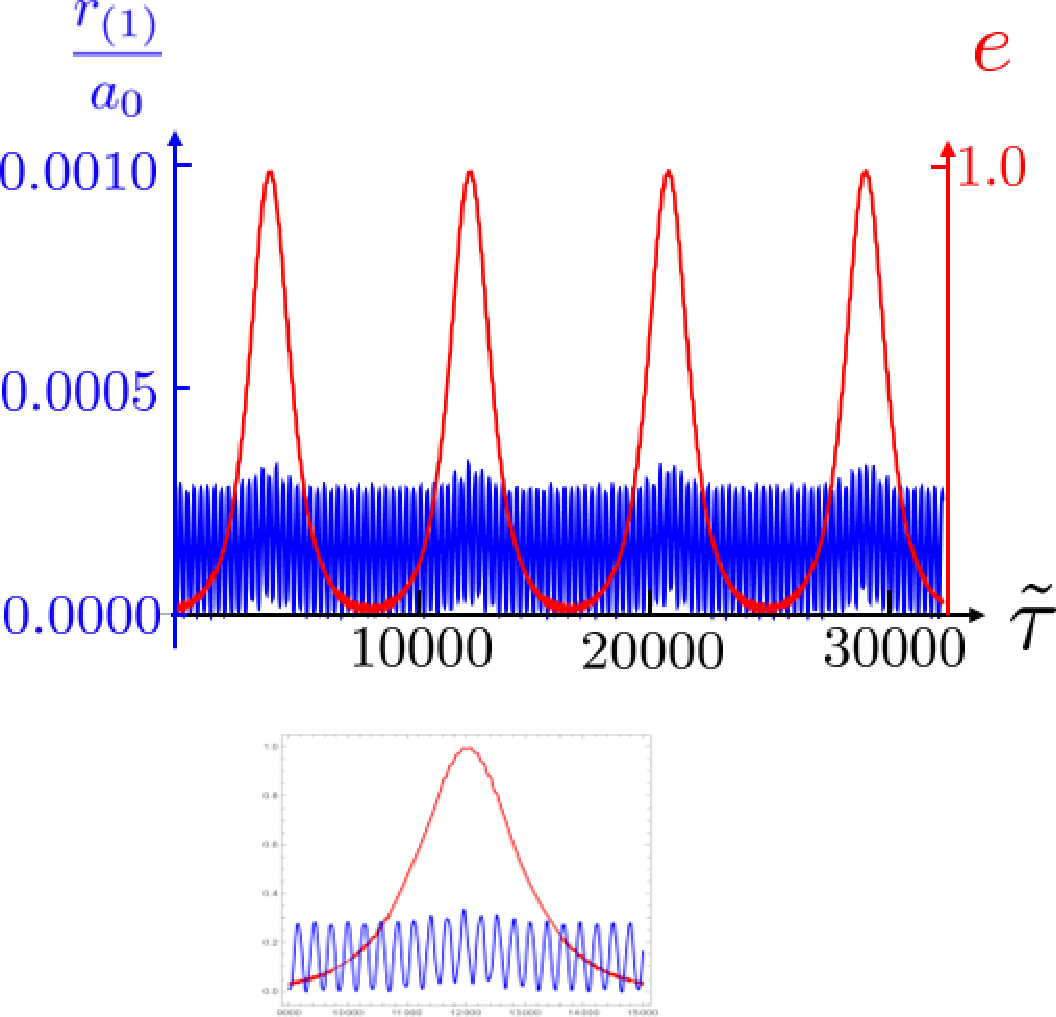}
\hskip 1cm
\includegraphics[width=6.5cm]{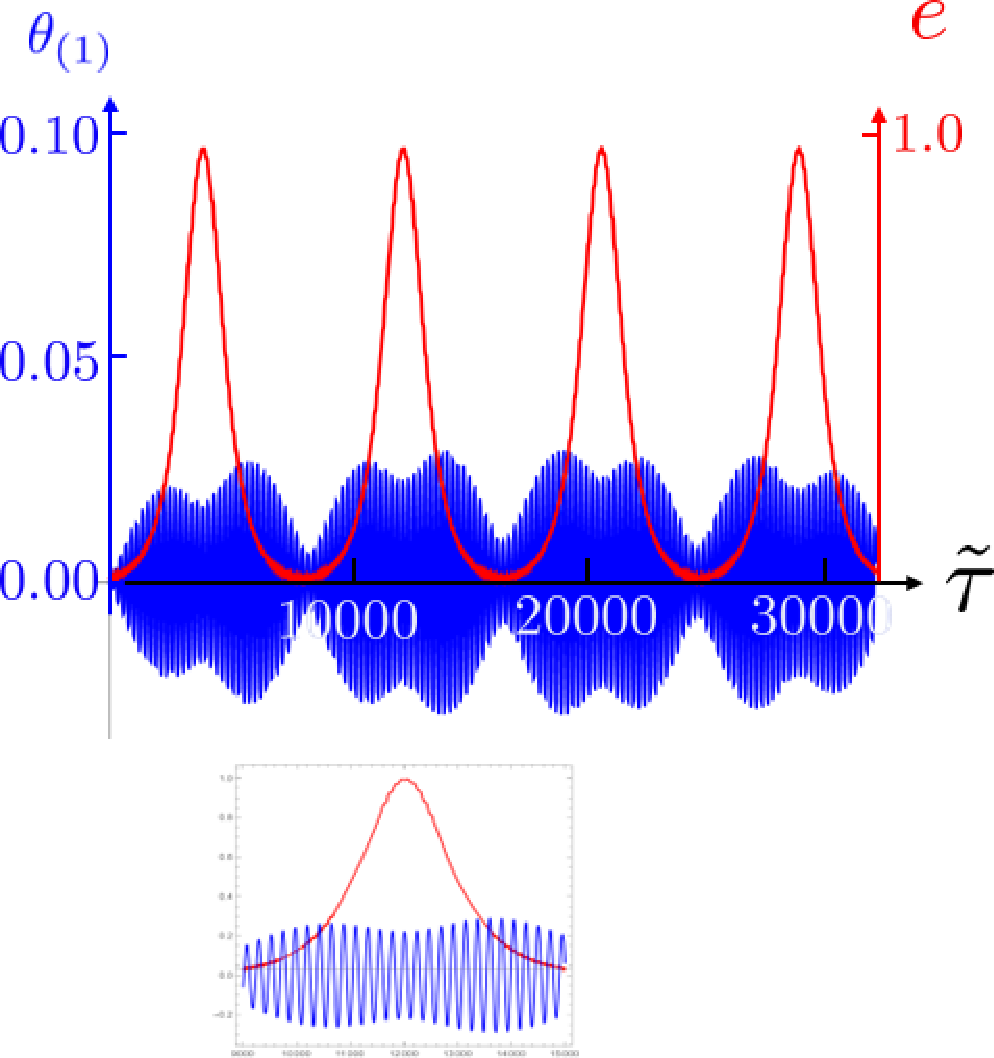}
\caption{The time evolution of 
$\mathfrak{r}_{(1)}$ (left)
and $\mathfrak{\theta}_{(1)}$ (right)
for Model I ($a=0.9M, \mathfrak{r}_0=10M$, and $\mathfrak{a}_0=0.005M$). 
The bottom figures
show the enlargement of the period of $\tilde \tau=9000-15000$.
We also show the evolution of the eccentricity (red curves) 
to see the correlation.}
\label{fig:stability_CM0}
\end{center}
\end{figure}
\begin{figure}[htbp]
\begin{center}
\includegraphics[width=7cm]{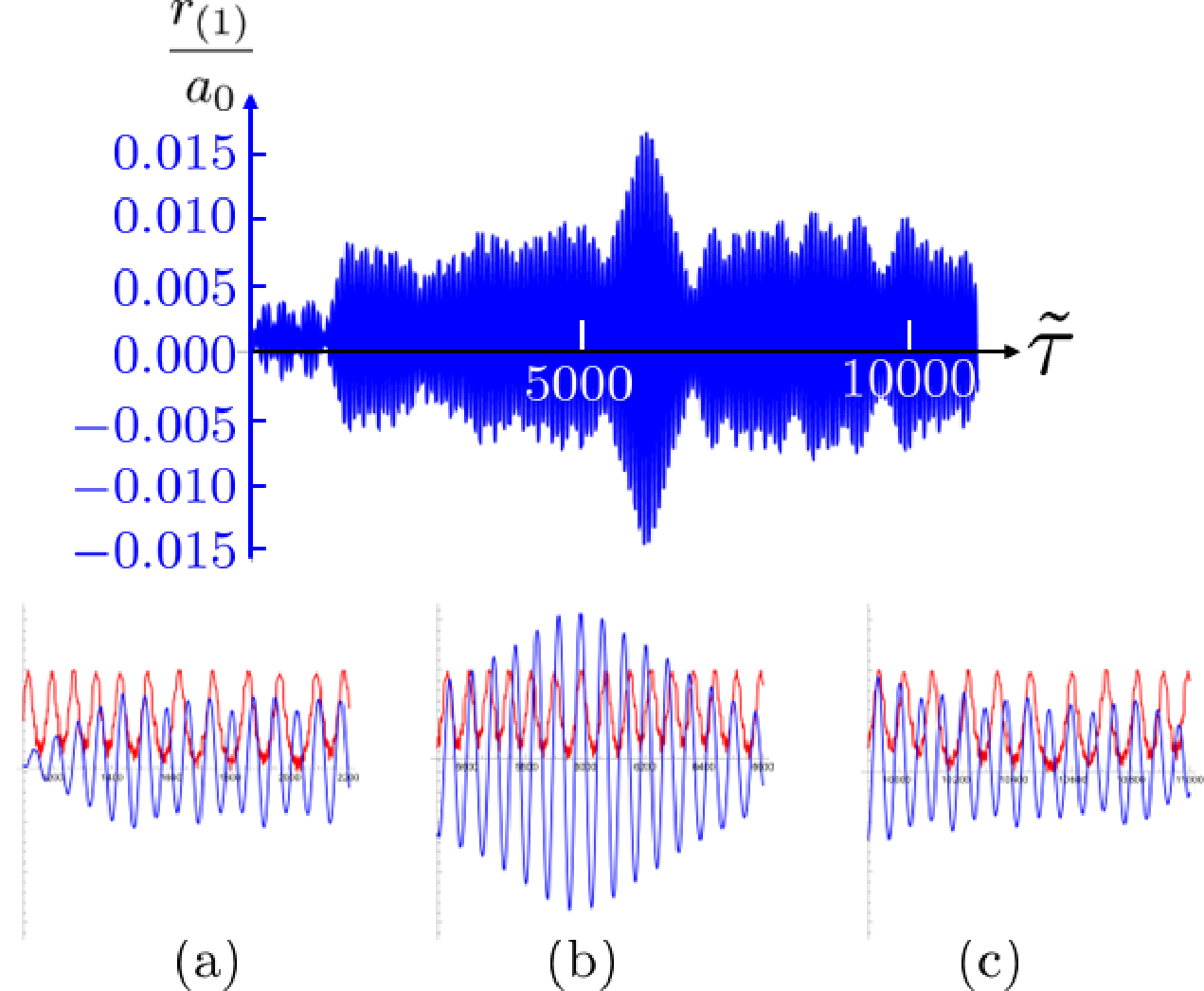}
\hskip 1cm
\includegraphics[width=6.5cm]{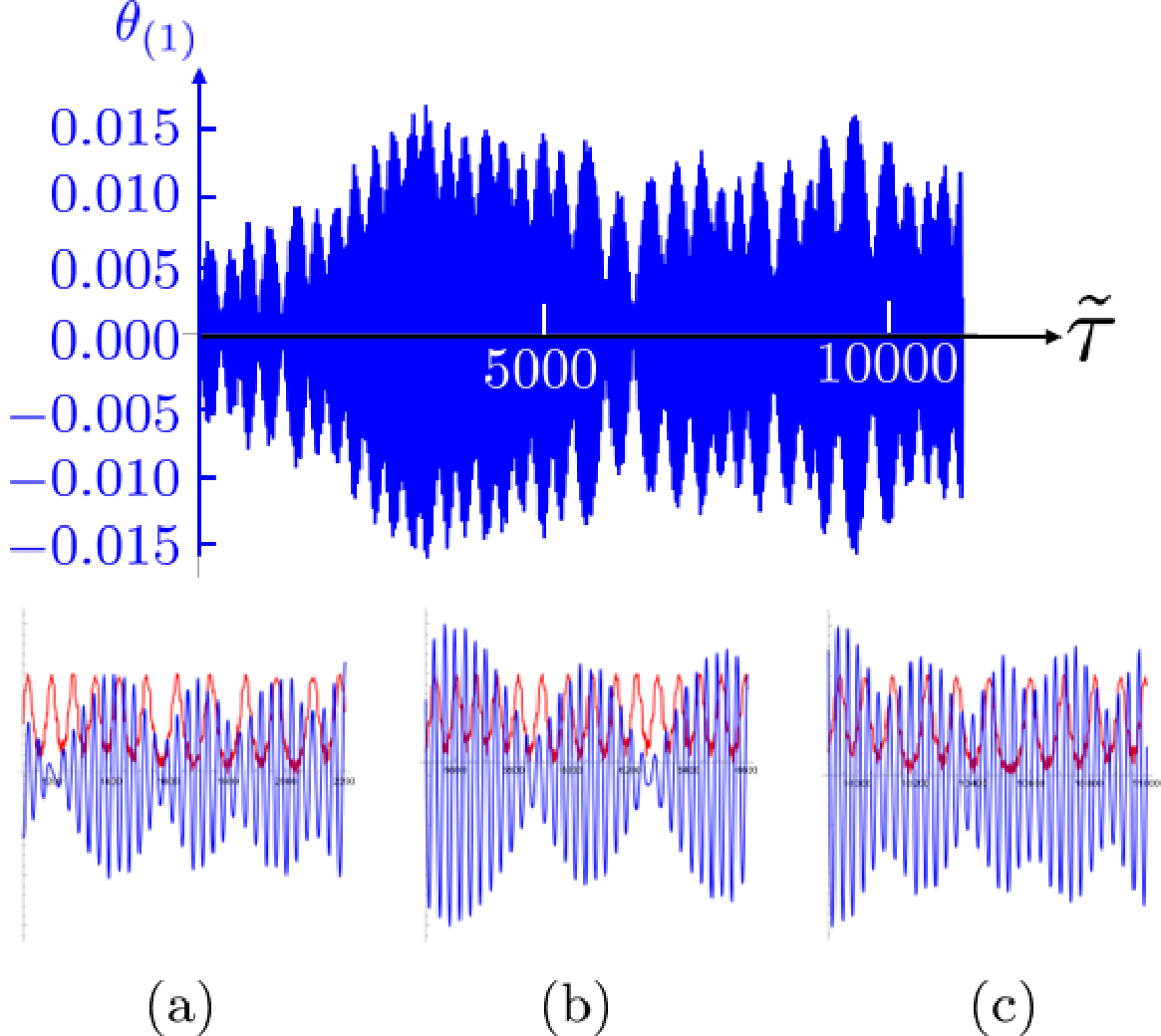}
\caption{The time evolution of 
$\mathfrak{r}_{(1)}$ (left)
and $\mathfrak{\theta}_{(1)}$ (right)
for the model with $a=0.9M, \mathfrak{r}_0=3.2M$, and $\mathfrak{a}_0=0.005M$. 
The bottom figures (a), (b), and (c) 
depict the first tenth, the middle tenth and the last tenth of the period, respectively.
We show the evolution of the eccentricity (red curves).
There is no correlation between the eccentricity ($e$) and the deviations ($\mathfrak{r}_{(1)}, \mathfrak{\theta}_{(1)}$).}
\label{fig:stability_CM1}
\end{center}
\end{figure}

In Figs. \ref{fig:stability_CM0} and \ref{fig:stability_CM1} , we show the time evolution of 
$\mathfrak{r}_{(1)}$ 
and $\mathfrak{\theta}_{(1)}$ for
Model I ($a=0.9M, \mathfrak{r}_0=10M$, and $\mathfrak{a}_0=0.005M$) and Model III ($a=0.9M, \mathfrak{r}_0=3.2M$, and $\mathfrak{a}_0=0.005M$).
We assume that the initial values of 
$\mathfrak{r}_{(1)}$ and $\mathfrak{\theta}_{(1)}$
and their time derivatives 
vanish.

Since Model I shows a regular vZLK oscillation, from Fig.~\ref{fig:stability_CM0}, 
we find good correlation between the eccentricity and the deviations from a circular orbit.
$\mathfrak{r}_{(1)}$ oscillates around small non-zero positive value
with small amplitude. The oscillation center slightly increases when the eccentricity becomes large, but the oscillation amplitude does not change.
For $\mathfrak{\theta}_{(1)}$, the oscillation center is almost zero, and oscillation amplitude changes in time. Although there is no correlation with the eccentricity, the oscillation pattern is periodic and the period is the same as that of the vZLK oscillations.
Note that the oscillation amplitude in the radial direction ($|\mathfrak{r}_{(1)}|\sim 10^{-4}$) is much smaller than that in the $\theta$ direction ($|\mathfrak{\theta}_{(1)}|\sim 10^{-2}$).

For Model III (Fig.~\ref{fig:stability_CM1}),
the binary system is close to the chaotic boundary.
In this chaotic vZLK oscillation model, we find the motion of the CM also becomes irregular.
There is no correlation between 
the evolution of the eccentricity 
(red curves) and the oscillations of  $\mathfrak{r}_{(1)}$ and $\mathfrak{\theta}_{(1)}$
(blue curves), as shown in the bottom figures of \ref{fig:stability_CM1}.
The oscillation amplitudes of the radial direction and $\theta$ direction are almost the same in this chaotic case ($|\mathfrak{r}_{(1)}|\sim |\mathfrak{\theta}_{(1)}|\sim 10^{-2}$).
\\


\section{Lagrange planetary equations for a binary system near SMBH}
\label{planetary_equations}

To comprehend our numerical findings better, we should consider the Lagrange planetary equations. These equations provide the evolution of orbital parameters, such as the semi-major axis, eccentricity, and inclination. To derive these planetary equations, we work with the proper Hamiltonian, where the mass parameter is set to $\mu=1$. The proper Hamiltonian is defined as follows:
\beann
\bar{{\cal H}}=\bar{{\cal H}}_0+\bar{{\cal H}}_1,
\enann
where
\beann
\bar{{\cal H}}_0
&=&{1\over 2} \bar{\vect{\mathsf{p}}}^2- {G (m_1+m_2)\over \mathsf{r}},
\\
\bar{{\cal H}}_1
&=&\sigma \mathfrak{w}_{\rm P}\left(\bar{\mathsf{p}}_{\mathsf{y}}\mathsf{x}-\bar{\mathsf{p}}_{\mathsf{x}}\mathsf{y}\right)+
{M\over 2\mathfrak{r}_0^3} 
\left[\mathsf{r}^2+{3\over F^2_\sigma(\mathfrak{r}_0)}\left(-\Delta(\mathfrak{r}_0)\left(\mathsf{x}\cos\mathfrak{w}_{\rm R}   \tau -\mathsf{y} \sin \mathfrak{w}_{\rm R}  \tau\right)^2
+(\sigma\sqrt{M\mathfrak{r}_0}-a)^2 \mathsf{z}^2\right)
 \right]
 \enann

\end{widetext}
The position $\bf{\mathsf{r}}=(\mathsf{x},\mathsf{y},\mathsf{z})$ 
of a binary 
should be described in the non-rotating proper reference frame.
\\

The unperturbed Hamiltonian, denoted as $\bar{{\cal H}}_0$, is equivalent to that of a binary system in Newtonian dynamics. It leads to an elliptical orbit, described by the equation
\bea
\mathsf{r}= {\mathfrak{a}(1-e^2)\over 1+e\cos f}
\label{elliptic_r}
\,,
\ena
Here, $\mathsf{r}$ represents the radial distance from the center of mass, while $a$, $e$, and $f$ are the semi-major axis, eccentricity, and true anomaly, respectively. This orbital plane is inclined at an angle $I$ relative to the equatorial plane in the proper reference frame. Consequently, the relative position vector $\bf{\mathsf{r}}=(\mathsf{x},\mathsf{y},\mathsf{z})$ of the binary system can be determined by the orbital parameters $(\omega\,,\Omega\,,  \mathfrak{a}\,, e\,, I\,, f)$ as described in Eq. (\ref{orbital_parameters}) with Eq. (\ref{elliptic_r}). The introduction of Delaunay variables further refines this description as follows:

\vskip -.5cm
\beann
\left\{
\begin{array}{l}
\mathfrak{l}=n(t-t_0)
\\
\mathfrak{g}=\omega
\\
\mathfrak{h}=\Omega
\\
\end{array}
\right.
\hskip 3.3cm
\\
{\rm and}\hskip 6cm
\\
\left\{
\begin{array}{l}
\mathfrak{L}=\sqrt{G(m_1+m_2)\mathfrak{a}}
\\
\mathfrak{G}=\sqrt{G(m_1+m_2)\mathfrak{a}(1-e^2)}
\\
\mathfrak{H}=\sqrt{G(m_1+m_2)\mathfrak{a}(1-e^2)}\cos I
\,,
\\
\end{array}
\right.
\enann
where
\beann
n\equiv {2\pi \over  P}=\sqrt{{G(m_1+m_2)\over \mathfrak{a}^3}},
\enann
is the mean motion,
we find new unperturbed Hamiltonian as
\beann
\widetilde{\bar{\cal H}}_0=-{G^2(m_1+m_2)^2\over 2\mathfrak{L}^2}.
\enann
\\[-1em]

Including the perturbations $\bar{\cal H}_1$,
we obtain the Hamiltonian the Delaunay variables as
\beann
\widetilde{\bar{\cal H}}=\widetilde{\bar{\cal H}}_0+\bar{\cal H}_1
\,.
\enann

The proper Hamiltonian is described by the orbital parameters by inserting the relation given in Eq.(\ref{orbital_parameters}) with Eq. (\ref{elliptic_r}).
We then find the perturbed Hamiltonian as
\beann
\bar{\cal H}_1=\bar{{\cal H}}_{1\mathchar`-{\rm P}}+\bar{{\cal H}}_{1\mathchar`-\bar{\cal R}}
\,,
\enann
where
\begin{widetext}
\bea
\bar{{\cal H}}_{1\mathchar`-{\rm P}}&=&
\sigma \mathfrak{w}_{\rm P}\mathsf{r}^2(\mathfrak{a}, e, f)
\left\{n\cos I (1-e^2)^{-3/2}(1+e\cos f)^2 -\mathfrak{w}_{\rm P}\left(\cos^2(\omega+f)+\sin^2(\omega+f)\cos^2 I \right)\right\},~~~~~~~
\label{H1_P}\\
\bar{{\cal H}}_{1\mathchar`-\bar{\cal R}}&=&{M\over 2\mathfrak{r}_0^3}
\mathsf{r}^2(\mathfrak{a}, e, f)\Big\{1-{3\Delta(\mathfrak{r}_0)\over F^2_\sigma(\mathfrak{r}_0)}
\left[\cos (\Omega+\mathfrak{w}_{\rm R} \tau)
\cos(\omega+f)-\sin(\Omega+\mathfrak{w}_{\rm R} \tau)\sin(\omega+f)\cos I 
\right]^2
\nn
&&~~~~~~~~~~~~~~~~~~~~~~~
+{3 (\sigma\sqrt{M\mathfrak{r}_0}-a)^2 \over F^2_\sigma(\mathfrak{r}_0)} \sin^2(\omega+f)\sin^2 I\Big\}
\,.
\label{H1_R}
\ena
We then obtain the planetary equations for the present hierarchical triple system, which is mathematically equivalent to our basic equations in the text.

\subsection{Double-averaging (DA) approach}

Rather than directly solving the Lagrange planetary equations, our approach involves averaging the perturbed Hamiltonian over two periods: the inner and outer orbital periods. This allows us to simplify the equations for analysis. We are interested in understanding the long-term behavior of the system, particularly phenomena like the vZLK mechanism.

The doubly-averaged Hamiltonian is defined by
\beann
\langle\langle\bar{\cal H}_1 \rangle\rangle\equiv {1\over 2\pi}\int_0^{2\pi}
d\mathfrak{l}_{\rm out}\left( {1\over 2\pi}\int_0^{2\pi}
d\mathfrak{l}\, \bar{\cal H}_1\right)
\enann
Since the outer orbit is circular, we find that $\mathfrak{l}_{\rm out}=f_{\rm out}=\mathfrak{w}_0\tau$.
We also have
\beann
d\mathfrak{l}={1\over \sqrt{1-e^2}}\left({\mathsf{r}\over \mathfrak{a}}\right)^2 df
\,.
\enann
Inserting Eqs. (\ref{H1_P}) and (\ref{H1_R}) into the above integrals, 
we find the doubly-averaged Hamiltonian as
\bea
\langle\langle\bar{\cal H}_1\rangle\rangle
&=&\sigma\mathfrak{w}_{\rm P}n\mathfrak{a}^2 \sqrt{1-e^2}\cos I
-{\mathfrak{a}^2\over 8}\Big\{
(2+3e^2)\Big[\mathfrak{w}_{\rm P}^2(3+\cos 2I)+
{M(\mathfrak{r}_0^2+3a^2-4a\sigma\sqrt{M\mathfrak{r}_0})\over 4F^2_\sigma(\mathfrak{r}_0)\mathfrak{r}_0^3}(1+3\cos 2I)\Big]
\nn
&&
~~~~~~~~~~~~~~~~~~~~~~~~~~~~~~~~~
+10e^2\sin^2 I\cos 2\omega\left[
\mathfrak{w}_{\rm P}^2+{3M(\mathfrak{r}_0^2
+3a^2-4a\sigma\sqrt{M\mathfrak{r}_0})\over 4F^2_\sigma(\mathfrak{r}_0) \mathfrak{r}_0^3}
\right]
\Big\}.
\label{DA_perturbed_H1}
\ena
Using the double-averaged Hamiltonian Eq.(\ref{DA_perturbed_H1}), we obtain the double-averaged Lagrange planetary equations as
\bea
\dot e&=&{5\over 4}  \left(\mathfrak{w}_{\rm P}^2+{3M(\mathfrak{r}_0^2
+3a^2-4a\sigma\sqrt{M\mathfrak{r}_0})\over 4F^2_\sigma(\mathfrak{r}_0) \mathfrak{r}_0^3}
\right){e\sqrt{1-e^2}\over n}\left(1-\cos 2I\right)\sin (2\omega),
\label{PE_Sch_e}
\\
\dot I&=&-{5\over 4}  \left(\mathfrak{w}_{\rm P}^2+{3M(\mathfrak{r}_0^2
+3a^2-4a\sigma\sqrt{M\mathfrak{r}_0})\over 4F^2_\sigma(\mathfrak{r}_0) \mathfrak{r}_0^3}
\right){e^2\over n\sqrt{1-e^2}}\sin 2I \sin(2\omega),
\label{PE_Sch_I}
\\
\dot \omega&=&
{1\over 4n}\left(\mathfrak{w}_{\rm P}^2+{3M(\mathfrak{r}_0^2
+3a^2-4a\sigma\sqrt{M\mathfrak{r}_0})\over 4F^2_\sigma(\mathfrak{r}_0) \mathfrak{r}_0^3}
\right)
\Big{[}\sqrt{1-e^2} [3+5\cos 2I +5\left(1-\cos 2I \right)\cos 2\omega]
\nonumber
\\
&&+{5e^2\over \sqrt{1-e^2} }(1+\cos 2I)(1-\cos2\omega)\Big{]}
+{2\sqrt{1-e^2} \over n}\mathfrak{w}_{\rm P}^2,
\label{PE_Sch_omega}
\\
\dot \Omega&=&\sigma\mathfrak{w}_{\rm P}+
{\cos I\over 2n\sqrt{1-e^2}}\left(\mathfrak{w}_{\rm P}^2+{3M(\mathfrak{r}_0^2
+3a^2-4a\sigma\sqrt{M\mathfrak{r}_0})\over 4F^2_\sigma(\mathfrak{r}_0) \mathfrak{r}_0^3}
\right)
\left[-(2+3e^2)
+5e^2\cos(2\omega)\right].
\label{PE_Sch_Omega}
\ena
\end{widetext}
The semi-major axis $\mathfrak{a}$ is constant in the present approximation.
Also, from Eqs. (\ref{PE_Sch_e}) and (\ref{PE_Sch_I}), 
we can easily check the conservation equation such that
\beann
{d\over d\tau}\left(\sqrt{1-e^2}\cos I\right)=0
\,,
\enann
which corresponds to conservation of 
the $\mathsf{z}$-component of the angular momentum.

\subsection{vZLK oscillations}
\label{DA_KL_oscillations}

Introducing a ``potential" by
$V_S\equiv - \langle\langle\bar{\cal H}_1\rangle\rangle
$,
we rewrite the above planetary equations as
\bea
\dot{e}&=&-{\sqrt{1-e^2}\over n\mathfrak{a}^2 e}{\partial {V_S}\over \partial \omega},
\label{LPeq_e1}
\\
\dot I&=&{\cos I\over n\mathfrak{a}^2\sin I\sqrt{1-e^2}}
{\partial {V_S}\over \partial \omega},
\label{LPeq_I1}
\\
\dot \omega&=&{\sqrt{1-e^2}\over n\mathfrak{a}^2 e}{\partial {V_S}\over \partial e}-{\cos I\over n\mathfrak{a}^2\sin I\sqrt{1-e^2}}{\partial {V_S}\over \partial I},~~~~~~~
\label{LPeq_om1}
\\
\dot \Omega&=&{1\over n\mathfrak{a}^2\sin I\sqrt{1-e^2}}{\partial {V_S}\over \partial I}
\label{LPeq_Om1}
\,.
\ena
We derive closed-form differential equations for the variables $e$, $I$, and $\omega$ using equations (\ref{LPeq_e1}), (\ref{LPeq_I1}), and (\ref{LPeq_om1}). These equations provide insights into various properties of vZLK oscillations, including the oscillation amplitude of eccentricity and the oscillation time scale. This analysis is consistent with previous studies on Newtonian and 1PN hierarchical triple systems, as discussed in~\cite{Suzuki:2020zbg}. 

The potential is written by use of 
$\eta \equiv  \sqrt{1-e^2}$ and $\mu_I \equiv  \cos I$
as
\beann
V_S&\equiv &
-\langle\langle\bar{\cal H}_1\rangle\rangle
\nn
&=& {\mathfrak{a}^2 M (\mathfrak{r}_0^2
+3a^2-4a\sigma\sqrt{M\mathfrak{r}_0})\over 16 F^2_\sigma(\mathfrak{r}_0)  \mathfrak{r}_0^3}v_S(\eta, \mu_I),
\enann
where
\beann
v_S(\eta, \mu_I)&\equiv & 2(-1+3\mu_I^2\eta^2)(1+\alpha_{\rm P})+12 C_{\rm vZLK}
\\
&&
+4\alpha_{\rm P}\left(2-{3\sigma n\over\mathfrak{w}_{\rm P}}\mu_I\eta\right),
\enann
with
\beann
\alpha_{\rm P}&\equiv&{4 \mathfrak{w}_{\rm P}^2 F^2_\sigma(\mathfrak{r}_0)  \mathfrak{r}_0^3
\over 3M (\mathfrak{r}_0^2
+3a^2-4a\sigma\sqrt{M\mathfrak{r}_0})},
\\
C_{\rm vZLK}&\equiv& (1-\eta^2)
\Big{[}(1+2\alpha_{\rm P})
\\
&&
-{5\over 2}(1+\alpha_{\rm P})(1-\mu_I^2)\sin^2\omega \Big{]}.
\enann

Note that when $\alpha_{\rm P}=0$, we find the same equations for Newtonian hierarchical triple system with quadrupole approximation.
The terms with $\alpha_{\rm P}$ give relativistic corrections.
\\

Introducing the normalized time $\tilde \tau$, which is defined by
\beann
\tilde \tau\equiv
{\tau \over \tau_{\rm vZLK}} ,
\enann
with the typical vZLK time scale 
\beann
\tau_{\rm vZLK}\equiv {16 n F^2_\sigma(\mathfrak{r}_0)  \mathfrak{r}_0^3\over M(\mathfrak{r}_0^2
+3a^2-4a\sigma\sqrt{M\mathfrak{r}_0})}
\,,
\enann
the above planetary equation is rewritten as
\beann
{d\eta\over d\tilde \tau}&=&{\partial {v_S}\over \partial \omega},
\\
{1\over \mu_I}{d\mu_I\over d\tilde \tau}&=&
-{1\over \eta}{\partial {v_S}\over \partial \omega},
\\
{d\omega\over d\tilde \tau}&=&
-{\partial {v_S}\over \partial \eta}+{\mu_I\over \eta}
{\partial {v_S}\over \partial \mu_I}.
\enann
From these equations, we can easily show that 
\beann
{d(\mu_I\eta) \over d\tilde \tau}=0
~\,,~~~
{dv_S\over d\tilde \tau}=0
\,,
\enann
\\[-1em]

\noindent
which means there exist two conserved quantities $\vartheta\equiv \mu_I\eta$ and $C_{\rm vZLK}$ just as the Newtonian and 1PN hierarchical triple system under dipole approximation.
Using these two conserved quantities, we obtain a single equation for $\eta$ as
\beann
{d\eta^2\over d\tilde \tau}=-24\sqrt{2}\sqrt{f(\eta^2)g(\eta^2)},
\enann
with
\beann
f(\eta^2)&\equiv&(1+2\alpha_{\rm P})(1-\eta^2)-C_{\rm vZLK},
\\
g(\eta^2)&\equiv&-5(1+\alpha_{\rm P})\vartheta^2
-(3+\alpha_{\rm P})\eta^4.
\\
&&
+\left[5(1+\alpha_{\rm P})\vartheta^2+3+\alpha_{\rm P}+2C_{\rm KvZLK}\right]\eta^2
\enann

\begin{widetext}

Setting $\xi=\eta^2$, we find
\beann
{d\xi\over d\tilde \tau}=-24\sqrt{2(1+2\alpha_{\rm P})(3+\alpha_{\rm P})}\sqrt{(\xi-\xi_0)(\xi-\xi_+)(\xi-\xi_-)},
\enann
where
\beann
\xi_0&=&1-{C_{\rm vZLK}\over 1+2\alpha_{\rm P}},
\\
\xi_\pm&=&{1\over 2}\Big[\left(1+{5(1+\alpha_{\rm P})\over 3+\alpha_{\rm P}}
\vartheta^2+{2\over 3+\alpha_{\rm P}}C_{\rm vZLK}\right)
\pm \sqrt{\left(1+{5(1+\alpha_{\rm P})\over 3+\alpha_{\rm P}}
\vartheta^2+{2\over 3+\alpha_{\rm P}}C_{\rm vZLK}\right)^2-{20(1+\alpha_{\rm P})\over 3+\alpha_{\rm P}}\vartheta^2
}\Big],
\enann
are the solutions of $f(\xi)=0$ and $g(\xi)=0$, respectively.
\\

\end{widetext}

We can find  the relativistic 
corrections with $\alpha_{\rm P}$, which is evaluated as 
\beann
\alpha_{\rm P}&=&
{4(\mathfrak{r}_0-F_\sigma(\mathfrak{r}_0)  )^2\over 3(\mathfrak{r}_0^2
+3a^2-4a\sigma\sqrt{M\mathfrak{r}_0})}
\,.
\enann
This constant $\alpha_{\rm P}$ is small when rotation of SMBH is small 
(the maximum value for Schwarzschild BH is $0.11$), 
but it can be large when SMBH is rotating rapidly and the CM is near the ISCO radius.
For example, when $\mathfrak{a}=0.9M (0.99M)$, $\alpha_{\rm P}\approx 0.88 (5.56)$ at the ISCO radius, and
it diverges as $\mathfrak{a}\rightarrow M$.
It is because the denominator vanishes in this limit.

Analyzing the above equation, we find that there exists vZLK oscillations in this system just the same as in Newtonian hierarchical triple system, and we can classify the vZLK oscillations by the sign of $C_{\rm vZLK}$ into two cases:\\[.5em] 
\indent $a$. $C_{\rm vZLK}>0$ (rotation) \\[.5em] 
\indent $b$. $C_{\rm vZLK}<0$ (libration).

\subsubsection{$C_{\rm vZLK}>0$ {\rm (rotation)}}
In this case, $0<\xi_-<1<\xi_+$ and $0<\xi_0<1$.
This is possible if 
\beann
0<C_{\rm vZLK}<1+2\alpha_{\rm P}
\,.
\enann

Hence we find the maximum and minimum values of the eccentricity as
\beann
e_{\rm max}=\sqrt{1-\xi_-}
~~,~~
e_{\rm min}=\sqrt{1-\xi_0}.
\enann

The vZLK oscillation  time scale is given by 
\bea
T_{\rm vZLK}=\tau_{\rm vZLK}\, \mathfrak{T}_{\rm vZLK}^{\rm (rot)},
\label{TKL_rotation}
\ena
where
\beann
\mathfrak{T}_{\rm vZLK}^{\rm (rot)}&\equiv& 
{K\left(\sqrt{\xi_0-\xi_-\over \xi_+-\xi_-}\right)\over 6\sqrt{2(1+2\alpha_{\rm P})(3+\alpha_{\rm P})(\xi_+-\xi_-)}}
\,.
\enann
$K(k)$ is the complete elliptic integral of the first kind with the elliptic modulus $k$.

\subsubsection{$C_{\rm vZLK}<0$ {\rm (libration)}}
Since $0<\xi_-<\xi_+<1$ and $\xi_0<0$ in this case, 
 we find 
\beann
e_{\rm max}=\sqrt{1-\xi_-}\,,~~~
e_{\rm min}=\sqrt{1-\xi_+}.
\enann
It occurs when 
\beann
&&
 -{3+\alpha_{\rm P}\over 2}<C_{\rm vZLK}<0\,,~
\enann
 and
\beann
\vartheta<{(\sqrt{3+\alpha_{\rm P}}-\sqrt{-2C_{\rm vZLK}})\over \sqrt{5(1+\alpha_{\rm P})}}
\,.
\enann

The vZLK time scale is given by 
\bea
T_{\rm vZLK}=\tau_{\rm vZLK}\, \mathfrak{T}_{\rm vZLK}^{\rm (lib)}\,,
\label{TKL_libration}
\ena
where
\beann
\mathfrak{T}_{\rm vZLK}^{\rm (lib)}&\equiv& 
{K\left(\sqrt{\xi_+-\xi_-\over \xi_0-\xi_-}\right)\over 6\sqrt{2(1+2\alpha_{\rm P})(3+\alpha_{\rm P})(\xi_0-\xi_-)}}
\enann

The maximum and minimum values of the eccentricity in the vZLK oscillations
are determined by two conserved parameters, $\vartheta$ and $C_{\rm vZLK}$.
Note that the maximum eccentricity in vZLK oscillations is important, especially when we discuss emission of GWs.

\begin{figure}[htbp]
\begin{center}
\includegraphics[width=6.5cm]{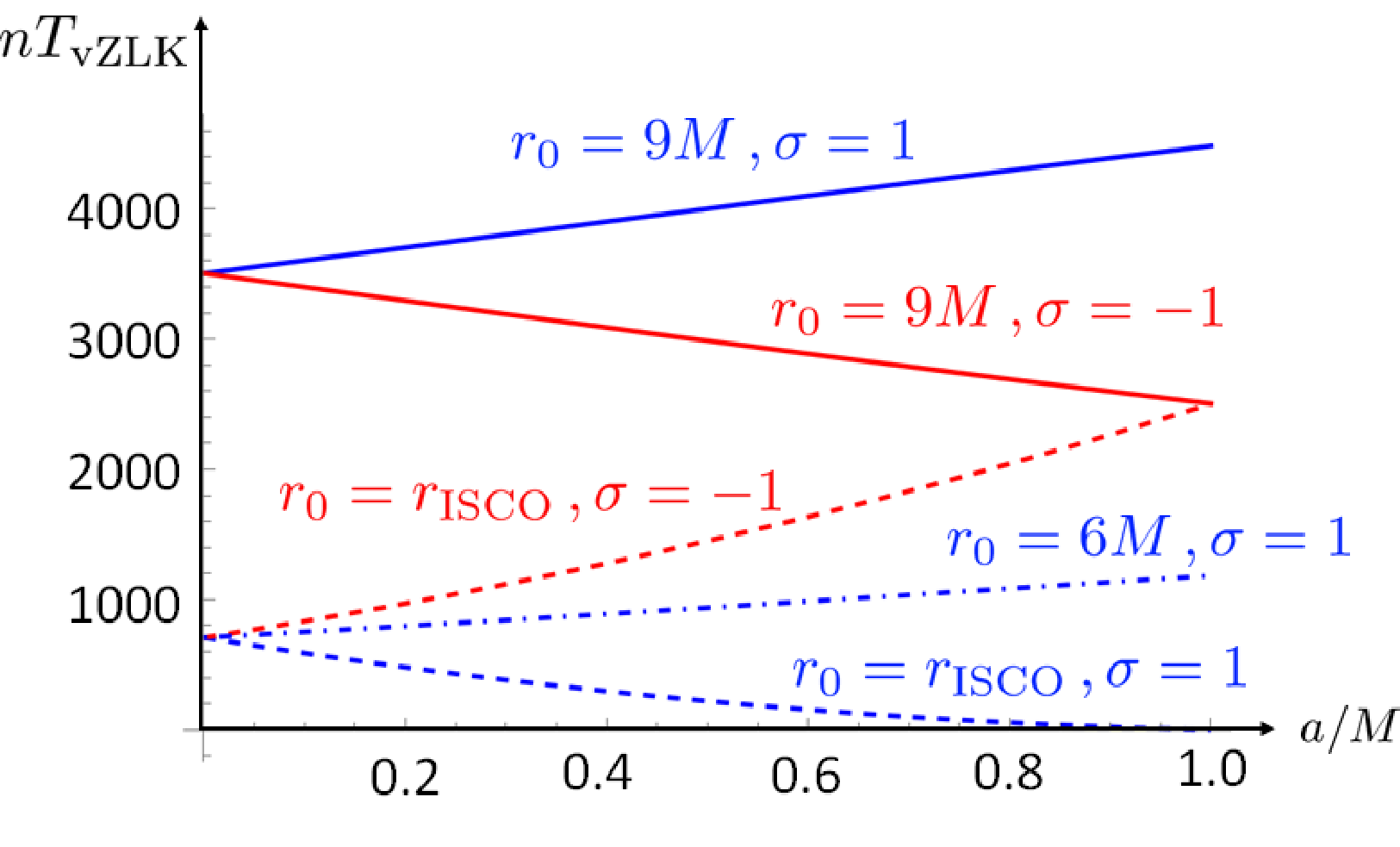}
\caption{The period of vZLK oscillations $T_{\rm vZLK}$ is shown in terms of the Kerr rotation parameter $a$.
The blue and red curves correspond to the prograde ($\sigma=1$) and retrograde ($\sigma=-1$) orbits.
The solid curves show the case of $\mathfrak{r}_0=9M$, while the dot-dashed one and dashed ones are $\mathfrak{r}_0=6M$ and $\mathfrak{r}_0=\mathfrak{r}_{\rm ISCO}$,
respectively. The semi-major axis is chosen as $\mathfrak{a}=0.005M$.
}
\label{fig:vZLK_timescale}
\end{center}
\end{figure}

The time scale of the vZLK oscillations is also important for 
observation of the gravitational waves.
Since 
$\mathfrak{T}_{\rm vZLK}^{\rm (rot)}$ and $\mathfrak{T}_{\rm vZLK}^{\rm (lib)}$
are order of unity, the time scale is almost determined by $\tau_{\rm vZLK}$,
which is rewritten by
\bea
n\tau_{\rm vZLK}=16 \mathfrak{f}\,{(\mathfrak{r}_0^2-3M\mathfrak{r}_0+2a\sigma\sqrt{M\mathfrak{r}_0})
\over 
(\mathfrak{r}_0^2+3a^2-4a\sigma\sqrt{M\mathfrak{r}_0}
)}
\,.
\ena
Some example of the exact values of 
$T_{\rm vZLK}$ is given in Fig. \ref{fig:vZLK_timescale}.

\begin{figure}[htbp]
\begin{center}
\includegraphics[width=6.5cm]{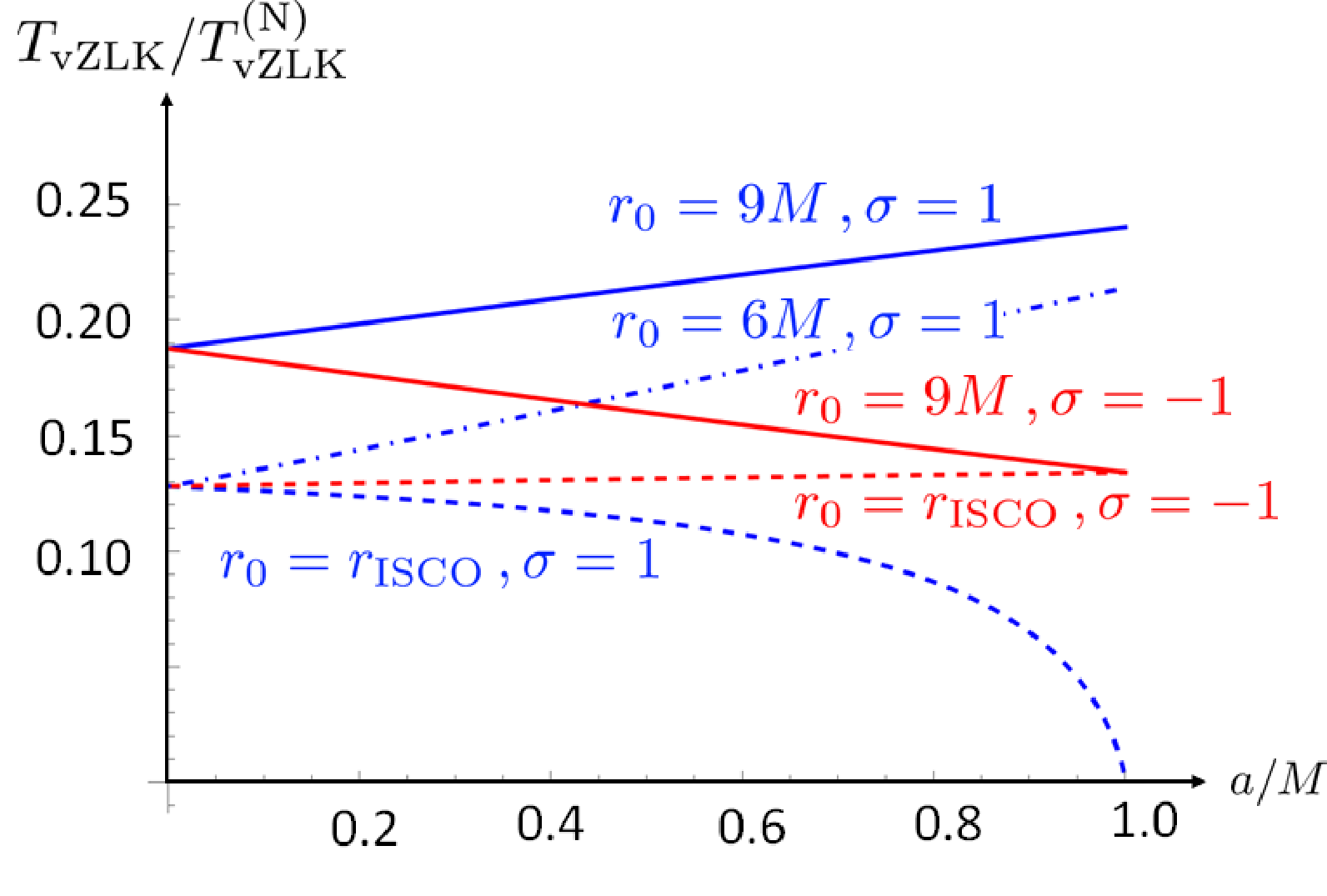}
\caption{The ratio of vZLK oscillation period  $T_{\rm vZLK}$ to its Newtonian value is shown in terms of the Kerr rotation parameter $a$.
The colors and types of the curves are 
the same as those in Fig.\ref{fig:vZLK_timescale}. The semi-major axis is chosen as $\mathfrak{a}=0.005M$.
}
\label{fig:TKLvsNTKL}
\end{center}
\end{figure}

We then may evaluate 
the relativistic effects
(including de Sitter precession)
by comparison with the Newtonian value
$T_{\rm vZLK}^{\rm (N)}$
as shown in Fig. \ref{fig:TKLvsNTKL}.

Fig. \ref{fig:TKLvsNTKL} shows that 
the relativistic effects reduce the vZLK time scale by a factor $0.1-0.25$.
Although the ratio vanishes 
at the ISCO radius 
in the limit of 
$a\rightarrow M$, 
the double-averaging approximation
is no longer valid in this limit.

\subsubsection{\rm Critical inclination angle}
\label{critical_inclination_angle_in_DA}
\begin{figure}[htbp]
\begin{center}
\includegraphics[width=5.5cm]{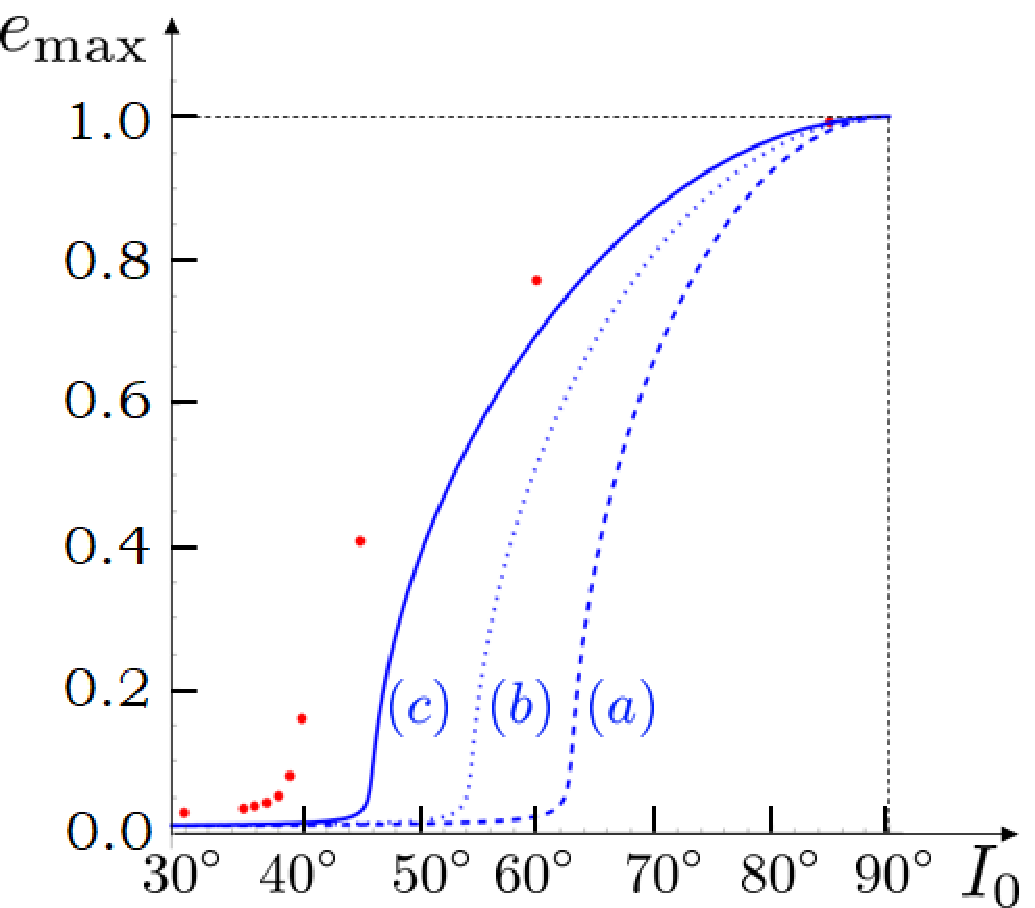}
\caption{ The maximum values of the eccentricity in vZLK oscillations based on the double-averaging approximation: (a) $C_{\rm chaotic}=1$ (the dashed curve),\,(b) $C_{\rm chaotic}=3$ (the dotted curve),\, (c) $C_{\rm chaotic}=5$ (the solid curve).
 We choose 
 $\mathfrak{r}_0=r_{\rm ISCO}$ and 
$a=0.999 M$. The red dots denote the result obtained by the direct integration of the equations of motion.
}
\label{fig:I0cr_DA999}
\end{center}
\end{figure}

We can also 
evaluate a critical inclination angle, 
beyond which the vZLK oscillation occurs even when the initial eccentricity is very small. It is given by the condition for a bifurcation point with 
$C_{\rm vZLK}=0$ with $\omega=90^\circ$. 
Setting 
\beann
(1+2\alpha_{\rm P})-{5\over 2}(1+\alpha_{\rm P})\sin^2 I_{\rm crit}=0
\,,
\enann
we obtain 
\beann
I_{\rm crit}=\sin^{-1}\sqrt{2(1+2\alpha_{\rm P})\over 5(1+\alpha_{\rm P})}
\,.
\enann
We find that the critical inclination angle 
changes from $63.4^\circ$ ($a=M$) [$41.6^\circ$ ($a=0$)] to the Newtonian value $I_{\rm crit}^{\rm (N)}=\sin^{-1}\sqrt{2/5}\approx 39.2^\circ$
as $\mathfrak{r}_0$ increases from the ISCO radius to infinity.
However the critical value depends on the orbital parameters, especially on 
 $\mathfrak{r}_0$ and $\mathfrak{a}_0$.
 Assuming $a=0.999M$,  $\mathfrak{r}_0=r_{\rm ISCO}$, and  $e_0=0.01$, we evaluate the maximum value of the eccentricity in  vZLK oscillations, which is given in Fig. \ref{fig:I0cr_DA999}.
 In the case of $C_{\rm chaotic}=1$, which corresponds to 
 the firmness parameter $\mathfrak{f}=1$, the critical inclination angle 
 is slightly larger than $60^\circ$. 
 However, this models suffers from chaotic instability.
 As we discussed in the text, stability against chaotic perturbations
 must satisfy $C_{\rm chaotic}\gsim 3-4$.
 In that case, the critical inclination angle becomes smaller 
 as $40^\circ-50^\circ$.
 The red dots denote the result obtained by the direct integration of the equations of motion, which shows that the true critical inclination angle is about $40^\circ$ even for $a=0.999M$.

\newpage
\bibliography{refer}

\begin{thebibliography}{76}
\expandafter\ifx\csname natexlab\endcsname\relax\def\natexlab#1{#1}\fi
\expandafter\ifx\csname bibnamefont\endcsname\relax
  \def\bibnamefont#1{#1}\fi
\expandafter\ifx\csname bibfnamefont\endcsname\relax
  \def\bibfnamefont#1{#1}\fi
\expandafter\ifx\csname citenamefont\endcsname\relax
  \def\citenamefont#1{#1}\fi
\expandafter\ifx\csname url\endcsname\relax
  \def\url#1{\texttt{#1}}\fi
\expandafter\ifx\csname urlprefix\endcsname\relax\def\urlprefix{URL }\fi
\providecommand{\bibinfo}[2]{#2}
\providecommand{\eprint}[2][]{\url{#2}}

\bibitem[{\citenamefont{Abbott et~al.}(2020)\citenamefont{Abbott, Abbott,
  Abbott, Abraham, Acernese, Ackley, Adams, Adya, Affeldt, and
  et~al.}}]{Abbott_2020}
\bibinfo{author}{\bibfnamefont{B.~P.} \bibnamefont{Abbott}},
  \bibinfo{author}{\bibfnamefont{R.}~\bibnamefont{Abbott}},
  \bibinfo{author}{\bibfnamefont{T.~D.} \bibnamefont{Abbott}},
  \bibinfo{author}{\bibfnamefont{S.}~\bibnamefont{Abraham}},
  \bibinfo{author}{\bibfnamefont{F.}~\bibnamefont{Acernese}},
  \bibinfo{author}{\bibfnamefont{K.}~\bibnamefont{Ackley}},
  \bibinfo{author}{\bibfnamefont{C.}~\bibnamefont{Adams}},
  \bibinfo{author}{\bibfnamefont{V.~B.} \bibnamefont{Adya}},
  \bibinfo{author}{\bibfnamefont{C.}~\bibnamefont{Affeldt}}, \bibnamefont{and}
  \bibinfo{author}{\bibnamefont{et~al.}}, \bibinfo{journal}{Living Reviews in
  Relativity} \textbf{\bibinfo{volume}{23}} (\bibinfo{year}{2020}), ISSN
  \bibinfo{issn}{1433-8351}.

\bibitem[{\citenamefont{Abbott et~al.}(2021)\citenamefont{Abbott, Abbott,
  Abbott, Abraham, Acernese, Ackley, Adams, Adya, Affeldt, and
  et~al.}}]{Abbott_2021}
\bibinfo{author}{\bibfnamefont{B.~P.} \bibnamefont{Abbott}},
  \bibinfo{author}{\bibfnamefont{R.}~\bibnamefont{Abbott}},
  \bibinfo{author}{\bibfnamefont{T.~D.} \bibnamefont{Abbott}},
  \bibinfo{author}{\bibfnamefont{S.}~\bibnamefont{Abraham}},
  \bibinfo{author}{\bibfnamefont{F.}~\bibnamefont{Acernese}},
  \bibinfo{author}{\bibfnamefont{K.}~\bibnamefont{Ackley}},
  \bibinfo{author}{\bibfnamefont{C.}~\bibnamefont{Adams}},
  \bibinfo{author}{\bibfnamefont{V.~B.} \bibnamefont{Adya}},
  \bibinfo{author}{\bibfnamefont{C.}~\bibnamefont{Affeldt}}, \bibnamefont{and}
  \bibinfo{author}{\bibnamefont{et~al.}}, \bibinfo{journal}{SoftwareX}
  \textbf{\bibinfo{volume}{13}}, \bibinfo{pages}{100658}
  (\bibinfo{year}{2021}), ISSN \bibinfo{issn}{2352-7110}.

\bibitem[{\citenamefont{Collaboration and Collaboration}(2016)}]{GW150914}
\bibinfo{author}{\bibfnamefont{L.~S.} \bibnamefont{Collaboration}}
  \bibnamefont{and}
  \bibinfo{author}{\bibfnamefont{V.}~\bibnamefont{Collaboration}},
  \bibinfo{journal}{Phys. Rev. Lett.} \textbf{\bibinfo{volume}{116}},
  \bibinfo{pages}{061102} (\bibinfo{year}{2016}).

\bibitem[{\citenamefont{Collaboration and Collaboration}(2017)}]{GW170817}
\bibinfo{author}{\bibfnamefont{L.~S.} \bibnamefont{Collaboration}}
  \bibnamefont{and}
  \bibinfo{author}{\bibfnamefont{V.}~\bibnamefont{Collaboration}},
  \bibinfo{journal}{Phys. Rev. Lett.} \textbf{\bibinfo{volume}{119}},
  \bibinfo{pages}{161101} (\bibinfo{year}{2017}).

\bibitem[{\citenamefont{Collaboration and the
  Virgo~Collaboration}(2019)}]{test1}
\bibinfo{author}{\bibfnamefont{T.~L.~S.} \bibnamefont{Collaboration}}
  \bibnamefont{and} \bibinfo{author}{\bibnamefont{the Virgo~Collaboration}},
  \bibinfo{journal}{Phys. Rev. D} \textbf{\bibinfo{volume}{100}},
  \bibinfo{pages}{104036} (\bibinfo{year}{2019}).

\bibitem[{\citenamefont{Collaboration and the
  Virgo~Collaboration}(2020)}]{test2}
\bibinfo{author}{\bibfnamefont{T.~L.~S.} \bibnamefont{Collaboration}}
  \bibnamefont{and} \bibinfo{author}{\bibnamefont{the Virgo~Collaboration}},
  \bibinfo{journal}{Astrophys. J.} \textbf{\bibinfo{volume}{900}},
  \bibinfo{pages}{L13} (\bibinfo{year}{2020}).

\bibitem[{\citenamefont{Collaboration and the
  Virgo~Collaboration}(2021{\natexlab{a}})}]{test3}
\bibinfo{author}{\bibfnamefont{T.~L.~S.} \bibnamefont{Collaboration}}
  \bibnamefont{and} \bibinfo{author}{\bibnamefont{the Virgo~Collaboration}},
  \bibinfo{journal}{Phys. Rev. D} \textbf{\bibinfo{volume}{103}}
  (\bibinfo{year}{2021}{\natexlab{a}}).

\bibitem[{\citenamefont{Collaboration and the
  Virgo~Collaboration}(2021{\natexlab{b}})}]{test4}
\bibinfo{author}{\bibfnamefont{T.~L.~S.} \bibnamefont{Collaboration}}
  \bibnamefont{and} \bibinfo{author}{\bibnamefont{the Virgo~Collaboration}},
  \bibinfo{journal}{Astrophys. J. Letters} \textbf{\bibinfo{volume}{913}},
  \bibinfo{pages}{L7} (\bibinfo{year}{2021}{\natexlab{b}}).

\bibitem[{\citenamefont{Collaboration et~al.}(2021)\citenamefont{Collaboration,
  the Virgo~Collaboration, and the KAGRA~collaboration}}]{test5}
\bibinfo{author}{\bibfnamefont{T.~L.~S.} \bibnamefont{Collaboration}},
  \bibinfo{author}{\bibnamefont{the Virgo~Collaboration}}, \bibnamefont{and}
  \bibinfo{author}{\bibnamefont{the KAGRA~collaboration}},
  \emph{\bibinfo{title}{Constraints on the cosmic expansion history from
  gwtc-3}} (\bibinfo{year}{2021}), \eprint{arXiv: 2111.03604}.

\bibitem[{\citenamefont{Heggie}(1975)}]{Heggie1975}
\bibinfo{author}{\bibfnamefont{D.~C.} \bibnamefont{Heggie}},
  \bibinfo{journal}{Mon. Not. R. Astron. Soc.} \textbf{\bibinfo{volume}{173}},
  \bibinfo{pages}{729} (\bibinfo{year}{1975}).

\bibitem[{\citenamefont{Hut}(1993)}]{Hut1993}
\bibinfo{author}{\bibfnamefont{P.}~\bibnamefont{Hut}},
  \bibinfo{journal}{Astrophys. J.} \textbf{\bibinfo{volume}{403}},
  \bibinfo{pages}{256} (\bibinfo{year}{1993}).

\bibitem[{\citenamefont{Samsing et~al.}(2014)\citenamefont{Samsing, MacLeod,
  and Ramirez-Ruiz}}]{Samsing2014}
\bibinfo{author}{\bibfnamefont{J.}~\bibnamefont{Samsing}},
  \bibinfo{author}{\bibfnamefont{M.}~\bibnamefont{MacLeod}}, \bibnamefont{and}
  \bibinfo{author}{\bibfnamefont{E.}~\bibnamefont{Ramirez-Ruiz}},
  \bibinfo{journal}{Astrophys. J.} \textbf{\bibinfo{volume}{784}},
  \bibinfo{pages}{71} (\bibinfo{year}{2014}).

\bibitem[{\citenamefont{Riddle~et al.}(2015)}]{Riddle2015}
\bibinfo{author}{\bibfnamefont{R.~L.} \bibnamefont{Riddle~et al.}},
  \bibinfo{journal}{Astrophys. J.} \textbf{\bibinfo{volume}{799}},
  \bibinfo{pages}{4} (\bibinfo{year}{2015}).

\bibitem[{\citenamefont{Antonini
  et~al.}(2016{\natexlab{a}})\citenamefont{Antonini, Chattejee, Rodriguez,
  Morscher, Pattabiraman, Kalogera, and Rasio}}]{Fabio2016}
\bibinfo{author}{\bibfnamefont{F.}~\bibnamefont{Antonini}},
  \bibinfo{author}{\bibfnamefont{S.}~\bibnamefont{Chattejee}},
  \bibinfo{author}{\bibfnamefont{C.}~\bibnamefont{Rodriguez}},
  \bibinfo{author}{\bibfnamefont{M.}~\bibnamefont{Morscher}},
  \bibinfo{author}{\bibfnamefont{B.}~\bibnamefont{Pattabiraman}},
  \bibinfo{author}{\bibfnamefont{V.}~\bibnamefont{Kalogera}}, \bibnamefont{and}
  \bibinfo{author}{\bibfnamefont{F.}~\bibnamefont{Rasio}},
  \bibinfo{journal}{Astrophys. J.} \textbf{\bibinfo{volume}{816}},
  \bibinfo{pages}{2} (\bibinfo{year}{2016}{\natexlab{a}}).

\bibitem[{\citenamefont{{Stephan} et~al.}(2019)\citenamefont{{Stephan}, {Naoz},
  {Ghez}, {Morris}, {Ciurlo}, {Do}, {Breivik}, {Coughlin}, and
  {Rodriguez}}}]{stephan2019}
\bibinfo{author}{\bibfnamefont{A.~P.} \bibnamefont{{Stephan}}},
  \bibinfo{author}{\bibfnamefont{S.}~\bibnamefont{{Naoz}}},
  \bibinfo{author}{\bibfnamefont{A.~M.} \bibnamefont{{Ghez}}},
  \bibinfo{author}{\bibfnamefont{M.~R.} \bibnamefont{{Morris}}},
  \bibinfo{author}{\bibfnamefont{A.}~\bibnamefont{{Ciurlo}}},
  \bibinfo{author}{\bibfnamefont{T.}~\bibnamefont{{Do}}},
  \bibinfo{author}{\bibfnamefont{K.}~\bibnamefont{{Breivik}}},
  \bibinfo{author}{\bibfnamefont{S.}~\bibnamefont{{Coughlin}}},
  \bibnamefont{and} \bibinfo{author}{\bibfnamefont{C.~L.}
  \bibnamefont{{Rodriguez}}}, \bibinfo{journal}{\apj}
  \textbf{\bibinfo{volume}{878}}, \bibinfo{pages}{58} (\bibinfo{year}{2019}).

\bibitem[{\citenamefont{Mapelli et~al.}(2021)\citenamefont{Mapelli,
  Santoliquido, Bouffanais, Arca~Sedda, Artale, and Ballone}}]{sym13091678}
\bibinfo{author}{\bibfnamefont{M.}~\bibnamefont{Mapelli}},
  \bibinfo{author}{\bibfnamefont{F.}~\bibnamefont{Santoliquido}},
  \bibinfo{author}{\bibfnamefont{Y.}~\bibnamefont{Bouffanais}},
  \bibinfo{author}{\bibfnamefont{M.}~\bibnamefont{Arca~Sedda}},
  \bibinfo{author}{\bibfnamefont{M.~C.} \bibnamefont{Artale}},
  \bibnamefont{and} \bibinfo{author}{\bibfnamefont{A.}~\bibnamefont{Ballone}},
  \bibinfo{journal}{Symmetry} \textbf{\bibinfo{volume}{13}}
  (\bibinfo{year}{2021}), ISSN \bibinfo{issn}{2073-8994},
  \urlprefix\url{https://www.mdpi.com/2073-8994/13/9/1678}.

\bibitem[{\citenamefont{Gayathri et~al.}(2020)\citenamefont{Gayathri, Bartos,
  Haiman, Klimenko, Kocsis, M{\'{a}}rka, and Yang}}]{Gayathri_2020}
\bibinfo{author}{\bibfnamefont{V.}~\bibnamefont{Gayathri}},
  \bibinfo{author}{\bibfnamefont{I.}~\bibnamefont{Bartos}},
  \bibinfo{author}{\bibfnamefont{Z.}~\bibnamefont{Haiman}},
  \bibinfo{author}{\bibfnamefont{S.}~\bibnamefont{Klimenko}},
  \bibinfo{author}{\bibfnamefont{B.}~\bibnamefont{Kocsis}},
  \bibinfo{author}{\bibfnamefont{S.}~\bibnamefont{M{\'{a}}rka}},
  \bibnamefont{and} \bibinfo{author}{\bibfnamefont{Y.}~\bibnamefont{Yang}},
  \bibinfo{journal}{Astrophys. J.} \textbf{\bibinfo{volume}{890}},
  \bibinfo{pages}{L20} (\bibinfo{year}{2020}).

\bibitem[{\citenamefont{Gerosa and Fishbach}(2021)}]{Gerosa_2021}
\bibinfo{author}{\bibfnamefont{D.}~\bibnamefont{Gerosa}} \bibnamefont{and}
  \bibinfo{author}{\bibfnamefont{M.}~\bibnamefont{Fishbach}},
  \bibinfo{journal}{Nature Astronomy} \textbf{\bibinfo{volume}{5}},
  \bibinfo{pages}{749} (\bibinfo{year}{2021}).

\bibitem[{\citenamefont{von Zeipel}(1910)}]{vonZeipel10}
\bibinfo{author}{\bibfnamefont{H.}~\bibnamefont{von Zeipel}},
  \bibinfo{journal}{Astronomische Nachrichten} \textbf{\bibinfo{volume}{183}},
  \bibinfo{pages}{345–418} (\bibinfo{year}{1910}).

\bibitem[{\citenamefont{Lidov}(1962)}]{Lidov62}
\bibinfo{author}{\bibfnamefont{M.}~\bibnamefont{Lidov}},
  \bibinfo{journal}{Planet. Space Sci.} \textbf{\bibinfo{volume}{9}},
  \bibinfo{pages}{719} (\bibinfo{year}{1962}).

\bibitem[{\citenamefont{Kozai}(1962)}]{Kozai62}
\bibinfo{author}{\bibfnamefont{Y.}~\bibnamefont{Kozai}},
  \bibinfo{journal}{Astron. J.} \textbf{\bibinfo{volume}{67}},
  \bibinfo{pages}{591} (\bibinfo{year}{1962}).

\bibitem[{\citenamefont{Shevchenko}(Springer, 2017)}]{Shevchenko17}
\bibinfo{author}{\bibfnamefont{I.}~\bibnamefont{Shevchenko}},
  \bibinfo{journal}{{\it The Lidov-Kozai Effect - Applications in Exoplanet
  Research and Dynamical Astronomy}}  (\bibinfo{year}{Springer, 2017}).

\bibitem[{\citenamefont{{Amaro-Seoane}
  et~al.}(2010)\citenamefont{{Amaro-Seoane}, {Sesana}, {Hoffman},
  {Benacquista}, {Eichhorn}, {Makino}, and {Spurzem}}}]{Amaro-Seoane2010}
\bibinfo{author}{\bibfnamefont{P.}~\bibnamefont{{Amaro-Seoane}}},
  \bibinfo{author}{\bibfnamefont{A.}~\bibnamefont{{Sesana}}},
  \bibinfo{author}{\bibfnamefont{L.}~\bibnamefont{{Hoffman}}},
  \bibinfo{author}{\bibfnamefont{M.}~\bibnamefont{{Benacquista}}},
  \bibinfo{author}{\bibfnamefont{C.}~\bibnamefont{{Eichhorn}}},
  \bibinfo{author}{\bibfnamefont{J.}~\bibnamefont{{Makino}}}, \bibnamefont{and}
  \bibinfo{author}{\bibfnamefont{R.}~\bibnamefont{{Spurzem}}},
  \bibinfo{journal}{Mon. Not. R. Astron. Soc.} \textbf{\bibinfo{volume}{402}},
  \bibinfo{pages}{2308} (\bibinfo{year}{2010}).

\bibitem[{\citenamefont{Antonini and Perets}(2012)}]{Antonini2012}
\bibinfo{author}{\bibfnamefont{F.}~\bibnamefont{Antonini}} \bibnamefont{and}
  \bibinfo{author}{\bibfnamefont{H.}~\bibnamefont{Perets}},
  \bibinfo{journal}{Astrophys. J.} \textbf{\bibinfo{volume}{757}},
  \bibinfo{pages}{27} (\bibinfo{year}{2012}).

\bibitem[{\citenamefont{Hoang and Naoz}(2018)}]{hoang18}
\bibinfo{author}{\bibfnamefont{B.}~\bibnamefont{Hoang}} \bibnamefont{and}
  \bibinfo{author}{\bibfnamefont{S.}~\bibnamefont{Naoz}},
  \bibinfo{journal}{Astrophys. J.} \textbf{\bibinfo{volume}{852}},
  \bibinfo{pages}{2} (\bibinfo{year}{2018}).

\bibitem[{\citenamefont{Antonini
  et~al.}(2016{\natexlab{b}})\citenamefont{Antonini, Chatterjee, Rodriguez,
  Morscher, and Pattabiraman}}]{Antonini2016}
\bibinfo{author}{\bibfnamefont{F.}~\bibnamefont{Antonini}},
  \bibinfo{author}{\bibfnamefont{S.}~\bibnamefont{Chatterjee}},
  \bibinfo{author}{\bibfnamefont{C.}~\bibnamefont{Rodriguez}},
  \bibinfo{author}{\bibfnamefont{M.}~\bibnamefont{Morscher}}, \bibnamefont{and}
  \bibinfo{author}{\bibfnamefont{B.}~\bibnamefont{Pattabiraman}},
  \bibinfo{journal}{Astrophys. J.} \textbf{\bibinfo{volume}{816}},
  \bibinfo{pages}{2} (\bibinfo{year}{2016}{\natexlab{b}}).

\bibitem[{\citenamefont{Meiron et~al.}(2017)\citenamefont{Meiron, Kocsis, and
  Loeb}}]{Meiron2017}
\bibinfo{author}{\bibfnamefont{Y.}~\bibnamefont{Meiron}},
  \bibinfo{author}{\bibfnamefont{B.}~\bibnamefont{Kocsis}}, \bibnamefont{and}
  \bibinfo{author}{\bibfnamefont{A.}~\bibnamefont{Loeb}},
  \bibinfo{journal}{Astrophys. J.} \textbf{\bibinfo{volume}{84}},
  \bibinfo{pages}{2} (\bibinfo{year}{2017}).

\bibitem[{\citenamefont{Robson et~al.}(2018)\citenamefont{Robson, Cornish,
  Tamanini, and Toonen}}]{Robson2018}
\bibinfo{author}{\bibfnamefont{T.}~\bibnamefont{Robson}},
  \bibinfo{author}{\bibfnamefont{N.}~\bibnamefont{Cornish}},
  \bibinfo{author}{\bibfnamefont{N.}~\bibnamefont{Tamanini}}, \bibnamefont{and}
  \bibinfo{author}{\bibfnamefont{S.}~\bibnamefont{Toonen}},
  \bibinfo{journal}{Phys. Rev. D} \textbf{\bibinfo{volume}{98}},
  \bibinfo{pages}{064012} (\bibinfo{year}{2018}).

\bibitem[{\citenamefont{Randall and Xianyu}(2019{\natexlab{a}})}]{Lisa2018}
\bibinfo{author}{\bibfnamefont{L.}~\bibnamefont{Randall}} \bibnamefont{and}
  \bibinfo{author}{\bibfnamefont{Z.-Z.} \bibnamefont{Xianyu}},
  \bibinfo{journal}{Astrophys. J.} \textbf{\bibinfo{volume}{878}},
  \bibinfo{pages}{75} (\bibinfo{year}{2019}{\natexlab{a}}).

\bibitem[{\citenamefont{Randall and Xianyu}(2019{\natexlab{b}})}]{Lisa2019}
\bibinfo{author}{\bibfnamefont{L.}~\bibnamefont{Randall}} \bibnamefont{and}
  \bibinfo{author}{\bibfnamefont{Z.-Z.} \bibnamefont{Xianyu}},
  \bibinfo{journal}{arXiv:1902.08604 [astro-ph.HE]}
  (\bibinfo{year}{2019}{\natexlab{b}}).

\bibitem[{\citenamefont{{Hoang} et~al.}(2019)\citenamefont{{Hoang}, {Naoz},
  {Kocsis}, {Farr}, and {McIver}}}]{Hoang2019}
\bibinfo{author}{\bibfnamefont{B.-M.} \bibnamefont{{Hoang}}},
  \bibinfo{author}{\bibfnamefont{S.}~\bibnamefont{{Naoz}}},
  \bibinfo{author}{\bibfnamefont{B.}~\bibnamefont{{Kocsis}}},
  \bibinfo{author}{\bibfnamefont{W.~M.} \bibnamefont{{Farr}}},
  \bibnamefont{and} \bibinfo{author}{\bibfnamefont{J.}~\bibnamefont{{McIver}}},
  \bibinfo{journal}{Astrophys. J. Lett.} \textbf{\bibinfo{volume}{875}},
  \bibinfo{pages}{L31} (\bibinfo{year}{2019}).

\bibitem[{\citenamefont{Emami and Loeb}(2019)}]{Loeb2019}
\bibinfo{author}{\bibfnamefont{R.}~\bibnamefont{Emami}} \bibnamefont{and}
  \bibinfo{author}{\bibfnamefont{A.}~\bibnamefont{Loeb}},
  \bibinfo{journal}{arXiv:1910.04828[astro-ph.GA]}  (\bibinfo{year}{2019}).

\bibitem[{\citenamefont{Gupta et~al.}(2020)\citenamefont{Gupta, Suzuki, Okawa,
  and Maeda}}]{Gupta_2020}
\bibinfo{author}{\bibfnamefont{P.}~\bibnamefont{Gupta}},
  \bibinfo{author}{\bibfnamefont{H.}~\bibnamefont{Suzuki}},
  \bibinfo{author}{\bibfnamefont{H.}~\bibnamefont{Okawa}}, \bibnamefont{and}
  \bibinfo{author}{\bibfnamefont{K.}~\bibnamefont{Maeda}},
  \bibinfo{journal}{Phys. Rev. D} \textbf{\bibinfo{volume}{101}}
  (\bibinfo{year}{2020}).

\bibitem[{\citenamefont{Kuntz and Leyde}(2022)}]{kuntz2022transverse}
\bibinfo{author}{\bibfnamefont{A.}~\bibnamefont{Kuntz}} \bibnamefont{and}
  \bibinfo{author}{\bibfnamefont{K.}~\bibnamefont{Leyde}},
  \emph{\bibinfo{title}{Transverse doppler effect and parameter estimation of
  lisa three-body systems}} (\bibinfo{year}{2022}), \eprint{2212.09753}.

\bibitem[{\citenamefont{Naoz et~al.}(2013b)\citenamefont{Naoz, Kocsis, Loeb,
  and Yunes}}]{naoz13b}
\bibinfo{author}{\bibfnamefont{S.}~\bibnamefont{Naoz}},
  \bibinfo{author}{\bibfnamefont{B.}~\bibnamefont{Kocsis}},
  \bibinfo{author}{\bibfnamefont{A.}~\bibnamefont{Loeb}}, \bibnamefont{and}
  \bibinfo{author}{\bibfnamefont{N.}~\bibnamefont{Yunes}},
  \bibinfo{journal}{Astrophys. J.} \textbf{\bibinfo{volume}{773}},
  \bibinfo{pages}{187} (\bibinfo{year}{2013b}).

\bibitem[{\citenamefont{Naoz et~al.}(2012)\citenamefont{Naoz, Farr, and
  Rasio}}]{Naoz12}
\bibinfo{author}{\bibfnamefont{S.}~\bibnamefont{Naoz}},
  \bibinfo{author}{\bibfnamefont{W.}~\bibnamefont{Farr}}, \bibnamefont{and}
  \bibinfo{author}{\bibfnamefont{F.}~\bibnamefont{Rasio}},
  \bibinfo{journal}{Astrophys. J.} \textbf{\bibinfo{volume}{754}},
  \bibinfo{pages}{L36} (\bibinfo{year}{2012}).

\bibitem[{\citenamefont{Naoz}(2016)}]{Naoz16}
\bibinfo{author}{\bibfnamefont{S.}~\bibnamefont{Naoz}},
  \bibinfo{journal}{Annual Review of Astronomy and Astrophysics}
  \textbf{\bibinfo{volume}{54}}, \bibinfo{pages}{441} (\bibinfo{year}{2016}).

\bibitem[{\citenamefont{Naoz et~al.}(2019)\citenamefont{Naoz, Will,
  Ramirez-Ruiz, Hees, Ghez, and Do}}]{Naoz2020}
\bibinfo{author}{\bibfnamefont{S.}~\bibnamefont{Naoz}},
  \bibinfo{author}{\bibfnamefont{C.~M.} \bibnamefont{Will}},
  \bibinfo{author}{\bibfnamefont{E.}~\bibnamefont{Ramirez-Ruiz}},
  \bibinfo{author}{\bibfnamefont{A.}~\bibnamefont{Hees}},
  \bibinfo{author}{\bibfnamefont{A.~M.} \bibnamefont{Ghez}}, \bibnamefont{and}
  \bibinfo{author}{\bibfnamefont{T.}~\bibnamefont{Do}},
  \bibinfo{journal}{Astrophys. J.} \textbf{\bibinfo{volume}{888}},
  \bibinfo{pages}{L8} (\bibinfo{year}{2019}).

\bibitem[{\citenamefont{Teyssandier et~al.}(2013)\citenamefont{Teyssandier,
  Naoz, Lizarraga, and Rasio}}]{tey13}
\bibinfo{author}{\bibfnamefont{J.}~\bibnamefont{Teyssandier}},
  \bibinfo{author}{\bibfnamefont{S.}~\bibnamefont{Naoz}},
  \bibinfo{author}{\bibfnamefont{I.}~\bibnamefont{Lizarraga}},
  \bibnamefont{and} \bibinfo{author}{\bibfnamefont{F.~A.} \bibnamefont{Rasio}},
  \bibinfo{journal}{Astrophys. J.} \textbf{\bibinfo{volume}{779}},
  \bibinfo{pages}{169} (\bibinfo{year}{2013}).

\bibitem[{\citenamefont{Li et~al.}(2015)\citenamefont{Li, Naoz, Kocsis, and
  Loeb}}]{Li15}
\bibinfo{author}{\bibfnamefont{G.}~\bibnamefont{Li}},
  \bibinfo{author}{\bibfnamefont{S.}~\bibnamefont{Naoz}},
  \bibinfo{author}{\bibfnamefont{B.}~\bibnamefont{Kocsis}}, \bibnamefont{and}
  \bibinfo{author}{\bibfnamefont{A.}~\bibnamefont{Loeb}},
  \bibinfo{journal}{Mon. Not. R. Astron. Soc.} \textbf{\bibinfo{volume}{451}},
  \bibinfo{pages}{1341} (\bibinfo{year}{2015}).

\bibitem[{\citenamefont{Will}(2014{\natexlab{a}})}]{Will14a}
\bibinfo{author}{\bibfnamefont{C.}~\bibnamefont{Will}}, \bibinfo{journal}{Phys.
  Rev. D} \textbf{\bibinfo{volume}{89}}, \bibinfo{pages}{044043}
  (\bibinfo{year}{2014}{\natexlab{a}}).

\bibitem[{\citenamefont{Will}(2014{\natexlab{b}})}]{Will14b}
\bibinfo{author}{\bibfnamefont{C.}~\bibnamefont{Will}},
  \bibinfo{journal}{Class. Quantum Grav.} \textbf{\bibinfo{volume}{31}},
  \bibinfo{pages}{244001} (\bibinfo{year}{2014}{\natexlab{b}}).

\bibitem[{\citenamefont{Suzuki et~al.}(2019)\citenamefont{Suzuki, Gupta, Okawa,
  and Maeda}}]{Haruka2019}
\bibinfo{author}{\bibfnamefont{H.}~\bibnamefont{Suzuki}},
  \bibinfo{author}{\bibfnamefont{P.}~\bibnamefont{Gupta}},
  \bibinfo{author}{\bibfnamefont{H.}~\bibnamefont{Okawa}}, \bibnamefont{and}
  \bibinfo{author}{\bibfnamefont{K.}~\bibnamefont{Maeda}},
  \bibinfo{journal}{Mon. Not. R. Astron. Soc.:Letters}
  \textbf{\bibinfo{volume}{486, 1}} (\bibinfo{year}{2019}).

\bibitem[{\citenamefont{Suzuki et~al.}(2020)\citenamefont{Suzuki, Gupta, Okawa,
  and Maeda}}]{Suzuki:2020zbg}
\bibinfo{author}{\bibfnamefont{H.}~\bibnamefont{Suzuki}},
  \bibinfo{author}{\bibfnamefont{P.}~\bibnamefont{Gupta}},
  \bibinfo{author}{\bibfnamefont{H.}~\bibnamefont{Okawa}}, \bibnamefont{and}
  \bibinfo{author}{\bibfnamefont{K.}~\bibnamefont{Maeda}},
  \bibinfo{journal}{Mon. Not. Roy. Astron. Soc.}
  \textbf{\bibinfo{volume}{500}}, \bibinfo{pages}{1645} (\bibinfo{year}{2020}),
  \eprint{2006.11545}.

\bibitem[{\citenamefont{Liu et~al.}(2019)\citenamefont{Liu, Lai, and
  Wang}}]{Liu:2019tqr}
\bibinfo{author}{\bibfnamefont{B.}~\bibnamefont{Liu}},
  \bibinfo{author}{\bibfnamefont{D.}~\bibnamefont{Lai}}, \bibnamefont{and}
  \bibinfo{author}{\bibfnamefont{Y.-H.} \bibnamefont{Wang}},
  \bibinfo{journal}{Astrophys. J. Lett.} \textbf{\bibinfo{volume}{883}},
  \bibinfo{pages}{L7} (\bibinfo{year}{2019}), \eprint{1906.07726}.

\bibitem[{\citenamefont{Liu and Lai}(2022)}]{Liu:2021uam}
\bibinfo{author}{\bibfnamefont{B.}~\bibnamefont{Liu}} \bibnamefont{and}
  \bibinfo{author}{\bibfnamefont{D.}~\bibnamefont{Lai}},
  \bibinfo{journal}{Astrophys. J.} \textbf{\bibinfo{volume}{924}},
  \bibinfo{pages}{127} (\bibinfo{year}{2022}), \eprint{2105.02230}.

\bibitem[{\citenamefont{Lim and Rodriguez}(2020)}]{Lim2020}
\bibinfo{author}{\bibfnamefont{H.}~\bibnamefont{Lim}} \bibnamefont{and}
  \bibinfo{author}{\bibfnamefont{C.~L.} \bibnamefont{Rodriguez}},
  \bibinfo{journal}{Phys. Rev. D} \textbf{\bibinfo{volume}{102}},
  \bibinfo{pages}{064033} (\bibinfo{year}{2020}).

\bibitem[{\citenamefont{Fang and Huang}(2019)}]{Fang_2019a}
\bibinfo{author}{\bibfnamefont{Y.}~\bibnamefont{Fang}} \bibnamefont{and}
  \bibinfo{author}{\bibfnamefont{Q.-G.} \bibnamefont{Huang}},
  \bibinfo{journal}{Phys. Rev. D} \textbf{\bibinfo{volume}{99}}
  (\bibinfo{year}{2019}).

\bibitem[{\citenamefont{Fang et~al.}(2019)\citenamefont{Fang, Chen, and
  Huang}}]{Fang_2019b}
\bibinfo{author}{\bibfnamefont{Y.}~\bibnamefont{Fang}},
  \bibinfo{author}{\bibfnamefont{X.}~\bibnamefont{Chen}}, \bibnamefont{and}
  \bibinfo{author}{\bibfnamefont{Q.-G.} \bibnamefont{Huang}},
  \bibinfo{journal}{Astrophys. J.} \textbf{\bibinfo{volume}{887}},
  \bibinfo{pages}{210} (\bibinfo{year}{2019}).

\bibitem[{\citenamefont{{Manasse} and {Misner}}(1963)}]{1963JMP.....4..735M}
\bibinfo{author}{\bibfnamefont{F.~K.} \bibnamefont{{Manasse}}}
  \bibnamefont{and} \bibinfo{author}{\bibfnamefont{C.~W.}
  \bibnamefont{{Misner}}}, \bibinfo{journal}{Journal of Mathematical Physics}
  \textbf{\bibinfo{volume}{4}}, \bibinfo{pages}{735} (\bibinfo{year}{1963}).

\bibitem[{\citenamefont{Nesterov}(1999)}]{Nesterov_1999}
\bibinfo{author}{\bibfnamefont{A.~I.} \bibnamefont{Nesterov}},
  \bibinfo{journal}{Classical and Quantum Gravity}
  \textbf{\bibinfo{volume}{16}}, \bibinfo{pages}{465} (\bibinfo{year}{1999}).

\bibitem[{\citenamefont{Delva and Angonin}(2012)}]{Delva:2011abw}
\bibinfo{author}{\bibfnamefont{P.}~\bibnamefont{Delva}} \bibnamefont{and}
  \bibinfo{author}{\bibfnamefont{M.~C.} \bibnamefont{Angonin}},
  \bibinfo{journal}{Gen. Rel. Grav.} \textbf{\bibinfo{volume}{44}},
  \bibinfo{pages}{1} (\bibinfo{year}{2012}), \eprint{0901.4465}.

\bibitem[{\citenamefont{Banerjee et~al.}(2019)\citenamefont{Banerjee, Paul,
  Shaikh, and Sarkar}}]{Banerjee_2019}
\bibinfo{author}{\bibfnamefont{P.}~\bibnamefont{Banerjee}},
  \bibinfo{author}{\bibfnamefont{S.}~\bibnamefont{Paul}},
  \bibinfo{author}{\bibfnamefont{R.}~\bibnamefont{Shaikh}}, \bibnamefont{and}
  \bibinfo{author}{\bibfnamefont{T.}~\bibnamefont{Sarkar}},
  \bibinfo{journal}{Physics Letters B} \textbf{\bibinfo{volume}{795}},
  \bibinfo{pages}{29} (\bibinfo{year}{2019}).

\bibitem[{\citenamefont{Ishii et~al.}(2005)\citenamefont{Ishii, Shibata, and
  Mino}}]{PhysRevD.71.044017}
\bibinfo{author}{\bibfnamefont{M.}~\bibnamefont{Ishii}},
  \bibinfo{author}{\bibfnamefont{M.}~\bibnamefont{Shibata}}, \bibnamefont{and}
  \bibinfo{author}{\bibfnamefont{Y.}~\bibnamefont{Mino}},
  \bibinfo{journal}{Phys. Rev. D} \textbf{\bibinfo{volume}{71}},
  \bibinfo{pages}{044017} (\bibinfo{year}{2005}).

\bibitem[{\citenamefont{Cheng and Evans}(2013)}]{Cheng_2013}
\bibinfo{author}{\bibfnamefont{R.~M.} \bibnamefont{Cheng}} \bibnamefont{and}
  \bibinfo{author}{\bibfnamefont{C.~R.} \bibnamefont{Evans}},
  \bibinfo{journal}{Phys. Rev. D} \textbf{\bibinfo{volume}{87}}
  (\bibinfo{year}{2013}).

\bibitem[{\citenamefont{Kuntz et~al.}(2021)\citenamefont{Kuntz, Serra, and
  Trincherini}}]{Kuntz_2021}
\bibinfo{author}{\bibfnamefont{A.}~\bibnamefont{Kuntz}},
  \bibinfo{author}{\bibfnamefont{F.}~\bibnamefont{Serra}}, \bibnamefont{and}
  \bibinfo{author}{\bibfnamefont{E.}~\bibnamefont{Trincherini}},
  \bibinfo{journal}{Phys. Rev. D} \textbf{\bibinfo{volume}{104}}
  (\bibinfo{year}{2021}).

\bibitem[{\citenamefont{Gorbatsievich and Bobrik}(2010)}]{Gorbatsievich_Bobrik}
\bibinfo{author}{\bibfnamefont{A.}~\bibnamefont{Gorbatsievich}}
  \bibnamefont{and} \bibinfo{author}{\bibfnamefont{A.}~\bibnamefont{Bobrik}},
  \bibinfo{journal}{AIP Conference Proceedings}
  \textbf{\bibinfo{volume}{1205}}, \bibinfo{pages}{87} (\bibinfo{year}{2010}).

\bibitem[{\citenamefont{Chen and Zhang}(2022)}]{Chen_Zhang}
\bibinfo{author}{\bibfnamefont{X.}~\bibnamefont{Chen}} \bibnamefont{and}
  \bibinfo{author}{\bibfnamefont{Z.}~\bibnamefont{Zhang}},
  \bibinfo{journal}{Phys. Rev. D} \textbf{\bibinfo{volume}{106}}
  (\bibinfo{year}{2022}).

\bibitem[{\citenamefont{Camilloni et~al.}(2023)\citenamefont{Camilloni,
  Grignani, Harmark, Oliveri, Orselli, and Pica}}]{camilloni2023tidal}
\bibinfo{author}{\bibfnamefont{F.}~\bibnamefont{Camilloni}},
  \bibinfo{author}{\bibfnamefont{G.}~\bibnamefont{Grignani}},
  \bibinfo{author}{\bibfnamefont{T.}~\bibnamefont{Harmark}},
  \bibinfo{author}{\bibfnamefont{R.}~\bibnamefont{Oliveri}},
  \bibinfo{author}{\bibfnamefont{M.}~\bibnamefont{Orselli}}, \bibnamefont{and}
  \bibinfo{author}{\bibfnamefont{D.}~\bibnamefont{Pica}},
  \emph{\bibinfo{title}{Tidal deformations of a binary system induced by an
  external kerr black hole}} (\bibinfo{year}{2023}), \eprint{2301.04879}.

\bibitem[{\citenamefont{Maeda et~al.}(1972)\citenamefont{Maeda, Gupta, and
  Okawa}}]{Maeda:2023tao}
\bibinfo{author}{\bibfnamefont{K.}~\bibnamefont{Maeda}},
  \bibinfo{author}{\bibfnamefont{P.}~\bibnamefont{Gupta}}, \bibnamefont{and}
  \bibinfo{author}{\bibfnamefont{H.}~\bibnamefont{Okawa}},
  \bibinfo{journal}{Astrophys. J.} \textbf{\bibinfo{volume}{178}},
  \bibinfo{pages}{347} (\bibinfo{year}{1972}).

\bibitem[{\citenamefont{{Misner} et~al.}(1973)\citenamefont{{Misner}, {Thorne},
  and {Wheeler}}}]{Misner1973}
\bibinfo{author}{\bibfnamefont{C.~W.} \bibnamefont{{Misner}}},
  \bibinfo{author}{\bibfnamefont{K.~S.} \bibnamefont{{Thorne}}},
  \bibnamefont{and} \bibinfo{author}{\bibfnamefont{J.~A.}
  \bibnamefont{{Wheeler}}}, \emph{\bibinfo{title}{{Gravitation}}}
  (\bibinfo{publisher}{San Francisco: W.H.~Freeman and Co.},
  \bibinfo{year}{1973}).

\bibitem[{\citenamefont{Mathisson}(1937)}]{Mathisson1937}
\bibinfo{author}{\bibfnamefont{M.}~\bibnamefont{Mathisson}},
  \bibinfo{journal}{Acta Phys. Polon.} \textbf{\bibinfo{volume}{6}},
  \bibinfo{pages}{163} (\bibinfo{year}{1937}).

\bibitem[{\citenamefont{Papapetrou}(1951)}]{Papapetrou1951}
\bibinfo{author}{\bibfnamefont{A.}~\bibnamefont{Papapetrou}},
  \bibinfo{journal}{Proc. R. Soc. London A} \textbf{\bibinfo{volume}{209}},
  \bibinfo{pages}{248} (\bibinfo{year}{1951}).

\bibitem[{\citenamefont{Dixon}(1974)}]{Dixon1974}
\bibinfo{author}{\bibfnamefont{W.}~\bibnamefont{Dixon}},
  \bibinfo{journal}{Philos. Trans. R. Soc. London A}
  \textbf{\bibinfo{volume}{277}}, \bibinfo{pages}{59} (\bibinfo{year}{1974}).

\bibitem[{\citenamefont{Bardeen et~al.}(2023)\citenamefont{Bardeen, Press, and
  Teukolsky}}]{Bardeen1972}
\bibinfo{author}{\bibfnamefont{J.~M.} \bibnamefont{Bardeen}},
  \bibinfo{author}{\bibfnamefont{W.~H.} \bibnamefont{Press}}, \bibnamefont{and}
  \bibinfo{author}{\bibfnamefont{S.~A.} \bibnamefont{Teukolsky}}
  (\bibinfo{year}{2023}), \eprint{2303.16553}.

\bibitem[{\citenamefont{Will}(2019)}]{Will_2019}
\bibinfo{author}{\bibfnamefont{C.~M.} \bibnamefont{Will}},
  \bibinfo{journal}{Classical and Quantum Gravity}
  \textbf{\bibinfo{volume}{36}}, \bibinfo{pages}{195013}
  (\bibinfo{year}{2019}).

\bibitem[{\citenamefont{Mardling and Aarseth}(2001)}]{mardling2001tidal}
\bibinfo{author}{\bibfnamefont{R.~A.} \bibnamefont{Mardling}} \bibnamefont{and}
  \bibinfo{author}{\bibfnamefont{S.~J.} \bibnamefont{Aarseth}},
  \bibinfo{journal}{Monthly Notices of the Royal Astronomical Society}
  \textbf{\bibinfo{volume}{321}}, \bibinfo{pages}{398} (\bibinfo{year}{2001}).

\bibitem[{\citenamefont{Myll{\"a}ri et~al.}(2018)\citenamefont{Myll{\"a}ri,
  Valtonen, Pasechnik, and Mikkola}}]{myllari2018stability}
\bibinfo{author}{\bibfnamefont{A.}~\bibnamefont{Myll{\"a}ri}},
  \bibinfo{author}{\bibfnamefont{M.}~\bibnamefont{Valtonen}},
  \bibinfo{author}{\bibfnamefont{A.}~\bibnamefont{Pasechnik}},
  \bibnamefont{and} \bibinfo{author}{\bibfnamefont{S.}~\bibnamefont{Mikkola}},
  \bibinfo{journal}{Monthly Notices of the Royal Astronomical Society}
  \textbf{\bibinfo{volume}{476}}, \bibinfo{pages}{830} (\bibinfo{year}{2018}).

\bibitem[{\citenamefont{Maggiore}(2007)}]{MMbook}
\bibinfo{author}{\bibfnamefont{M.}~\bibnamefont{Maggiore}},
  \emph{\bibinfo{title}{{Gravitational Waves: Volume 1: Theory and
  Experiments}}} (\bibinfo{publisher}{Oxford University Press},
  \bibinfo{year}{2007}), ISBN \bibinfo{isbn}{9780198570745}.

\bibitem[{\citenamefont{Peters and Mathews}(1963)}]{Peters-Mathews}
\bibinfo{author}{\bibfnamefont{P.~C.} \bibnamefont{Peters}} \bibnamefont{and}
  \bibinfo{author}{\bibfnamefont{J.}~\bibnamefont{Mathews}},
  \bibinfo{journal}{Phys. Rev.} \textbf{\bibinfo{volume}{131}},
  \bibinfo{pages}{435} (\bibinfo{year}{1963}).

\bibitem[{\citenamefont{Cardoso et~al.}(2021)\citenamefont{Cardoso, Duque, and
  Khanna}}]{PhysRevD.103.L081501}
\bibinfo{author}{\bibfnamefont{V.}~\bibnamefont{Cardoso}},
  \bibinfo{author}{\bibfnamefont{F.}~\bibnamefont{Duque}}, \bibnamefont{and}
  \bibinfo{author}{\bibfnamefont{G.}~\bibnamefont{Khanna}},
  \bibinfo{journal}{Phys. Rev. D} \textbf{\bibinfo{volume}{103}},
  \bibinfo{pages}{L081501} (\bibinfo{year}{2021}).

\bibitem[{\citenamefont{Lithwick and Naoz}(2011)}]{Lithwick:2011hh}
\bibinfo{author}{\bibfnamefont{Y.}~\bibnamefont{Lithwick}} \bibnamefont{and}
  \bibinfo{author}{\bibfnamefont{S.}~\bibnamefont{Naoz}},
  \bibinfo{journal}{Astrophys. J.} \textbf{\bibinfo{volume}{742}},
  \bibinfo{pages}{94} (\bibinfo{year}{2011}), \eprint{1106.3329}.

\bibitem[{\citenamefont{Katz et~al.}(2011)\citenamefont{Katz, Dong, and
  Malhotra}}]{Katz:2011hn}
\bibinfo{author}{\bibfnamefont{B.}~\bibnamefont{Katz}},
  \bibinfo{author}{\bibfnamefont{S.}~\bibnamefont{Dong}}, \bibnamefont{and}
  \bibinfo{author}{\bibfnamefont{R.}~\bibnamefont{Malhotra}},
  \bibinfo{journal}{Phys. Rev. Lett.} \textbf{\bibinfo{volume}{107}},
  \bibinfo{pages}{181101} (\bibinfo{year}{2011}), \eprint{1106.3340}.

\bibitem[{\citenamefont{Naoz et~al.}(2013)\citenamefont{Naoz, Farr, Lithwick,
  Rasio, and Teyssandier}}]{Naoz:2011mb}
\bibinfo{author}{\bibfnamefont{S.}~\bibnamefont{Naoz}},
  \bibinfo{author}{\bibfnamefont{W.~M.} \bibnamefont{Farr}},
  \bibinfo{author}{\bibfnamefont{Y.}~\bibnamefont{Lithwick}},
  \bibinfo{author}{\bibfnamefont{F.~A.} \bibnamefont{Rasio}}, \bibnamefont{and}
  \bibinfo{author}{\bibfnamefont{J.}~\bibnamefont{Teyssandier}},
  \bibinfo{journal}{Mon. Not. Roy. Astron. Soc.}
  \textbf{\bibinfo{volume}{431}}, \bibinfo{pages}{2155} (\bibinfo{year}{2013}),
  \eprint{1107.2414}.

\bibitem[{\citenamefont{{Li} et~al.}(2014)\citenamefont{{Li}, {Naoz}, {Kocsis},
  and {Loeb}}}]{Li2014ApJ}
\bibinfo{author}{\bibfnamefont{G.}~\bibnamefont{{Li}}},
  \bibinfo{author}{\bibfnamefont{S.}~\bibnamefont{{Naoz}}},
  \bibinfo{author}{\bibfnamefont{B.}~\bibnamefont{{Kocsis}}}, \bibnamefont{and}
  \bibinfo{author}{\bibfnamefont{A.}~\bibnamefont{{Loeb}}},
  \bibinfo{journal}{\apj} \textbf{\bibinfo{volume}{785}}, \bibinfo{eid}{116}
  (\bibinfo{year}{2014}), \eprint{1310.6044}.

\bibitem[{\citenamefont{{Liu} et~al.}(2015)\citenamefont{{Liu}, {Mu{\~n}oz},
  and {Lai}}}]{Liu2015MNRAS}
\bibinfo{author}{\bibfnamefont{B.}~\bibnamefont{{Liu}}},
  \bibinfo{author}{\bibfnamefont{D.~J.} \bibnamefont{{Mu{\~n}oz}}},
  \bibnamefont{and} \bibinfo{author}{\bibfnamefont{D.}~\bibnamefont{{Lai}}},
  \bibinfo{journal}{mnras} \textbf{\bibinfo{volume}{447}}, \bibinfo{pages}{747}
  (\bibinfo{year}{2015}), \eprint{1409.6717}.

\end{thebibliography}

\end{document}